\documentstyle[12pt]{article}

\newcommand{\eps}{\varepsilon}

\def\N{{\rm I\kern-.1567em N}}

\def\R{{\rm I\kern-.1567em R}}
\def\C{{\rm C\kern-4.7pt
\vrule height 7.7pt width 0.4pt depth -0.5pt \phantom {.}}}
\def\Z{{\sf Z\kern-4.5pt Z}}

\def\eps{\varepsilon}

\begin{document}

\newtheorem{theorem}{Theorem}[section]
\renewcommand{\thetheorem}{\arabic{section}.\arabic{theorem}}
\newtheorem{definition}[theorem]{Definition}
\newtheorem{deflem}[theorem]{Definition and Lemma}
\newtheorem{lemma}[theorem]{Lemma}
\newtheorem{example}[theorem]{Example}
\newtheorem{remark}[theorem]{Remark}
\newtheorem{remarks}[theorem]{Remarks}
\newtheorem{cor}[theorem]{Corollary}
\newtheorem{pro}[theorem]{Proposition}
\newtheorem{proposition}[theorem]{Proposition}
 
\renewcommand{\theequation}{\thesection.\arabic{equation}}

\begin{titlepage}
\vspace*{3cm}
\begin{center}
{\Large\bf Post-Coulombian Dynamics at Order $\mathbf{1.5}$}\\
\vspace{3cm}
{\large Markus Kunze}\medskip\\
        Zentrum Mathematik, TU M\"unchen \\
        D-80290 M\"unchen, Germany\\
        email: mkunze@mathematik.tu-muenchen.de\bigskip\\
{\large Herbert Spohn}\medskip\\
        Zentrum Mathematik and Physik Department, TU M\"unchen\\
        D-80290 M\"unchen, Germany\\
        email: spohn@mathematik.tu-muenchen.de\bigskip\\
\end{center}
\date{\today}
\vspace{3cm}

\begin{abstract}\noindent We study the dynamics of $N$ charges
interacting with the Maxwell field. If their initial velocities are
small compared to the velocity of light, $c$, then in lowest order
their motion is governed by the static Coulomb Lagrangian. We
investigate higher order corrections with an explicit control on the
error terms. The Darwin correction, order $|v/c|^{2}$, has been proved
previously. In this contribution we obtain the dissipative corrections
due to radiation damping, which are of order $|v/c|^{3}$ relative to
the Coulomb dynamics. If all particles have the same charge-to-mass ratio,
the dissipation would vanish at that order.
\end{abstract}

\end{titlepage}


\section{Introduction}

Experimental general relativity is at the edge of taking the lead
in ultraprecision, surpassing even the famous measurements
of the anomalous magnetic moment of the electron \cite{brown}.
The best-studied test case is provided by the Hulse-Taylor binary pulsar
PSR 1913+16, which consists of two neutron stars,
each roughly of one solar mass and with a radius of $10$ km.
The stars are a distance $2.6 \times 10^{6}$ km apart and revolve around their
common center of mass with a period of about $7$h $45$min at a speed of
$|v/c|\cong 10^{-3}$ (with $c$ the velocity of light). One of the stars rotates around
its own body axis and emits a precisely pulsed radio wave which can be detected
on earth, thus providing an indirect measurement of the orbital motion
\cite{hulse,taylor}.

On the theoretical side one has to solve Einstein's equations with matter
such that the mass is well-concentrated in the two neutron stars. Since
$|v/c|\ll 1$, a natural strategy is to expand the metric
in powers of $|v/c|$. The zero order contribution corresponds to the
non-relativistic limit, where the stars move on the Kepler orbits of the
Newtonian theory of gravity. Therefore higher order corrections are commonly
called ``post-Newtonian'' and they are counted in powers of $|v/c|^2$.
To first post-Newtonian order, $|v/c|^2$, the motion of the binary pulsar
is governed by additional velocity dependent forces, and this order is followed
by corrections, order $|v/c|^4$, which are still of conservative
(Hamiltonian) nature. Damping through the emission of gravitational waves
appears at order $|v/c|^5$, which roughly means a correction of order
$10^{-15}$  relative to the Kepler orbit. The predictions of the theory
and the observed minute shrinking of the orbit agree within $0.3 \%$.
It is even claimed that improved experimental devices would yield a precision
of order $|v/c|^{11}$, see \cite{will}. At present theoretical studies
of $3.5$ post-Newtonian dynamics are available, cf.~\cite{dam-ja-sch}
and the references therein.

{}From a mathematical point of view one would like to establish that the true
orbit of the neutron stars, as governed by Einstein's equations,
is well-approximated by the solution of the effective second order differential
equation at the appropriate post-Newtonian order. While order zero has been
accomplished in \cite{rendall} for asymptotically flat geometry,
any further progress seems difficult at this point. In fact, since
in the relativistic context there is no sufficiently general theory
for the existence of solutions, even the notion of a true orbit is somewhat vague.
Therefore in the present paper we propose to investigate a very similar,
but considerably less involved problem where the neutron stars are replaced
by charges and the gravitational field is replaced by the Maxwell field.
Of course the physics  is then completely different, but as a theoretical problem
most qualitative features are maintained with the welcome simplification that the
equations for the electromagnetic field are linear, in contrast to Einstein's equations.
Moreover the matter field can be modeled through a rigid charge distribution.
As the only drawback, there seems to exist no obvious experimental realization
of the model. For a single charge in a Penning trap the radiation damping is measured
through the shrinking amplitude of the cyclotron mode
\cite{brown,spohn}. On the other hand, two charges of opposite sign
will rapidly form a neutral atom which is governed by the laws of quantum mechanics.
Despite this fact we believe that our approach will improve on the understanding
of how radiation damping emerges from a fully microscopic Hamiltonian system
for the interaction of matter with a wave field.

The dynamical system under study consists of Maxwell's equations
for the electromagnetic field,
\begin{eqnarray}
   &&  c^{-1}\frac{\partial}{\partial t}  B(x, t)= -\nabla \wedge E(x, t),
   \quad c^{-1}\frac{\partial}{\partial t} E(x, t)=\nabla\wedge B(x, t)
   -c^{-1} j(x, t),  \label{system0int} \\[1ex]
   && \nabla\cdot E(x, t)=\rho(x, t),\quad \nabla \cdot B(x, t)=0, \label{system1int}
\end{eqnarray}
and the Lorentz force equations for the charges,
\begin{equation}\label{lofo1int}
   \frac{d}{dt}\Big(m_{{\rm b}\alpha}\gamma_\alpha v_\alpha(t)\Big)
   =e_\alpha\Big(E_{\varphi}(q_\alpha(t),t)+c^{-1}\,v_\alpha(t)\wedge
   B_{\varphi}(q_\alpha(t), t)\Big),
   \quad\alpha=1, \ldots, N.
\end{equation}
Our above example corresponds to two charges, but a general number $N\ge 1$
will make little difference, except that some solutions of the Coulomb dynamics
may fail to exist globally in time. Concerning our notation,
$E$ is the electric and $B$ the magnetic field. Moreover $q_\alpha(t)$
denotes the position of particle $\alpha$, and its velocity
is $v_\alpha(t)=\dot{q}_\alpha(t)$. The particle has (bare) mass
$m_{{\rm b}\alpha}$, charge $e_\alpha$, and a relativistic kinetic energy
with $\gamma_\alpha={(1-(v_\alpha/c)^2)}^{-1/2}$. Each particle
carries a rigid charge distribution as given by the form factor
$\varphi$, which we assume to be smooth, radial, compactly supported,
and normalized, i.e.,
$$ 0\le\varphi\in C_0^\infty(\R^3)\,,\quad\varphi(x)=\varphi_r(|x|)\,,\quad
   \varphi(x)=0\,\,\,\,\,\mbox{for}\,\,\,\,\,|x|\ge R_{\varphi}\,,
   \quad\int d^3x\,\varphi(x)=1\,. \eqno{(C)} $$
Then the charge and current densities generated by the charges are given by
\begin{equation}\label{rho-j-def}
   \rho(x, t)=\sum_{\alpha=1}^N e_\alpha\varphi(x-q_\alpha(t))
   \quad\mbox{and}\quad
   j(x, t)=\sum_{\alpha=1}^N e_\alpha\varphi(x-q_\alpha(t))v_\alpha(t),
\end{equation}
which determine the source terms in  Maxwell equations
and thereby couple (\ref{system0int}), (\ref{system1int}), and
(\ref{lofo1int}). The functions $E_{\varphi}$ and $B_{\varphi}$ in (\ref{lofo1int})
are the fields smeared out by $\varphi$, i.e., we introduce $E_{\varphi}(x,t)
=\int \varphi(x-x')E(x',t)d^{3}x'$ and $B_{\varphi}(x,t)
=\int\varphi(x-x')B(x',t)d^{3}x'$. The coupled equations (\ref{system0int}),
(\ref{system1int}), (\ref{lofo1int}),  and (\ref{rho-j-def})
define the Abraham model.

Another variant of interest is to have large $N$ and to use a fluid-like
description for the particles in terms of a distribution function
$f_{\alpha}(q,v,t)$ for species $\alpha$, which is then governed by the
Liouville equation corresponding to (\ref{lofo1int}). Together with
(\ref{system0int}), (\ref{system1int}), and the continuum analogue
of (\ref{rho-j-def}) one arrives at the Vlasov-Maxwell system. In the limit
of small velocities (corresponding to $c\to\infty$) this system
is well approximated by the Vlasov-Poisson equations \cite{Degond,Schaeffer}.
Corrections due to radiation damping are studied in \cite{MK-AR}.

For charges interacting with the radiation field, as modeled by
Abraham, the zero order effective dynamics is just the Coulomb dynamics,
with the Darwin term appearing as the first post-Coulombian ($1$PC) correction.
Both are conveniently summarized through the Lagrangian function
\begin{eqnarray}\label{LCD}
   {\cal L}_{{\rm D}}(r, u) & = & \sum_{\alpha=1}^N \Big(\frac{1}{2}
   m_\alpha u_\alpha^2+\frac{1}{8c^2}\,m_\alpha^\ast
   u_\alpha^4\Big)-\frac{1}{2}
   \sum_{\stackrel{\alpha, \beta=1}{\alpha\neq\beta}}^N
   \frac{e_\alpha e_\beta}{4\pi |r_\alpha-r_\beta|}\nonumber
   \\ & & +\,\frac{1}{4c^2}\sum_{\stackrel{\alpha, \beta=1}{\alpha\neq\beta}}^N
   \frac{e_\alpha e_\beta}{4\pi |r_\alpha-r_\beta|}
   \Big(u_\alpha\cdot u_\beta + {|r_\alpha-r_\beta|}^{-2}
   (u_\alpha\cdot [r_\alpha-r_\beta])(u_\beta\cdot [r_\alpha-r_\beta])\Big).
   \qquad
\end{eqnarray}
Here $r=(r_1, \ldots, r_N)$ and $u=(u_1, \ldots, u_N)$
denote position and velocity of the particles in the approximating
system, to notationally distinguish them from the ``true''
positions and velocities as governed by (\ref{system0int}), (\ref{system1int}),
(\ref{lofo1int}), and (\ref{rho-j-def}).

In (\ref{LCD}) the kinetic energy is necessarily expanded in $|v/c|$.
For a relativistic particle with rest mass $m_0$ we have the kinetic
energy $T(v) = m_0(1- \gamma^{-1})\simeq m_0\Big(\frac{1}{2}(v/c)^{2}
+\frac{1}{8} (v/c)^{4}+{\cal O}((v/c)^6)\Big)$. Within the Abraham model
the situation is slightly more complicated. Firstly, the comoving Coulomb field
carries some inertia and the bare kinetic energy is renormalized to
an effective kinetic energy $T_{{\rm eff}}$. In addition, since the
charge distribution is rigid in the given rest frame, $T_{{\rm eff}}$
cannot be of relativistic form. There is a recent proposal for a relativistic model
of extended charges coupled to the radiation field, which necessarily also includes
their inner rotation, cf.~\cite{miki}. If one would carry out the small velocity expansion
for such a fully relativistic model, $T_{{\rm eff}}$ has to be relativistic.
Instead, for the Abraham model one obtains
\begin{equation}\label{me-def}
   m_\alpha=m_{{\rm b}\alpha}+\frac{4}{3}\,e_\alpha^2
   m_{{\rm e}}\quad\mbox{and}\quad m_\alpha^\ast=m_{{\rm b}\alpha}+\frac{16}{15}\,
   e_\alpha^2 m_{{\rm e}},\quad\mbox{with}\quad m_{{\rm e}}
   =\frac{1}{2c^{2}}\int d^3k\,{|\hat{\varphi}(k)|}^2 k^{-2}.
\end{equation}

The task ahead is to improve the Lagrangian effective equations of motion
\begin{equation}\label{ottf}
   \frac{d}{dt}\bigg(\frac{\partial {\cal L}_{{\rm D}}}
   {\partial u_\alpha}\bigg)=\frac{\partial {\cal L}_{{\rm D}}}
   {\partial r_\alpha},\quad\alpha=1, \ldots, N,
\end{equation}
associated to ${\cal L}_{{\rm D}}$ from $1$PC to $1.5$PC.
This cannot be a mere addition of extra terms to ${\cal L}_{{\rm D}}$,
since at $1.5$PC the charges loose energy through dipole radiation,
which must be reflected by dissipative contributions appearing in the dynamics.
A formal expansion in $|v/c|$, some details of which are explained
in Section \ref{main-sect} below, yields the next-order approximate equations of motion
\begin{equation}\label{LD-dyn-eff}
   \frac{d}{dt}\bigg(\frac{\partial {\cal L}_{{\rm D}}}
   {\partial u_\alpha}\bigg)=\frac{\partial {\cal L}_{{\rm D}}}
   {\partial r_\alpha}+\frac{e_\alpha}{6\pi c^3}\sum^N_{\beta =1}
   e_{\beta}\ddot{u}_{\beta},\quad\alpha=1, \ldots, N.
\end{equation}

The presence of the $\ddot{u}_{\beta}$-terms
means that the dimension of the phase space is increased from $6N$
for (\ref{ottf}) to $9N$ for (\ref{LD-dyn-eff}).
As in the case of a single particle, the equations of motion at $1.5$PC
are of third order and admit unphysical runaway solutions with velocities
which grow exponentially fast in time. To obtain the effective dynamics
free from runaway solutions, the standard (formal) practice,
also used in the analogous general relativity setting, is to regard
the $\ddot{u}_{\beta}$-terms in (\ref{LD-dyn-eff}) as a small perturbation.
Therefore one may think of differentiating (the explicit form of)
Eq.~(\ref{ottf}) with respect to $t$ and of substituting the resulting expression
for $\ddot{u}_{\beta}$ back into (\ref{LD-dyn-eff}), at the same time dropping
all higher than second derivatives of the $r_{\alpha}$'s. The resulting equation is
\begin{eqnarray}\label{vogar}
   \frac{d}{dt}\bigg(\frac{\partial {\cal L}_{{\rm D}}}
   {\partial u_\alpha}\bigg) & = & \frac{\partial {\cal L}_{{\rm D}}}
   {\partial r_\alpha}+\frac{e_\alpha}{12\pi c^3}
   \sum_{\stackrel{\beta, \beta'=1}{\beta\neq\beta'}}^N\frac{e_\beta e_{\beta'}}{4\pi}
   \bigg(\frac{e_\beta}{m_\beta}-\frac{e_{\beta'}}{m_{\beta'}}\bigg)
   \nonumber \\ & & \hspace{5em}
   \times\bigg[\frac{1}{|r_{\beta}-r_{\beta'}|^3}(u_\beta-u_{\beta'})
   -\frac{3(r_{\beta}-r_{\beta'})\cdot (u_\beta-u_{\beta'})}
   {|r_{\beta}-r_{\beta'}|^5}\,(r_{\beta}-r_{\beta'})\bigg]\qquad
\end{eqnarray}
for $\alpha = 1,\ldots, N$. Thus our goal is to prove that
the true solution $q_\alpha$, $v_\alpha$ is well-approximated by a solution
of (\ref{vogar}), up to errors of order $|v/c|^4$.
We note in passing that (\ref{vogar}) implies there is no radiation damping
at $1.5$PC in case the charge-to-mass ratios $e_\beta/m_\beta$ are independent
of $\beta$.

The substitution described above looks like a magic trick, since there is
no a priori reason to expect that the terms dropped in the end are really
of higher order. However, as has been recognized for some time, cf.~\cite{MK-S-1,MK-S-2},
the substitution can be justified rigorously in the framework of singular (geometric)
perturbation theory. The basic observation is that in (\ref{LD-dyn-eff}),
transformed to the appropriate dimensionless scale,
the highest derivatives $\ddot{u}_{\beta}$ carry a small prefactor.
The solution flow then admits for a repulsive manifold which can be addressed
as center-like, since it contains the true effective dynamics. On this center manifold
there is slow motion corresponding to the physical relevant solutions.
For initial conditions away from the center manifold the solution trajectory
runs off to infinity exponentially fast. Hence (\ref{vogar}) has to be
interpreted as the lowest order approximation to the motion on the center manifold.
There is one caveat, however: the matrix of coefficients for the $\ddot{u}$-term,
i.e., the map $(\ddot{u}_1, \ldots, \ddot{u}_N)\mapsto ((e_\alpha/6\pi c^3)
\sum^N_{\beta=1}e_{\beta}\ddot{u}_{\beta})_{1\le\alpha\le N}$,
is not invertible. Such a case does not seem to be covered by the
standard geometric theory of singular perturbations, and we therefore had to supply
the missing pieces.

In \cite{Nteil} we proved that the motion according to the Darwin Lagrangian
well approximates the true orbit to this order, provided that the
initial electromagnetic field minimizes the field energy at $\rho$, $j$
given through the initial positions and velocities of the charges,
cf.~(\ref{ini-bed}) below. As pointed out by V.~Imaikin we handled the initial
slip somewhat loosely. This is no problem at order $0$PC, however at order
$1$PC the desired precision requires us in fact to adjust the data
for the charges at some later time (rather than $t=0$) which is short
on the Coulomb time scale but long on the microscopic time scale,
but still the bounds on the time-derivatives of the solution are needed starting
from $t=0$. We use the occasion to supply a complete proof, see Lemma \ref{q3esti}
and Lemma \ref{q4esti} in Section \ref{main-sect}.

On a formal level, without control of the error term, $2$PC has been
computed by Damour and Sch\"afer \cite{dam-sch}.
Their starting point is the Wheeler-Feynman action, truncated at $1/c^4$, with
point-like charges. The equations of motion are for the charges only
and derive from a higher order Lagrangian. Working out the same order for the
Abraham model, which non-rigorously could be handled with little extra effort,
one would obtain additional terms reflecting the finite size
of the charge distribution.

These formal expansions assume implicitly that for $1\le\alpha\le N$
\begin{equation}\label{110}
   |v_\alpha(t)-u_\alpha(t)|\le {\rm const.}\,|v/c|^5
\end{equation}
at $2$PC, say. Here $v_\alpha$ is the true solution and $u_\alpha$ is the
$2$PC approximate solution, with $|v|$ denoting some average initial
velocity. The approximation is supposed to be valid over many Coulomb periods.
There are two difficulties associated with (\ref{110}). Firstly, one has to
specify for which initial conditions this estimate holds, and secondly $v_\alpha(t)$
depends as well on the initial data for the Maxwell field. Does (\ref{110})
mean that for generic initial data of the Abraham model there is {\em some} solution
of the comparison dynamics such that (\ref{110}) holds? In fact we are not able
to come even close to (\ref{110}). At $1.5$PC we will roughly prove a
precise estimate in the radial direction of the form
\[ |v_\alpha(t)^2-u_\alpha(t)^2|\le {\rm const.}\,|v/c|^4, \]
for $\alpha = 1,\ldots, N$, cf.~(\ref{H-diff}) below,
whereas for the phase we only have
\[ |v_\alpha(t)-u_\alpha(t)|\le {\rm const.}\,|v/c|^3. \]
Thus on the shell of constant kinetic energy our rigorous estimate
is not improved as compared to $1$PC. It might be the case that our method
is not powerful enough to distinguish such fine details. But even
on a theoretical physics level it would be of interest to better understand
the precise claim hidden behind (\ref{110}).

In the present paper we will establish the $1.5$PC approximation, where the
main effort goes into a control of the error terms. From our analysis a rather
general pattern (presumably valid to any order) emerges. (i) At each order
higher derivatives in $t$ and higher powers of $|r_i-r_j|^{-1}$ do appear,
as dictated by their dimension. For example, at $1.5$PC a term like $\ddot{u}_\alpha$
is dimensionally admitted. Of course each term comes with a prefactor
which has to be computed from the Abraham model and which also might vanish,
usually because of symmetry. (ii) From step (i) in the effective equations
of motion necessarily higher
time derivatives come up  and the phase space
for the approximate dynamics is of dimension larger than $6N$.
However, since in the
appropriate dimensionless form the higher
time derivatives carry a small prefactor, the solution flow
has a $6N$ dimensional repulsive center manifold. The motion on this center
manifold can be approximated by a second order equation which is the desired comparison
dynamics at the given PC approximation.


\setcounter{equation}{0}

\section{Scales}

As in any multiscale problem we have to first identify the relevant scale
parameter. The dynamical equations (\ref{system0int}), (\ref{system1int}),
(\ref{lofo1int}), and (\ref{rho-j-def}) are written on the microscopic scale
for which distance is measured in units of $R_\varphi$, the radius of the
charge distribution from $(C)$, and time is measured in units
of $t_\varphi=R_\varphi/c$. On this scale we require the particles to be far
apart initially, which means
\begin{equation}\label{bopi}
   |q_\alpha(0)-q_\beta(0)|\cong\eps^{-1}R_\varphi\quad (\alpha\neq\beta),
\end{equation}
and this defines the dimensionless parameter $\eps>0$, $\eps\ll 1$.
It will be part of the proof to verify that (\ref{bopi}) is preserved
in the course of time. In addition we require $|v_\alpha(0)|/c$ to be small.
To find the right order in $\eps$, let us for the moment assume $|v_\alpha(0)|
=\eps^{\gamma}c$ with $\gamma>0$ to be determined. To see the changes due to
self-action and mutual interaction we have to follow the dynamics over long times
of some order $t=\eps^{-\delta}t_\varphi$. Over this time span we obtain a change
in position $\Delta q=\eps^{\gamma}c\,\eps^{-\delta}t_\varphi=\eps^{\gamma-\delta}
R_\varphi$, and this should equal $\Delta q=\eps^{-1}R_\varphi$ in view of (\ref{bopi}).
Anticipating that the force is proportional to the squared inverse distance,
multiplying with the dimensionally right factor we find for the change in velocity
that $\Delta v=(R_\varphi^3/t_\varphi^2){(\Delta q)}^{-2} t=\eps^{2-\delta}c$,
which should be of order $\eps^{\gamma}c$, by assumption.
Solving for $\gamma$ and $\delta$ we find $\gamma=\frac{1}{2}$
as well as $\delta=\frac{3}{2}$. Thus we require that initially
\[ |v_\alpha(0)|\cong\sqrt{\eps}c, \]
and we have to consider times of order
\begin{equation}\label{tispa}
   t\cong\eps^{-3/2}t_\varphi\,.
\end{equation}
Again it will be part of our proof to see that over the time span
(\ref{tispa}) the velocities remain of order $\sqrt{\eps}c$. In
passing, note that for the Hulse-Taylor pulsar $\eps\cong 10^{-6}$.

Having settled the initial conditions for the charges we turn
to the electromagnetic field. The initial field is assumed to have finite energy,
i.e.,
\[ \frac{1}{2}\int \Big({|E(x, 0)|}^2+{|B(x, 0)|}^2\Big)\,d^3x<\infty\,. \]
Our picture is that in the neighborhood of the charges through radiation
the electromagnetic field very rapidly (in a time of order $\eps^{-1}t_\varphi$)
reaches a state of minimal energy at the given constraint due to the presence of the
charges, the positions of which have been changing only little on the Coulomb scale
during this time span. In \cite{Ko-Spo} such a behavior
has been established for somewhat simpler situations.
Here we concentrate on longer times and merely assume an initial
electromagnetic field of low energy. For a charge at constant velocity $v$ the comoving
electric and magnetic fields are
\begin{equation}\label{EBv-def}
   E_v(x)=-\nabla\phi_v(x)+c^{-2}(v\cdot\nabla\phi_v(x))v
   \quad\mbox{and}\quad B_v(x)=-c^{-1}\,v\wedge\nabla\phi_v(x),
\end{equation}
where $\phi_v$ is defined through its Fourier transform
\begin{equation}\label{phiv-def}
   \hat{\phi}_v(k)=e\,\hat{\varphi}(k)/[k^2-c^{-2}{(k\cdot v)}^2],
\end{equation}
with $e=e_\alpha$ for $v=v_\alpha$. Such fields we call charge solitons
(centered at zero with velocity $v$). If $q^0=(q_1^0, \ldots, q_N^0)$
denotes the initial positions and $v^0=(v_1^0, \ldots, v_N^0)$ the initial velocities,
then we choose the initial fields as linear superposition of the form
\begin{equation}\label{ini-bed}
   E(x, 0)=E^0(x)=\sum_{\alpha=1}^N E_{v_\alpha^0}(x-q_\alpha^0)
   \quad\mbox{and}\quad
   B(x, 0)=B^0(x)=\sum_{\alpha=1}^N B_{v_\alpha^0}(x-q_\alpha^0).
\end{equation}
We can think of these fields as generated by the charges which have been forced
to move freely as $q_\alpha(t)=q_\alpha^0+v_\alpha^0 t$ for $-\infty<t\le 0$.

To summarize, we consider a situation where the charges are far apart
(on the scale $R_{\varphi}$) and move slowly (on the scale $c$) over
long times (on the scale $t_{\varphi}$). On a mathematical level this
means that the initial conditions and the time span under consideration
are $\eps$-dependent. As a consequence, also in the comparison dynamics
the initial conditions are $\eps$-dependent. Here $\eps$ is merely
a convenient device to order terms according to their magnitude. In
particular, $n$PC translates to the order $\eps^{2+n}$. An equivalent
approach, which will not be used here, would be to rewrite the equations of motion
on the Coulomb scale. Then the initial conditions are approximately
$\eps$-independent. However, the evolution equations pick up some
$\eps$-dependence, except for the pure Coulomb dynamics which is scale
invariant. Through the transformation to the Coulomb scale the order would be
reduced by a factor $\eps^2$, and $n$PC would correspond to order $\eps^n$.


\setcounter{equation}{0}

\section{Main results}
\label{main-sect}

In the following we use units for which $c=1$. We first summarize the dynamics
under consideration. The Maxwell equations are
\begin{equation}\label{system0}
   \frac{\partial}{\partial t} B(x, t)= -\nabla \wedge E(x, t), \quad
   \frac{\partial}{\partial t} E(x, t)= \nabla \wedge B(x, t)
   -\sum_{\alpha=1}^N \rho_\alpha(x-q_\alpha(t))v_\alpha(t),
\end{equation}
with the constraints
\begin{equation}\label{system1}
   \nabla\cdot E(x, t)=\sum_{\alpha=1}^N \rho_\alpha(x-q_\alpha(t)),
   \quad \nabla \cdot B(x, t)=0.
\end{equation}
Here we have introduced  the shorthand notation
\[ \rho_\alpha=e_\alpha\varphi, \]
and $\varphi$ is assumed to satisfy $(C)$. The Lorentz force equation is
\begin{equation}\label{system2}
   \frac{d}{dt}\Big(m_{{\rm b}\alpha}\gamma_\alpha v_\alpha(t)\Big)
   =\int d^3x\,\rho_\alpha (x-q_\alpha(t))
   \Big[E(x,t)+v_\alpha(t)\wedge B(x, t)\Big],\quad 1\le\alpha\le N,
\end{equation}
where $\gamma_\alpha={(1-v_\alpha^2)}^{-1/2}$. The initial conditions
for the electric and magnetic field are given by (\ref{ini-bed}),
and for the initial positions $q_\alpha^0=q_\alpha(0)$ resp.~the initial
velocities $v_\alpha(0)=v_\alpha^0$ we require
\begin{equation}\label{q-ini2}
   C_1\eps^{-1}\le |q_\alpha^0-q_\beta^0|\le C_2\eps^{-1}
   \quad (\alpha\neq\beta),
\end{equation}
for some constants $C_1, C_2>0$, as well as
\begin{equation}\label{v-ini2}
   |v_\alpha^0|\le C_3\sqrt{\eps}
\end{equation}
with $C_3>0$.

To order $0$PC the comparison system is governed by the Coulomb dynamics
\begin{equation}\label{C-dyn}
   \frac{d}{dt}\bigg(\frac{\partial {\cal L}_{{\rm C}}}
   {\partial u_\alpha}\bigg)=\frac{\partial {\cal L}_{{\rm C}}}
   {\partial r_\alpha},\quad\alpha=1, \ldots, N,
\end{equation}
with Coulomb Lagrange function
\[ {\cal L}_{{\rm C}}(r, u)=\sum_{\alpha=1}^N
   \frac{1}{2}m_\alpha u_\alpha^2
   -\frac{1}{2}\sum_{\stackrel{\alpha, \beta=1}{\alpha\neq\beta}}^N
   \frac{e_\alpha e_\beta}{4\pi |r_\alpha-r_\beta|}, \]
where $r=(r_1, \ldots, r_N)$ and $u=(u_1, \ldots, u_N)$.
Because of the Coulomb singularity the solution to (\ref{C-dyn}) may exist
only for a finite time, either because two particles collide or since one particle
is being expelled to infinity. To formalize this, for given data $(\bar{r}_\alpha^0,
\bar{u}_\alpha^0)$ we denote $\tau_{{\rm C}}\in ]0, \infty]$ the first time
for which either $\lim_{t\to\tau_{{\rm C}}^-}|\bar{r}_\alpha(t)-\bar{r}_\beta(t)|=0$
for some $\alpha\neq\beta$ or $\lim_{t\to\tau_{{\rm C}}^-}|\bar{r}_\alpha(t)|=\infty$
for some $\alpha$ holds for the corresponding solution $(\bar{r}(t), \bar{u}(t))$
of (\ref{C-dyn}) with data $(\bar{r}_\alpha^0, \bar{u}_\alpha^0)$.

Our first step is to conclude from (\ref{q-ini2})
and (\ref{v-ini2}) that if under the Coulomb dynamics (\ref{C-dyn})
there is no collision/expulsion, the same holds for the full system.

\begin{lemma}\label{C-diff-lem} Let the initial data for the Abraham
model satisfy (\ref{q-ini2}), (\ref{v-ini2}), and (\ref{ini-bed}).
We introduce
\begin{equation}\label{comp-data}
   \bar{r}_\alpha^0=\eps q_\alpha^0\quad\mbox{and}\quad
   \bar{u}_\alpha^0=\eps^{-1/2} v_\alpha^0,\quad\alpha=1, \ldots, N,
\end{equation}
as data for (\ref{C-dyn}). Moreover we fix $\delta_0\in ]0, \tau_{{\rm C}}[$ and $T_0>0$.
Then there exists a constant $C_\ast>0$ such that
\begin{equation}\label{low-bound}
   C_\ast\eps^{-1}\le \inf_{t\in [0, \min\{\tau_{{\rm C}}-\delta_0, T_0\}
   \eps^{-3/2}]}|q_\alpha(t)-q_\beta(t)|\quad (\alpha\ne\beta).
\end{equation}
\end{lemma}

See Appendix C, Section \ref{C-diff-sect}, for the proof. In the following we write
\begin{equation}\label{T-def}
   T=\min\{\tau_{{\rm C}}-\delta_0, T_0\}.
\end{equation}
Next we remind a result which has been obtained in \cite[Lemma 2.1]{Nteil}.

\begin{lemma}\label{esti} Let the initial data for the Abraham
model satisfy (\ref{q-ini2}), (\ref{v-ini2}), and (\ref{ini-bed}).
Moreover, assume that
\begin{equation}\label{low-bound2}
   C_\ast\eps^{-1}\le \inf_{t\in [0, T\eps^{-3/2}]}
   |q_\alpha(t)-q_\beta(t)|\quad (\alpha\ne\beta),
\end{equation}
with $T>0$ from (\ref{T-def}) (or any other $T$).
Then there exist constants $C^\ast, C_v>0$ such that
\begin{equation}\label{diff-bound}
   C_\ast\eps^{-1}\le \inf_{t\in [0,\,T\eps^{-3/2}]}
   |q_\alpha(t)-q_\beta(t)|,\quad\sup_{t\in [0,\,T\eps^{-3/2}]}
   |q_\alpha(t)-q_\beta(t)|\le C^\ast\eps^{-1}
   \quad (\alpha\ne\beta),
\end{equation}
and for $\alpha=1, \ldots, N$
\begin{equation}\label{v-bound}
   \sup_{t\in [0,\,T\eps^{-3/2}]}|v_\alpha(t)|\le C_v\sqrt{\eps}
\end{equation}
are satisfied. In particular, $\sup_{t\in [0, T\eps^{-3/2}]}|v_\alpha(t)|\le\bar{v}<1$
for some $\bar{v}$. Moreover, there is $C>0$ and $\bar{e}>0$ such that
for $\alpha=1, \ldots, N$ we have
\begin{equation}\label{wend}
   \sup_{t\in [0, T\eps^{-3/2}]}|\dot{v}_\alpha(t)|\le C\eps^2,
\end{equation}
provided that $|e_\alpha|\le\bar{e}$, $\alpha=1, \ldots, N$.
In (\ref{diff-bound}), (\ref{v-bound}), and (\ref{wend}), the constants
$C$ and $\bar{e}$ do depend only on $T$ and the bounds for the initial data,
but not on $\eps$.
\end{lemma}

In the situation described in Lemma \ref{esti} we set
\begin{equation}\label{t0-def}
   t_0=4(R_\varphi+C^\ast\eps^{-1}).
\end{equation}
The bounds from Lemma \ref{esti} also lead to a bound
on the $|q_\alpha(t)|$, since for $\eps\in ]0, 1]$ we have
\begin{eqnarray}\label{qbd}
   |q_\alpha(t)| & \le & |q_\alpha(t)-q_\alpha^0|+|q_\alpha^0|
   \le C_v\sqrt{\eps}t+|q_\alpha^0| \nonumber
   \\ & \le & \Big(C_vT+\max_{1\le\alpha\le N}|q_\alpha^0|\Big)\eps^{-1}
   =: C_q\,\eps^{-1},\quad t\in [0, T\eps^{-3/2}].
\end{eqnarray}

It is moreover possible to establish an a priori
estimate for the $\ddot{v}_\alpha(t)$. Defining
\begin{equation}\label{tauastast-def}
   \tau_{\ast\ast}=(C_\ast/8)\eps^{-1},
\end{equation}
we have the following result. We remark that $\tau_{\ast\ast}$
could be replaced by any other time of order ${\cal O}(\eps^{-1})$.

\begin{lemma}\label{q3esti} Under the assumptions of Lemma \ref{esti},
including the smallness hypothesis $|e_\alpha|\le\bar{e}$, $\alpha=1, \ldots, N$,
there exists a constant $C>0$ such that for $\alpha=1, \ldots, N$ we have
\[ \sup_{t\in [0, T\eps^{-3/2}]}|\ddot{v}_\alpha(t)|\le C\eps^{5/2}
   \quad\mbox{and}\quad
   \sup_{t\in [\tau_{\ast\ast}, T\eps^{-3/2}]}|\ddot{v}_\alpha(t)|\le C\eps^{7/2}. \]
\end{lemma}
{\bf Proof\,:} Our handling of this estimate in \cite{Nteil} was somewhat inaccurate,
since some expressions resulting from data terms at time $t=0$
do not vanish, which have been claimed to be zero; cf.~also the remarks
in the Introduction. It then becomes evident that the desired bound of order
${\cal O}(\eps^{7/2})$ can not hold directly from time $t=0$, but only after
some initial time $t\cong\eps^{-1}$, which is still enough for our purposes.
The argument will not be expanded here, as it is follows similar lines as the proof
of Lemma \ref{q4esti} below, where an analogous problem arises.
{\hfill$\Box$}\bigskip

In order to expand the dynamics up to the order of radiation reaction,
we need to have control of one further $t$-derivative.
Thus a main issue will be to verify the following lemma.

\begin{lemma}\label{q4esti} Under the assumptions of Lemma \ref{esti},
including $|e_\alpha|\le\bar{e}$, $\alpha=1, \ldots, N$,
there is a constant $C>0$ such that for $\alpha=1, \ldots, N$ we have
\[ \sup_{t\in [6\tau_{\ast\ast}, T\eps^{-3/2}]}|\stackrel{...}{v}_\alpha(t)|\le C\eps^5. \]
\end{lemma}
The rather technical proof is given in the Appendix A, Section \ref{append-sect}.
It turned out that the principal method used in \cite{MK-S-1,MK-S-2,Nteil}
needed to be improved in a substantial manner in order to be applied here as well,
which is mainly due to the ``bad'' decay properties of solutions to wave equations
in the vicinity of the light-cone.

Using Lemma \ref{q4esti} as a key ingredient, it will then follow
from Lemma \ref{force-esti} that
\begin{equation}\label{LD-dyn}
   \frac{d}{dt}\bigg(\frac{\partial {\cal L}_{{\rm D}}}
   {\partial v_\alpha}\bigg)=\frac{\partial {\cal L}_{{\rm D}}}
   {\partial q_\alpha}+\frac{e_\alpha}{6\pi}\sum^N_{\beta =1}
   e_{\beta}\ddot{v}_{\beta}+{\cal O}(\eps^4),\quad\alpha=1, \ldots, N,
\end{equation}
where ${\cal L}_{{\rm D}}$ is the Darwin Lagrangian
from (\ref{LCD}) which governs the system up to order ${\cal O}(\eps^3)$,
cf.~\cite{Nteil}. As comparison effective dynamics we hence introduce
(\ref{LD-dyn-eff}). It can be verified that along solutions of this system
(\ref{LD-dyn-eff}) the ``energy''
\[ {\cal H}_{{\rm RR}}(r, u, \dot{u})={\cal H}_{{\rm D}}(r, u)
   -\sum_{\alpha, \beta=1}^N \frac{e_\alpha e_\beta}{6\pi}
   \,u_\alpha\cdot\dot{u}_\beta, \]
with
\begin{eqnarray}\label{HD-def}
   {\cal H}_{{\rm D}}(r, u)
   & = & \sum_{\alpha=1}^N\Big(\frac{1}{2}m_\alpha u_\alpha^2
   +\frac{3}{8} m_\alpha^\ast u_\alpha^4\Big)
   +\frac{1}{2}\sum_{\stackrel{\alpha, \beta=1}{\alpha\neq\beta}}^N
   \frac{e_\alpha e_\beta}{4\pi |r_\alpha-r_\beta|} \nonumber
   \\ & & +\,\frac{1}{4}\sum_{\stackrel{\alpha, \beta=1}{\alpha\neq\beta}}^N
   \frac{e_\alpha e_\beta}{4\pi |r_\alpha-r_\beta|}\bigg(
   u_\alpha\cdot u_\beta+\frac{u_\alpha\cdot (r_\alpha-r_\beta)}
   {|r_\alpha-r_\beta|^2}\,u_\beta\cdot (r_\alpha-r_\beta)\bigg),
\end{eqnarray}
is decreasing, more precisely one obtains
\begin{equation}\label{energ-dec}
   \frac{d}{dt}{\cal H}_{{\rm RR}}=-\frac{1}{6\pi}\bigg(\sum_{\alpha=1}^N
   e_\alpha\dot{u}_\alpha\bigg)^2.
\end{equation}
Due to the presence of runaway solutions in (\ref{LD-dyn-eff}),
the data for (\ref{LD-dyn-eff}) in the data space
$\R^{9N}=\R^{3N}\times\R^{3N}\times\R^{3N}$ leading to physically reasonable
solutions have to be singled out. In Section \ref{ZM-sect} we will accordingly
construct a kind of center manifold ${\cal I}_\eps$ for (\ref{LD-dyn-eff})
on which the true effective dynamics takes place and which is locally invariant
for (\ref{LD-dyn-eff}), in the sense specified in Theorem \ref{invmfthm}.
It will moreover turn out that the true effective dynamics (on the center manifold)
of solutions to the full system is approximately described
by a second order system of ODEs, cf.~(\ref{LD-dyn-eff-3}) below.
Note that the existence of runaway solutions does not contradict (\ref{energ-dec}),
since ${\cal H}_{{\rm RR}}$ in general may be indefinite.

Our main result is the following theorem.

\begin{theorem}\label{main-thm} Assume the data $(q_\alpha^0, v_\alpha^0)$,
$\alpha=1, \ldots, N$, and $E^0(x)$ and $B^0(x)$ are such that
(\ref{q-ini2}), (\ref{v-ini2}), and (\ref{ini-bed}) are verified.
Define $\tau_{{\rm C}}$, $\delta_0$, and $T_0$ as in Lemma \ref{C-diff-lem},
and introduce $T$ through (\ref{T-def}). Then Lemma \ref{C-diff-lem}
implies the existence of $C_\ast>0$ such that (\ref{low-bound}) holds,
and this in turn yields the existence of suitable constants such
that the bounds from Lemmas \ref{esti}, \ref{q3esti}, and \ref{q4esti} are satisfied,
provided that $|e_\alpha|\le\bar{e}$, $1\le\alpha\le N$.
Moreover, $t_0$ from (\ref{t0-def}) is defined.

In this basic setup we denote
\[ K_\eps=\Big\{(r, u)\in\R^{3N}\times\R^{3N}:
       |r|\le 4C_q\eps^{-1},\,|u|\le 4C_v\sqrt{\eps}\Big\},  \]
with $C_q$ from (\ref{qbd}) and $C_v$ from (\ref{v-bound}), respectively.
Then there exists $\eps_1>0$ and for each $\eps\in ]0, \eps_1]$
a $C^4$-function $h_\eps: K_\eps\to\R^{3N}$ with the following property.
Consider the true solution $(q_\alpha(t), v_\alpha(t))$ resulting from
(\ref{rho-j-def}), (\ref{system0}), (\ref{system1}), and (\ref{system2}),
and the solution $(r_\alpha(t), u_\alpha(t))$ of (\ref{LD-dyn-eff})
with data
\begin{equation}\label{ZM-data}
   r_\alpha(t_0)=q_\alpha(t_0),\quad u_\alpha(t_0)=v_\alpha(t_0),\quad
   \mbox{and}\quad\dot{u}_\alpha(t_0)=h_\eps\Big(q_\alpha(t_0),
   v_\alpha(t_0)\Big),
\end{equation}
for $1\le\alpha\le N$. Then $(r_\alpha(t), u_\alpha(t))$ exists at least
for $t\in [t_0, T\eps^{-3/2}]$, and the estimates
\begin{equation}\label{zollv}
   |q_\alpha(t)-r_\alpha(t)|\le C\sqrt{\eps},
   \quad |v_\alpha(t)-u_\alpha(t)|\le C\eps^2,
   \quad |\dot{v}_\alpha(t)-\dot{u}_\alpha(t)|\le C\eps^{7/2},
\end{equation}
hold for $t\in [t_0, T\eps^{-3/2}]$. Moreover, along such solutions
of the effective equation (\ref{LD-dyn-eff}) we have
\begin{eqnarray}\label{LD-dyn-eff-3}
   \frac{d}{dt}\bigg(\frac{\partial {\cal L}_{{\rm D}}}
   {\partial u_\alpha}\bigg) & = & \frac{\partial {\cal L}_{{\rm D}}}
   {\partial r_\alpha}+\frac{e_\alpha}{12\pi}
   \sum_{\stackrel{\beta, \beta'=1}{\beta\neq\beta'}}^N\frac{e_\beta e_{\beta'}}{4\pi}
   \bigg(\frac{e_\beta}{m_\beta}-\frac{e_{\beta'}}{m_{\beta'}}\bigg)
   \nonumber \\ & & \hspace{5em}
   \times\bigg[\frac{1}{|\xi_{\beta\beta'}|^3}(u_\beta-u_{\beta'})
   -\frac{3}{|\xi_{\beta\beta'}|^5}\,\xi_{\beta\beta'}\cdot (u_\beta-u_{\beta'})
   \,\xi_{\beta\beta'}\bigg]+{\cal O}(\eps^{9/2})\qquad
\end{eqnarray}
for $\alpha=1, \ldots, N$ and $t\in [t_0, T\eps^{-3/2}]$. In addition,
with ${\cal H}_{{\rm D}}$ from (\ref{HD-def}) we obtain
\begin{equation}\label{H-diff}
   \Big|{\cal H}_{{\rm D}}(q(t), v(t))-{\cal H}_{{\rm D}}(r(t), u(t))\Big|
   \le C\eps^3,\quad t\in [t_0, T\eps^{-3/2}].
\end{equation}
\end{theorem}
\bigskip

\begin{remarks}{\rm (a) In fact for every $k\in\N$ with $k\ge 4$
it can be achieved that $h_\eps$ is of class $C^k$, with $\eps_1$ possibly
having to be decreased further. \medskip

\noindent
(b) The functions $h_\eps$ do depend only on the input constants
$C_1$, $C_2$, $C_3$, $\tau_{{\rm C}}$, $\delta_0$, and $T_0$.
\medskip
}
\end{remarks}


\setcounter{equation}{0}

\section{Expansion of the Lorentz force term}
\label{force-sect}

Due to the bound on $\stackrel{...}{v}_\alpha$ from Lemma \ref{q4esti}
it is possible to rigorously expand the Lorentz force
\begin{equation}\label{self-act}
   F_\alpha(t)=\int d^3x\,\rho_\alpha (x-q_\alpha(t))
   \big[E(x,t)+v_\alpha(t)\wedge B(x, t)\big]
\end{equation}
up to the order of radiation reaction. In view of (\ref{system0}) and
(\ref{system1}) we have
\[ E(x,t)=E^{(0)}(x, t)+E^{(r)}(x,t)\quad\mbox{and}\quad
   B(x,t)=B^{(0)}(x, t)+B^{(r)}(x,t), \]
with
\begin{eqnarray*}
   \hat{E}^{(0)}(k, t) & = & \cos |k|t\,\hat{E}(k, 0)
   -i\,\frac{\sin |k|t}{|k|}\,k\wedge\hat{B}(k, 0), \\
   \hat{B}^{(0)}(k, t) & = & \cos |k|t\,\hat{B}(k, 0)
   +i\,\frac{\sin |k|t}{|k|}\,k\wedge\hat{E}(k, 0), \\
   \hat{E}^{(r)}(k, t) & = & -\int_0^t ds\,\cos |k|(t-s)\,\hat{j}(k, s)
   +i\,\int_0^t ds\,\frac{\sin |k|(t-s)}{|k|}\,\hat{\rho}(k, s)k, \\
   \hat{B}^{(r)}(k, t) & = & -i\,\int_0^t ds\,\frac{\sin |k|(t-s)}{|k|}\,
   k\wedge\hat{j}(k, s),
\end{eqnarray*}
where $j(x, t)$ and $\rho(x, t)$ are given by (\ref{rho-j-def}),
with $c=1$. Accordingly we write $F_\alpha(t)$ from (\ref{self-act}) as
\begin{eqnarray}
   F_\alpha(t) & = & \int d^3x\,\rho_\alpha(x-q_\alpha(t))[E^{(0)}(x, t)
   +v_\alpha (t)\wedge B^{(0)}(x, t)] \nonumber
   \\ & & +\,\int d^3x\,\rho_\alpha(x-q_\alpha(t))[E^{(r)}(x, t)+v_\alpha(t)
   \wedge B^{(r)}(x, t)]
   \nonumber \\ & =: & F^{(0)}_\alpha (t)+F^{(r)}_\alpha(t).
   \label{zerle}
\end{eqnarray}
With $t_0=4(R_\varphi+C^\ast\eps^{-1})$ from (\ref{t0-def})
we first recall from \cite[Lemma 3.1]{Nteil} the following result
concerning $F^{(0)}_\alpha (t)$.

\begin{lemma}\label{Falph0} For $t\in [t_0, T\eps^{-3/2}]$
we have $F^{(0)}_\alpha (t)=0$.
\end{lemma}
Applying the Fourier transform and noting $\rho_\alpha=e_\alpha\varphi$,
it is moreover seen that the contribution $F^{(r)}_\alpha(t)$ to $F_\alpha(t)$
resulting from the retarded parts of the fields is
\begin{equation}\label{linny}
   F^{(r)}_\alpha (t)=e^2_\alpha F^{(r)}_{\alpha\alpha} (t)
   + \sum_{\stackrel{\beta=1}{\beta\neq\alpha}}^N
   e_\alpha e_\beta F^{(r)}_{\alpha\beta}(t),
\end{equation}
where
\begin{eqnarray}\label{fab}
   F_{\alpha\beta}^{(r)}(t) & = & \int_0^t ds\,
   \int d^3k\,{|\hat{\varphi}(k)|}^2
   e^{-ik\cdot[q_\alpha(t)-q_\beta(s)]}\,\Bigg\{
   -\cos|k|(t-s)\,v_\beta(s)+i\,\frac{\sin |k|(t-s)}{|k|}\,k
   \nonumber \\ & &  \hspace{15em} -i\,\frac{\sin
   |k|(t-s)}{|k|}\,v_\alpha(t)\wedge
   (k\wedge v_\beta(s))\Bigg\},
\end{eqnarray}
for $\alpha, \beta=1, \ldots, N$.

\subsection{Expansion of the self-force $F^{(r)}_{\alpha\alpha} (t)$}
\label{alphalph-sect}

For $t\in [t_0, T\eps^{-3/2}]$ we have
\begin{eqnarray}\label{charg}
   F_{\alpha\alpha}^{(r)}(t) & = & \int_0^{\infty} d\tau\,
   \int d^3k\,{|\hat{\varphi}(k)|}^2
   e^{-i(k\cdot v_\alpha)\tau}\Big(1+\frac{i}{2}(k\cdot\dot{v}_\alpha)
   \tau^2-\frac{i}{6}(k\cdot\ddot{v}_\alpha)\tau^3\Big)
   \nonumber \\ & & \hspace{4em}
   \times\,\Bigg\{-\cos|k|\tau\,[v_\alpha-\dot{v}_\alpha\tau
   +\frac{1}{2}\,\ddot{v}_\alpha\tau^2]+i\,\frac{\sin |k|\tau}{|k|}\,k
   \nonumber \\ & & \hspace{6em}
   -i\,\frac{\sin |k|\tau}{|k|}\,v_\alpha\wedge
   (k\wedge [v_\alpha-\dot{v}_\alpha\tau])\Bigg\}+{\cal O}(\eps^4),
\end{eqnarray}
with $v_\alpha=v_\alpha(t)$, etc. The proof of this formula will not be elaborated.
It proceeds similar to the proof of Lemma \ref{tayl} below,
cf.~also \cite{KKS,MK-S-2}. Arguing formally, we utilize
 \begin{eqnarray*}
    e^{-ik\cdot[q_\alpha(t)-q_\beta(s)]} & \cong &
    e^{-i(k\cdot v_\alpha)\tau}\Big(1+\frac{i}{2}(k\cdot\dot{v}_\alpha)
    \tau^2-\frac{i}{6}(k\cdot\ddot{v}_\alpha)\tau^3\Big)+{\cal O}(\eps^4), \\
    v_\alpha(s) & \cong & v_\alpha-\dot{v}_\alpha\tau
    +\frac{1}{2}\,\ddot{v}_\alpha\tau^2+{\cal O}(\eps^5),
\end{eqnarray*}
in (\ref{fab}) with $\alpha=\beta$, where $\tau=t-s$. To make this argument
rigorous, it is necessary to observe that $\int_0^t ds (\ldots)
=\int_{t-\bar{t}}^t ds (\ldots)=\int_0^{\bar{t}} d\tau (\ldots)$
for any $t, \bar{t}\ge\frac{2R_\varphi}{1-\bar{v}}$,
as it follows in case that $\alpha=\beta$ from condition $(C)$
by transforming (\ref{fab}) back to space variables, cf.~Lemma \ref{tayl}.
With the notation
\[ I_p=\int_0^{\bar{t}}\,d\tau\,\frac{\sin (|k|\tau)}{|k|}
   e^{-i (k\cdot v_\alpha)\tau}\,\tau^p,\quad
   J_p=\int_0^{\bar{t}}\,d\tau\cos(|k|\tau)
   e^{-i (k\cdot v_\alpha)\tau}\,\tau^p,\quad p\in\N_0, \]
we may reformulate (\ref{charg}) as
\begin{eqnarray}\label{alphalph}
   F_{\alpha\alpha}^{(r)}(t) & = & \lim_{\bar{t}\to\infty}\bigg(
   -\int d^3k\,{|\hat{\varphi}(k)|}^2\,
   \bigg\{v_\alpha J_0-\dot{v}_\alpha J_1+\frac{i}{2}(k\cdot\dot{v}_\alpha)
   v_\alpha J_2+\frac{1}{2}\,\ddot{v}_\alpha J_2\bigg\} \nonumber\\ & & \hspace{2.8em}
   + \int d^3k\,{|\hat{\varphi}(k)|}^2\,
   \bigg\{i\,[(1-v_\alpha^2)k + (k\cdot v_\alpha)v_\alpha] I_0
         +i\,[(v_\alpha\cdot\dot{v}_\alpha)k-(k\cdot
         v_\alpha)\dot{v}_\alpha]I_1 \nonumber\\ & & \hspace{9.8em}
         -\frac{1}{2}(k\cdot\dot{v}_\alpha)[(1-v_\alpha^2)k
         +(k\cdot v_\alpha)v_\alpha]I_2
         +\frac{1}{6}(k\cdot\ddot{v}_\alpha)kI_3\bigg\}\bigg)+{\cal O}(\eps^4).
         \nonumber \\ & =: & F_{\alpha\alpha, {\rm old}}^{(r)}(t)
         +F_{\alpha\alpha, {\rm new}}^{(r)}(t)+{\cal O}(\eps^4),
\end{eqnarray}
with
\[ F_{\alpha\alpha, {\rm new}}^{(r)}(t)
   =\lim_{\bar{t}\to\infty}\bigg(
   -\frac{1}{2}\,\ddot{v}_\alpha\int d^3k\,{|\hat{\varphi}(k)|}^2 J_2
   +\frac{1}{6}\int d^3k\,{|\hat{\varphi}(k)|}^2\,
   (k\cdot\ddot{v}_\alpha)kI_3\bigg). \]
Note that due to the fact that only terms up to order ${\cal O}(\eps^4)$
have to be taken into account some expressions appearing
in (\ref{charg}) have dropped out when passing to (\ref{alphalph}).
Compared to the expansion of $F_{\alpha\alpha}^{(r)}(t)$ up to
order ${\cal O}(\eps^{7/2})$ in \cite[(3.6)]{Nteil}, we have picked up
the two additional terms denoted $F_{\alpha\alpha, {\rm new}}^{(r)}(t)$,
which both contain $\ddot{v}_\alpha$. {}From \cite[(3.7), (3.8)]{Nteil}
we know that
\begin{equation}\label{alphalph-old}
   F_{\alpha\alpha, {\rm old}}^{(r)}(t)
   =-\bigg(\frac{4}{3}+\frac{8}{15}\,v_\alpha^2\bigg)m_{{\rm e}}\dot{v}_\alpha
   -\frac{16}{15}\,m_{{\rm e}}(v_\alpha\cdot\dot{v}_\alpha)v_\alpha+{\cal O}(\eps^4),
\end{equation}
where $m_{{\rm e}}=\frac{1}{2}\int d^3k\,{|\hat{\varphi}(k)|}^2 k^{-2}$,
cf.~(\ref{me-def}). To evaluate the contribution of the new term,
we recall from \cite[Lemma 4.3]{MK-S-2} that
\[ \lim_{\bar{t}\to\infty}\int d^3k\,{|\hat{\varphi}(k)|}^2 J_2
   =\int_0^\infty d\tau\tau^2\int d^3k\,{|\hat{\varphi}(k)|}^2
   \cos(|k|\tau)e^{-i (k\cdot v_\alpha)\tau}=-\frac{1}{2\pi}\,\gamma_\alpha^4, \]
where $\gamma_\alpha=(1-v_\alpha^2)^{-1/2}$. In addition, $(\xi\cdot\nabla_v)
\nabla_v I_1=-(k\cdot\xi)kI_3$ for $\xi\in\R^3$, and also
\[ \lim_{\bar{t}\to\infty}\int d^3k\,{|\hat{\varphi}(k)|}^2 I_1
   =\frac{1}{4\pi}\,\gamma_\alpha^2 \]
due to \cite[p.~637]{MK-S-2}. It follows that
\begin{eqnarray}\label{alphalph-new}
   F_{\alpha\alpha, {\rm new}}^{(r)}(t) & =& \frac{1}{4\pi}\,\gamma_\alpha^4\ddot{v}_\alpha
   -\frac{1}{24\pi}\,(\ddot{v}_\alpha\cdot\nabla_v)\nabla_v\gamma_\alpha^2
   =\frac{1}{4\pi}\,\gamma_\alpha^4\ddot{v}_\alpha
   -\frac{1}{12\pi}\Big(\gamma_\alpha^4\ddot{v}_\alpha
   +4\gamma_\alpha^6(v_\alpha\cdot\ddot{v}_\alpha)v_\alpha\Big) \nonumber \\
   & = & \frac{1}{6\pi}\,\ddot{v}_\alpha+{\cal O}(\eps^4),
\end{eqnarray}
the latter by expanding $\gamma_\alpha^4=1+{\cal O}(\eps)$.
We can summarize (\ref{alphalph-old}) and (\ref{alphalph-new})
in the following lemma.

\begin{lemma}\label{mota} For $t\in [t_0, T\eps^{-3/2}]$
we have
\[ F_{\alpha\alpha}^{(r)}(t) =
   -\bigg(\frac{4}{3}+\frac{8}{15}\,v_\alpha^2\bigg)m_{{\rm e}}\dot{v}_\alpha
   -\frac{16}{15}\,m_{{\rm e}}(v_\alpha\cdot\dot{v}_\alpha)v_\alpha
   +\frac{1}{6\pi}\,\ddot{v}_\alpha+{\cal O}(\eps^4). \]
\end{lemma}

\subsection{Expansion of the interaction force $F^{(r)}_{\alpha\beta}(t)$,
$\alpha\neq\beta$}

We return to (\ref{fab}) and consider $F^{(r)}_{\alpha\beta}(t)$
for $\alpha\neq\beta$. The main difference to Section \ref{alphalph-sect}
results from the fact that now $\xi_{\alpha\beta}:=q_\alpha(t)-q_\beta(t)
={\cal O}(\eps^{-1})$, cf.~Lemma \ref{esti}. This point in conjunction
with Lemma \ref{q4esti} also plays the key role in the proof
of the following technical lemma whose proof is postponed to Appendix B,
Section \ref{tayl-sect}.

\begin{lemma}\label{tayl} Let $1\le\alpha, \beta\le N$, $\alpha\neq\beta$.
For $t\in [t_0, T\eps^{-3/2}]$ we have
\begin{eqnarray*}
   & (a) & -\int_0^t ds\,\int d^3k\,{|\hat{\varphi}(k)|}^2
   e^{-ik\cdot[q_\alpha(t)-q_\beta(s)]}\,
   \cos|k|(t-s)\,v_\beta(s) \\[1ex]
   & & = -\int_0^\infty d\tau\,\int d^3k\,{|\hat{\varphi}(k)|}^2
   e^{-ik\cdot\xi_{\alpha\beta}}
   \cos|k|\tau\,\bigg\{v_\beta-i\tau (k\cdot v_\beta)v_\beta-\tau\dot{v}_\beta
   +\frac{i}{2}\tau^2(k\cdot \dot{v}_\beta)v_\beta \\[1ex]
   & & \hspace{15em} -\frac{1}{2}\tau^2(k\cdot v_\beta)^2 v_\beta
   +i\tau^2(k\cdot v_\beta)\dot{v}_\beta+\frac{1}{2}\tau^2\ddot{v}_\beta\bigg\}
   + {\cal O}(\eps^4), \\[1ex]
   & (b) & i\int_0^t ds\,\int d^3k\,{|\hat{\varphi}(k)|}^2
   e^{-ik\cdot[q_\alpha(t)-q_\beta(s)]}\,
   \frac{\sin|k|(t-s)}{|k|}k \\[1ex]
   & & = i\int_0^\infty d\tau\,\int d^3k\,{|\hat{\varphi}(k)|}^2
   e^{-ik\cdot\xi_{\alpha\beta}}\,
   \frac{\sin|k|\tau}{|k|}k\,\bigg\{1-ik\cdot \Big[\tau v_\beta
   -\frac{1}{2}\tau^2\dot{v}_\beta+\frac{1}{6}\tau^3\ddot{v}_\beta\Big]
   \\ & & \hspace{10.9em} -\frac{1}{2}\Big[\tau^2 {(k\cdot v_\beta)}^2
   -\tau^3(k\cdot v_\beta)(k\cdot\dot{v}_\beta)\Big]
   +\frac{i}{6}\tau^3(k\cdot v_\beta)^3\bigg\} + {\cal O}(\eps^4), \\[1ex]
   & (c) & (-i)\int_0^t ds\,\int d^3k\,{|\hat{\varphi}(k)|}^2
   e^{-ik\cdot[q_\alpha(t)-q_\beta(s)]}\,
   \frac{\sin|k|(t-s)}{|k|}\,v_\alpha(t)\wedge(k\wedge v_\beta(s)) \\[1ex]
   & & = (-i)\int_0^\infty d\tau\,\int d^3k\,{|\hat{\varphi}(k)|}^2
   e^{-ik\cdot\xi_{\alpha\beta}}\,
   \frac{\sin|k|\tau}{|k|}\,v_\alpha\wedge
   \Big(k\wedge \Big\{v_\beta-\tau\dot{v}_\beta-i\tau (k\cdot v_\beta)v_\beta\Big\}\Big)
   \\[1ex] & & \hspace{1em} +\,{\cal O}(\eps^4).
\end{eqnarray*}
Here $v_\alpha=v_\alpha(t)$, etc., and
$\xi_{\alpha\beta}=q_\alpha(t)-q_\beta(t)$.
\end{lemma}
These expressions will be substituted back into (\ref{fab}).
To simplify notation, we introduce for $p\in\N_0$
\[ A_p:=\int_0^\infty d\tau\int d^3k\,{|\hat{\varphi}(k)|}^2
   e^{-ik\cdot \xi_{\alpha\beta}}\,\frac{\sin|k|\tau}{|k|}\,\tau^p
   ={(4\pi)}^{-1}\,\int\int d^3x d^3y\,\varphi(x)\varphi(y)
   {|\xi_{\alpha\beta}+x-y|}^{p-1} \]
and
\begin{eqnarray*}
   B_p & := & \int_0^\infty d\tau\int d^3k\,{|\hat{\varphi}(k)|}^2
   e^{-ik\cdot \xi_{\alpha\beta}}\,\cos(|k|\tau)\,\tau^p
   \\ & = & (-p){(4\pi)}^{-1}\,\int\int d^3x d^3y\,\varphi(x)\varphi(y)
   {|\xi_{\alpha\beta}+x-y|}^{p-2}
   =(-p)A_{p-1}.
\end{eqnarray*}
Hence it follows from Lemma \ref{tayl} that for $\alpha\neq\beta$
and $t\in [t_0, T\eps^{-3/2}]$ we have
\begin{eqnarray}\label{kemp}
   F_{\alpha\beta}^{(r)}(t) & = & -v_\beta B_0-v_\beta
   (v_\beta\cdot\nabla_\xi)B_1+\dot{v}_\beta B_1
   +\frac{1}{2}(\dot{v}_\beta\cdot\nabla_\xi)B_2 v_\beta
   -\frac{1}{2}(v_\beta\cdot\nabla_\xi)^2 B_2 v_\beta
   +(v_\beta\cdot\nabla_\xi)B_2 \dot{v}_\beta \nonumber \\ & &
   -\frac{1}{2}B_2 \ddot{v}_\beta-\nabla_\xi A_0
   -(v_\beta\cdot\nabla_\xi)\nabla_\xi A_1
   +\frac{1}{2}(\dot{v}_\beta\cdot\nabla_\xi)\nabla_\xi A_2
   -\frac{1}{2} {(v_\beta\cdot\nabla_\xi)}^2\nabla_\xi A_2
   \nonumber \\ & & -\frac{1}{6}(\ddot{v}_\beta\cdot\nabla_\xi)\nabla_\xi A_3
   +\frac{1}{2}(v_\beta\cdot\nabla_\xi)
   (\dot{v}_\beta\cdot\nabla_\xi)\nabla_\xi A_3
   -\frac{1}{6}{(v_\beta\cdot\nabla_\xi)}^3\nabla_\xi A_3
   \nonumber \\ & & + v_\alpha\wedge (\nabla_\xi A_0\wedge v_\beta)
   -v_\alpha\wedge (\nabla_\xi A_1\wedge\dot{v}_\beta)
   +v_\alpha\wedge (\nabla_\xi\wedge (v_\beta\cdot\nabla_\xi)A_1 v_\beta)
   +{\cal O}(\eps^4) \nonumber \\ & = & -v_\beta
   (v_\beta\cdot\nabla_\xi)B_1+\dot{v}_\beta B_1
   -\frac{1}{2}B_2 \ddot{v}_\beta-\nabla_\xi A_0
   +\frac{1}{2}(\dot{v}_\beta\cdot\nabla_\xi)\nabla_\xi A_2
   -\frac{1}{2} {(v_\beta\cdot\nabla_\xi)}^2\nabla_\xi A_2
   \nonumber \\ & & -\frac{1}{6}(\ddot{v}_\beta\cdot\nabla_\xi)\nabla_\xi A_3
   +\frac{1}{2}(v_\beta\cdot\nabla_\xi)
   (\dot{v}_\beta\cdot\nabla_\xi)\nabla_\xi A_3
   -\frac{1}{6}{(v_\beta\cdot\nabla_\xi)}^3\nabla_\xi A_3
   \nonumber \\ & & +(v_\alpha\cdot v_\beta)\nabla_\xi A_0
   -v_\beta(v_\alpha\cdot\nabla_\xi)A_0+{\cal O}(\eps^4),
\end{eqnarray}
where in the last reduction we have used that $A_1=(4\pi)^{-1}$, $B_2=-(2\pi)^{-1}$,
and $B_0=0$, hence in particular $\nabla_\xi A_1=\nabla_\xi B_2=0$.
We rewrite (\ref{kemp}) as
\begin{equation}\label{alpbet-zerl}
   F_{\alpha\beta}^{(r)}(t)=F_{\alpha\beta, {\rm old}}^{(r)}(t)
   +F_{\alpha\beta, {\rm new}}^{(r)}(t)+{\cal O}(\eps^4),
\end{equation}
with
\begin{equation}\label{alpbet-new}
   F_{\alpha\beta, {\rm new}}^{(r)}(t)=-\frac{1}{2}B_2 \ddot{v}_\beta
   -\frac{1}{6}(\ddot{v}_\beta\cdot\nabla_\xi)\nabla_\xi A_3
   +\frac{1}{2}(v_\beta\cdot\nabla_\xi)(\dot{v}_\beta\cdot\nabla_\xi)\nabla_\xi A_3
   -\frac{1}{6}{(v_\beta\cdot\nabla_\xi)}^3\nabla_\xi A_3
\end{equation}
being the new radiation reaction contribution compared to our expansion
up to order ${\cal O}(\eps^3)$ in \cite{Nteil}. According to \cite[Section 3.2]{Nteil}
we have
\begin{equation}\label{alpbet-old}
   F_{\alpha\beta, {\rm old}}^{(r)}(t)=g_{\alpha\beta}(t)+{\cal O}(\eps^4),
\end{equation}
where
\begin{eqnarray}\label{galph-def}
   g_{\alpha\beta}(t) & = & \frac{\xi_{\alpha\beta}}{4\pi|\xi_{\alpha\beta}|^3}
   -\frac{1}{8\pi|\xi_{\alpha\beta}|}\,\dot{v}_\beta
   -\frac{(\dot{v}_\beta\cdot \xi_{\alpha\beta})}
   {8\pi|\xi_{\alpha\beta}|^3}\,\xi_{\alpha\beta}
   +\frac{v_\beta^2}{8\pi|\xi_{\alpha\beta}|^3}\,\xi_{\alpha\beta}
   -\frac{3{(v_\beta\cdot\xi_{\alpha\beta})}^2}{8\pi|\xi_{\alpha\beta}|^5}
   \,\xi_{\alpha\beta} \nonumber \\ & &
   -\frac{(v_\alpha\cdot v_\beta)}{4\pi|\xi_{\alpha\beta}|^3}\,
   \xi_{\alpha\beta}+\frac{(v_\alpha\cdot\xi_{\alpha\beta})}
   {4\pi|\xi_{\alpha\beta}|^3}\,v_\beta,
   \quad\mbox{with}\quad\xi_{\alpha\beta}=q_\alpha(t)-q_\beta(t).
\end{eqnarray}
We recall that for proving (\ref{alpbet-old}) the most difficult
part is to show that
\[ \bigg|\nabla_\xi A_0+\frac{\xi_{\alpha\beta}}
   {4\pi|\xi_{\alpha\beta}|^3}\bigg|
   =\frac{1}{4\pi|\xi_{\alpha\beta}|^2}\,
   \bigg|\int\int d^3x\,d^3y\,\varphi(x)\varphi(y)\Bigg(\frac{\vec{n}
   +\frac{x-y}{|\xi_{\alpha\beta}|}}{\big|\vec{n}
   +\frac{x-y}{|\xi_{\alpha\beta}|}\big|^3}-\vec{n}\Bigg)\bigg|
   ={\cal O}(\eps^4), \]
with $\vec{n}=\xi_{\alpha\beta}/|\xi_{\alpha\beta}|$.
However, this turns out to work well by expanding $\psi(R)=(\vec{n}+R)/|\vec{n}+R|^3
=\vec{n}+R-3(\vec{n}\cdot R)\vec{n}+{\cal O}(\eps^2)$,
where $R=(x-y)/|\xi_{\alpha\beta}|={\cal O}(\eps)$ for $|x|, |y|\le R_\varphi$,
and by noting that $\int\int d^3x\,d^3y\,\varphi(x)\varphi(y)(x-y)=0$.
Therefore we only have to consider the new part $F_{\alpha\beta, {\rm new}}^{(r)}(t)$
from (\ref{alpbet-new}). To begin with, we recall that $B_2=-2A_1=-(2\pi)^{-1}$.
The following lemma deals with the remaining terms.

\begin{lemma}\label{eckba} For $\alpha\neq\beta$ and $t\in [0, T\eps^{-3/2}]$
the following holds.
\begin{itemize}
\item[(a)] $\displaystyle\frac{1}{6}(\ddot{v}_\beta\cdot\nabla_\xi)\nabla_\xi A_3
=\frac{1}{12\pi}\,\ddot{v}_\beta$, and
\item[(b)] $(v_\beta\cdot\nabla_\xi)(\dot{v}_\beta\cdot\nabla_\xi)
\nabla_\xi A_3=0={(v_\beta\cdot\nabla_\xi)}^3\nabla_\xi A_3$.
\end{itemize}
\end{lemma}
{\bf Proof\,:} We have $A_3={(4\pi)}^{-1}\,\int\int d^3x d^3y\,\varphi(x)\varphi(y)
{|\xi_{\alpha\beta}+x-y|}^2$, and therefore
\[ \nabla_\xi A_3=\frac{1}{2\pi}\int\int d^3x d^3y\,\varphi(x)\varphi(y)
   \Big(\xi_{\alpha\beta}+x-y\Big)=\frac{1}{2\pi}\,\xi_{\alpha\beta}, \]
the latter in view of $\int\int d^3x d^3y\,\varphi(x)\varphi(y)(x-y)=0$
by the symmetry of $\varphi$, cf.~condition $(C)$. {\hfill$\Box$}\bigskip

\noindent
Turning back to $F_{\alpha\beta, {\rm new}}^{(r)}(t)$ from (\ref{alpbet-new})
we hence have shown that for $\alpha\neq\beta$
and $t\in [t_0, T\eps^{-3/2}]$ the estimate
\[ F_{\alpha\beta, {\rm new}}^{(r)}(t)=\frac{1}{4\pi}\,\ddot{v}_\beta
   -\frac{1}{12\pi}\,\ddot{v}_\beta+{\cal O}(\eps^4)
   =\frac{1}{6\pi}\,\ddot{v}_\beta+{\cal O}(\eps^4) \]
holds. In view of (\ref{alpbet-zerl}) and (\ref{alpbet-old})
we arrive at the following lemma.

\begin{lemma}\label{sailesa} For $\alpha\neq\beta$ and $t\in [t_0, T\eps^{-3/2}]$
we have
\[ F_{\alpha\beta}^{(r)}(t)=g_{\alpha\beta}(t)
   +\frac{1}{6\pi}\,\ddot{v}_\beta+{\cal O}(\eps^4), \]
with $g_{\alpha\beta}(t)$ given by (\ref{galph-def}).
\end{lemma}

\subsection{Summary of the expansion}

{}From (\ref{self-act}), (\ref{zerle}), and Lemma \ref{Falph0} we see that
$F_\alpha(t)=F_\alpha^{(r)}(t)$ for $t\in [t_0, T\eps^{-3/2}]$.
Thus (\ref{linny}), Lemma \ref{mota}, and Lemma \ref{sailesa}
can be summarized as follows. For $t\in [t_0, T\eps^{-3/2}]$
the Lorentz force $F_\alpha(t)$ from (\ref{self-act}) allows for the representation
\begin{eqnarray}\label{ga-def}
   F_\alpha(t) & = &
   -\bigg(\frac{4}{3}+\frac{8}{15}\,v_\alpha^2\bigg)e_\alpha^2 m_{{\rm e}}\dot{v}_\alpha
   -\frac{16}{15}\,e_\alpha^2 m_{{\rm e}}(v_\alpha\cdot\dot{v}_\alpha)v_\alpha
   +G_\alpha(q, v, \dot{v})
   +\frac{e_\alpha}{6\pi}\sum_{\beta=1}^N e_\beta\ddot{v}_\beta
   +{\cal O}(\eps^4), \nonumber \\[1ex]
   G_\alpha(q, v, \dot{v}) & = &
   \sum_{\stackrel{\beta=1}{\beta\neq\alpha}}^N e_\alpha e_\beta\,g_{\alpha\beta}(t)
   \nonumber \\ & = & \frac{e_\alpha}{4\pi}
   \sum_{\stackrel{\beta=1}{\beta\neq\alpha}}^N
   e_\beta\Bigg(\frac{\xi_{\alpha\beta}}{|\xi_{\alpha\beta}|^3}
   -\frac{1}{2|\xi_{\alpha\beta}|}\,\dot{v}_\beta
   -\frac{(\dot{v}_\beta\cdot \xi_{\alpha\beta})}
   {2|\xi_{\alpha\beta}|^3}\,\xi_{\alpha\beta}
   +\frac{v_\beta^2}{2|\xi_{\alpha\beta}|^3}\,\xi_{\alpha\beta}
   -\frac{3{(v_\beta\cdot\xi_{\alpha\beta})}^2}{2|\xi_{\alpha\beta}|^5}
   \,\xi_{\alpha\beta} \nonumber\\ & & \hspace{5.5em}
   -\frac{(v_\alpha\cdot v_\beta)}{|\xi_{\alpha\beta}|^3}\,
   \xi_{\alpha\beta}+\frac{(v_\alpha\cdot\xi_{\alpha\beta})}
   {|\xi_{\alpha\beta}|^3}\,v_\beta\Bigg),
\end{eqnarray}
with $t_0=4(R_\varphi+C^\ast\eps^{-1})$, cf.~(\ref{t0-def}),
and $\xi_{\alpha\beta}=q_\alpha(t)-q_\beta(t)$, $v_\alpha=v_\alpha(t)$,
$v_\beta=v_\beta(t)$, etc., and moreover
$m_{{\rm e}}=\frac{1}{2}\int d^3k\,{|\hat{\varphi}(k)|}^2 k^{-2}$.

Expanding $\gamma_\alpha=1+\frac{1}{2}v_\alpha^2+{\cal O}(\eps^2)$
and $\gamma_\alpha^3=1+{\cal O}(\eps)$ in the Lorentz equation
$\frac{d}{dt}(m_{{\rm b}\alpha}\gamma_\alpha v_\alpha)
=m_{{\rm b}\alpha}(\gamma_\alpha\dot{v}_\alpha+\gamma_\alpha^3
(v_\alpha\cdot\dot{v}_\alpha)v_\alpha)=F_\alpha(t)$, cf.~(\ref{system2}),
and recalling $m_\alpha=m_{{\rm b}\alpha}+\frac{4}{3}e_\alpha^2m_e$
as well as $m_\alpha^\ast=m_{{\rm b}\alpha}+\frac{16}{15}e_\alpha^2 m_e$,
we thus have deduced the following main lemma.

\begin{lemma}\label{force-esti} For $1\le\alpha\le N$ and $t\in [t_0, T\eps^{-3/2}]$
we have
\begin{equation}\label{bjack}
   M_\alpha(v_\alpha)\dot{v}_\alpha=G_\alpha(q, v, \dot{v})
   +\frac{e_\alpha}{6\pi}\sum_{\beta=1}^N e_\beta\ddot{v}_\beta
   +{\cal O}(\eps^4),
\end{equation}
where $G_\alpha$ is given by (\ref{ga-def}), and $M_\alpha(v)$ is the $(3\times 3)$-matrix
$M_\alpha(v)(z)=(m_\alpha+\frac{1}{2}m_\alpha^{\ast}v^2)z+m_\alpha^{\ast}(v\cdot z)v$
for $v, z\in\R^3$.
\end{lemma}

Note that (\ref{bjack}) agrees with (\ref{LD-dyn}), as may be verified
through explicit calculation. Introducing the transformation
\begin{eqnarray*}
   & & \bar{q}_\alpha(\tau)=\eps q_\alpha(\eps^{-3/2}\tau),\quad
   \bar{v}_\alpha(\tau)=\eps^{-1/2} v_\alpha(\eps^{-3/2}\tau), \\
   & & \dot{\bar{v}}_\alpha(\tau)=\eps^{-2} \dot{v}_\alpha(\eps^{-3/2}\tau),\quad
   \ddot{\bar{v}}_\alpha(\tau)=\eps^{-7/2} \ddot{v}_\alpha(\eps^{-3/2}\tau),
\end{eqnarray*}
we arrive at the following version of Lemma \ref{force-esti}.
Here and henceforth we drop the overbar for simplicity,
and $\tau$ will also be denoted by $t$.

\begin{lemma}\label{force-esti-2} For $1\le\alpha\le N$ and $t\in [\eps^{3/2} t_0, T]$
we have
\begin{equation}\label{wdsaig}
   M_{\alpha}(v_\alpha, \eps)\dot{v}_\alpha=G_{\alpha}(q, v,
   \dot{v}, \eps)+\eps^{3/2}\,\frac{e_\alpha}{6\pi}
   \sum_{\beta=1}^N e_\beta\ddot{v}_\beta+{\cal O}(\eps^2),
\end{equation}
where $G_{\alpha}(q, v, \dot{v}, \eps)\in\R^3$ is defined as
\begin{eqnarray}\label{Galph-def2}
   G_{\alpha}(q, v, \dot{v}, \eps) & = & \frac{e_\alpha}{4\pi}
   \sum_{\stackrel{\beta=1}{\beta\neq\alpha}}^N
   e_\beta\Bigg(\frac{\xi_{\alpha\beta}}{|\xi_{\alpha\beta}|^3}
   -\frac{\eps}{2|\xi_{\alpha\beta}|}\,\dot{v}_\beta
   -\frac{\eps (\dot{v}_\beta\cdot \xi_{\alpha\beta})}
   {2|\xi_{\alpha\beta}|^3}\,\xi_{\alpha\beta}
   +\frac{\eps v_\beta^2}{2|\xi_{\alpha\beta}|^3}\,\xi_{\alpha\beta}
   -\frac{3\eps {(v_\beta\cdot\xi_{\alpha\beta})}^2}{2|\xi_{\alpha\beta}|^5}
   \,\xi_{\alpha\beta} \nonumber\\ & & \hspace{5.5em}
   -\frac{\eps (v_\alpha\cdot v_\beta)}{|\xi_{\alpha\beta}|^3}\,
   \xi_{\alpha\beta}+\frac{\eps (v_\alpha\cdot\xi_{\alpha\beta})}
   {|\xi_{\alpha\beta}|^3}\,v_\beta\Bigg),
\end{eqnarray}
and $M_{\alpha}(v, \eps)$ is the $(3\times 3)$-matrix given by
\begin{equation}\label{Malph-def2}
   M_{\alpha}(v, \eps)(z)=(m_\alpha+\frac{\eps}{2}m_\alpha^{\ast}v^2)z
   +\eps m_\alpha^{\ast}(v\cdot z)v,\quad v, z\in\R^3.
\end{equation}
\end{lemma}

We remark that on the scale utilized in Lemma \ref{force-esti-2},
all quantities $\xi_{\alpha\beta}$, $v_\alpha$, $\dot{v}_\alpha$
and $\ddot{v}_\alpha$ are of the order ${\cal O}(1)$.


\setcounter{equation}{0}

\section{Construction of the center manifold}
\label{ZM-sect}

In this section we are going to construct
a kind of (locally) invariant manifold of solutions to the effective system
\begin{equation}\label{effect-equ}
   M_{\alpha}(u_\alpha, \eps)\dot{u}_\alpha=G_{\alpha}(r, u,
   \dot{u}, \eps)+\eps^{3/2}\,\frac{e_\alpha}{6\pi}\sum_{\beta=1}^N
   e_\beta\ddot{u}_\beta,\quad 1\le\alpha\le N,
\end{equation}
i.e., to (\ref{wdsaig}) without the error term ${\cal O}(\eps^2)$. We define
\begin{equation}\label{K0-def}
    K_0=\Big\{(r, u)\in\R^{3N}\times\R^{3N}: |r|\le 4C_q,\,|u|\le 4C_v\Big\},
\end{equation}
with $C_q$ and $C_v$ from (\ref{qbd}) and (\ref{v-bound}), respectively.
We are going to prove the following theorem concerning (\ref{effect-equ}).

\begin{theorem}\label{invmfthm} For every $k\in\N$ with $k\ge 4$
there exists $\eps_1>0$ and a $C^k$-function $\hat{h}: [0, \eps_1]\times K_0\to\R^{3N}$
such that
\begin{equation}\label{Ieps-def}
   {\cal I}_\eps=\Big\{(r, u, \dot{u}):\,\dot{u}=\hat{h}_\eps(r, u),\,
   (r, u)\in K_0\Big\}\subset\R^{3N}\times\R^{3N}\times\R^{3N}
\end{equation}
is locally invariant for (\ref{effect-equ}), in the following sense.
Consider given data $(r(\tau_0), u(\tau_0), \dot{u}(\tau_0))
\in\R^{3N}\times\R^{3N}\times\R^{3N}$ for (\ref{effect-equ}) such that
\begin{eqnarray*}
   & & \dot{u}(\tau_0)=\hat{h}_\eps\Big(r(\tau_0), u(\tau_0)\Big),\quad
   |r(\tau_0)|\le 2C_q,\quad |u(\tau_0)|\le 2C_v,\quad\mbox{and} \\
   & & (C_\ast/2)\le |r_\alpha(\tau_0)-r_\beta(\tau_0)|
   \le 2C^\ast\,\,\,(\alpha\neq\beta),
\end{eqnarray*}
hold. Then the corresponding solution $(r(t), u(t), \dot{u}(t))$
of (\ref{effect-equ}) with this data at $t=\tau_0$ exists at least until
$\tau_1>\tau_0$ and satisfies
\begin{equation}\label{uh-gl}
   \dot{u}(t)=\hat{h}_\eps\Big(r(t), u(t)\Big),\quad t\in [\tau_0, \tau_1],
\end{equation}
where $T\ge\tau_1>\tau_0$ denotes the longest time such that
\begin{equation}\label{samk}
   |r(t)|\le 3C_q,\quad |u(t)|\le 3C_v,\quad\mbox{and}\quad
   (C_\ast/3)\le |r_\alpha(t)-r_\beta(t)|\le 3C^\ast\,\,\,(\alpha\neq\beta),
\end{equation}
are valid simultaneously, for $t\in [\tau_0, \tau_1]$. Moreover
for $t\in [\tau_0, \tau_1]$ we also have
\begin{eqnarray}\label{sumdotu-form}
   \sum_{\beta=1}^N e_\beta\ddot{u}_\beta
   & = & \frac{1}{2}\sum_{\stackrel{\beta, \beta'=1}{\beta\neq\beta'}}^N
   \frac{e_\beta e_{\beta'}}{4\pi}
   \bigg(\frac{e_\beta}{m_\beta}-\frac{e_{\beta'}}{m_{\beta'}}\bigg)
   \bigg[\frac{1}{|\xi_{\beta\beta'}|^3}(u_\beta-u_{\beta'})
   \nonumber \\ & & \hspace{12em}
   -\frac{3}{|\xi_{\beta\beta'}|^5}\,\xi_{\beta\beta'}\cdot (u_\beta-u_{\beta'})
   \,\xi_{\beta\beta'}\bigg]+{\cal O}(\eps),\quad
\end{eqnarray}
recall $\xi_{\beta\beta'}=r_\beta-r_{\beta'}$. The constant $C>0$
defining ${\cal O}(\eps)$, i.e., $|\ldots|\le C\eps$, does depend only on the input
constants $C_q$, $C_v$, $C_\ast$, $C^\ast$, and $T$, but not on $\tau_1\le T$.
\end{theorem}

In order to build the center-like manifold described in the theorem
we cannot specify particular data, whence we are forced to a priori smoothen out
the Coulomb singularity and to introduce a bound for $u_\alpha$.
Thus rather than with (\ref{effect-equ}) we will be dealing with
the regularized problem
\begin{equation}\label{eff-regul}
   M_{\alpha}^{{\rm reg}}(u_\alpha, \eps)\dot{u}_\alpha
   =G_{\alpha}^{{\rm reg}}(r, u,
   \dot{u}, \eps)+\eps^{3/2}\,\frac{e_\alpha}{6\pi}\sum_{\beta=1}^N
   e_\beta\ddot{u}_\beta,\quad 1\le\alpha\le N,
\end{equation}
where $M_{\alpha}^{{\rm reg}}(u_\alpha, \eps)$ and $G_{\alpha}^{{\rm reg}}
(r, u, \dot{u}, \eps)$ are obtained
from $M_{\alpha}(u_\alpha, \eps)$ and $G_{\alpha}(r, u, \dot{u}, \eps)$
by replacing all $u_\alpha$ by $u_\alpha^{{\rm reg}}$
and all $\xi_{\alpha\beta}$ by $\xi_{\alpha\beta}^{{\rm reg}}$, respectively, with
\begin{equation}\label{reguluxi}
   u_\alpha^{{\rm reg}}=\chi_1(|u_\alpha|)u_\alpha\quad\mbox{and}\quad
   \xi_{\alpha\beta}^{{\rm reg}}=\chi_2(|\xi_{\alpha\beta}|)\xi_{\alpha\beta}.
\end{equation}
Here $\chi_1: [0, \infty[\to [0, 1]$ is a smooth function such that
$\chi_1(s)=1$ for $s\in [0, 3C_v]$ and $\chi_1(s)=0$ for $s\in [4C_v, \infty[$,
whereas $\chi_2: ]0, \infty[\to [0, \infty[$ is smooth and such that
$\chi_2(s)s=s$ for $s\in [C_\ast/3, 3C^\ast]$ as well as
$\chi_2(s)s\in [C_\ast/4, 4C^\ast]$ for $s\in [0, \infty[$;
the constants $C_\ast$, $C^\ast$, and $C_v$
are those appearing in Lemma \ref{esti}. We also note that
\begin{equation}\label{regul-bd}
   \Big|u_\alpha^{{\rm reg}}\Big|\le 4C_v\quad\mbox{and}\quad
   C_\ast/4\le\Big|\xi_{\alpha\beta}^{{\rm reg}}\Big|\le 4C^\ast.
\end{equation}

To rewrite (\ref{eff-regul}), we introduce the linear map
\[ P: (\R^3)^N\to (\R^3)^N,\quad Pz
   =\bigg(\frac{e_\alpha}{6\pi}\sum_{\beta=1}^N e_\beta
   z_\beta\bigg)_{1\le\alpha\le N}\quad\mbox{for}\quad
   z=(z_1, \ldots, z_N)\in (\R^3)^N. \]
Then (\ref{eff-regul}) reads as
\begin{equation}\label{eff-regul-P}
   \eps^{3/2}P\ddot{u}=M^{{\rm reg}}(u, \dot{u}, \eps)
   -G^{{\rm reg}}(r, u, \dot{u}, \eps),
\end{equation}
with
\begin{eqnarray}
   M^{{\rm reg}}(u, \dot{u}, \eps)
   & = & \bigg(M_{\alpha}^{{\rm reg}}(u_\alpha, \eps)
   \dot{u}_\alpha\bigg)_{1\le\alpha\le N}\in (\R^3)^N,\quad\mbox{and}
   \label{brien} \\ G^{{\rm reg}}(r, u, \dot{u}, \eps)
   & =& \bigg(G_{\alpha}^{{\rm reg}}(r, u, \dot{u}, \eps)\bigg)_{1\le\alpha\le N}
   \in (\R^3)^N. \label{gabb}
\end{eqnarray}
At this point we observe that (\ref{eff-regul-P}) does not present a singular
perturbation problem of standard form
\[ \dot{x}=f(x, y),\quad\eps\dot{y}=g(x, y, \eps), \]
where we think of $x=(r, u)\in\R^{3N}\times\R^{3N}$ and $y=\dot{u}\in\R^{3N}$,
due to the presence of $P$ which has $\dim({\rm range}(P))=3$ (assuming that
$e_\alpha\neq 0$ for all $\alpha$). However, (\ref{eff-regul-P}) can be suitably
transformed and cast in standard form, but with different variables.
To see this, we first diagonalize $P$ by means of the matrix $A: (\R^3)^N\to (\R^3)^N$,
\begin{equation}\label{Amart-def}
   Az=\Big(e_\alpha z_1\Big)_{1\le\alpha\le N}
   +\Big(e_2 z_2, e_3 z_3-e_1 z_2, e_4 z_4-e_2 z_3, \ldots,
   e_N z_N-e_{N-2}z_{N-1}, -e_{N-1}z_N\Big)
\end{equation}
for $z=(z_1, \ldots, z_N)\in (\R^3)^N$, being composed of the eigenvectors of $P$
as columns; note the eigenvalues are $\lambda=e^2:=\sum_{\alpha=1}^Ne_\alpha^2$
($3$ times) and $\lambda=0$ ($3N-3$ times). Then
\begin{equation}\label{At-def}
   A^t z=\Big(\sum_{\alpha=1}^N e_\alpha z_\alpha,
   e_2 z_1-e_1 z_2, e_3 z_2-e_2 z_3, \ldots, e_N z_{N-1}-e_{N-1}z_N\Big),
\end{equation}
and it can be verified that
\[ A^t PAz=\frac{e^4}{6\pi}\Big(z_1, 0, \ldots, 0\Big), \]
as a consequence of $\sum_{\beta=1}^N e_\beta (Az)_\beta=e^2 z_1$.
Hence we can introduce the equivalent variables
\begin{equation}\label{r-barr-trafo}
   r=A\underbar{r}\,,\quad u=A\underbar{u}\,,\quad\dot{u}=A\dot{\underbar{u}}\,,
   \quad\ddot{u}=A\ddot{\underbar{u}}\,,
\end{equation}
to transform (\ref{eff-regul-P}) to
\begin{equation}\label{eff-regul-bar}
   \eps^{3/2}\Big(\ddot{\underbar{u}}_1, 0, \ldots, 0\Big)
   =\Phi(\underbar{r}, \underbar{u}, \dot{\underbar{u}}, \eps),
\end{equation}
with
\begin{eqnarray}\label{Phi-defi}
   \Phi(\underbar{r}, \underbar{u}, \dot{\underbar{u}}, \eps)
   & = & 6\pi e^{-4}\,A^t\Big(M^{{\rm reg}}(u, \dot{u}, \eps)
   -G^{{\rm reg}}(r, u, \dot{u}, \eps)\Big) \nonumber \\
   & = & 6\pi e^{-4}\,A^t\Big(M^{{\rm reg}}(A\underbar{u}, A\dot{\underbar{u}},
   \eps)-G^{{\rm reg}}(A\underbar{r}, A\underbar{u}, A\dot{\underbar{u}},
   \eps)\Big).
\end{eqnarray}
Writing $\Phi={(\Phi_\alpha)}_{1\le\alpha\le N}\in (\R^3)^N$,
our strategy is now to solve the $(N-1)$ equations
\begin{equation}\label{2-N-eqs}
   0=\Phi_2(\underbar{r}, \underbar{u}, \dot{\underbar{u}}, \eps),
   \quad\ldots\,,\quad
   0=\Phi_N(\underbar{r}, \underbar{u}, \dot{\underbar{u}}, \eps),
\end{equation}
each in $\R^3$, for the $(N-1)$ variables $(\dot{\underbar{u}}_2, \ldots,
\dot{\underbar{u}}_N)$, also each in $\R^3$. This will yield a solution function
\[ \underbar{U}_{\,2N}: \R^{3N}\times\R^{3N}\times\R^3\times [0, \eps_0]
   \to\R^{3N-3},\quad (\dot{\underbar{u}}_2, \ldots,
   \dot{\underbar{u}}_N)=\underbar{U}_{\,2N}(\underbar{r}, \underbar{u},
   \dot{\underbar{u}}_1, \eps)  \]
i.e., we will have
\begin{equation}\label{augst}
   0=\Phi_2\Big(\underbar{r}, \underbar{u}, \dot{\underbar{u}}_1,
   \underbar{U}_{\,2N}(\underbar{r}, \underbar{u},
   \dot{\underbar{u}}_1, \eps), \eps\Big),\quad\ldots\,,\quad
   0=\Phi_N\Big(\underbar{r}, \underbar{u}, \dot{\underbar{u}}_1,
   \underbar{U}_{\,2N}(\underbar{r}, \underbar{u},
   \dot{\underbar{u}}_1, \eps), \eps\Big)
\end{equation}
for $(\underbar{r}, \underbar{u}, \dot{\underbar{u}}_1, \eps)
\in\R^{3N}\times\R^{3N}\times\R^3\times [0, \eps_0]$.
Setting $\underbar{x}=(\underbar{x}_1, \underbar{x}_2)=(\underbar{r}, \underbar{u})
\in\R^{3N}\times\R^{3N}$ and $\underbar{y}=\dot{\underbar{u}}_1\in\R^3$,
in view of (\ref{eff-regul-bar}) we then need to solve
\begin{eqnarray}
   \dot{\underbar{x}} & = & (\underbar{u}, \dot{\underbar{u}})
   =\Big(\underbar{u}, \dot{\underbar{u}}_1,
   \underbar{U}_{\,2N}(\underbar{r}, \underbar{u},
   \dot{\underbar{u}}_1, \eps)\Big)=\Big(\underbar{x}_2, \underbar{y},
   \underbar{U}_{\,2N}(\underbar{x}, \underbar{y}, \eps)\Big)
   =:f_1(\underbar{x}, \underbar{y}, \eps), \label{WAM}
   \\ \eps^{3/2}\dot{\underbar{y}} & = & \Phi_1\Big(\underbar{r}, \underbar{u},
   \dot{\underbar{u}}_1, \underbar{U}_{\,2N}(\underbar{r}, \underbar{u},
   \dot{\underbar{u}}_1, \eps), \eps\Big)=\Phi_1\Big(\underbar{x}, \underbar{y},
   \underbar{U}_{\,2N}(\underbar{x}, \underbar{y}, \eps), \eps\Big)
   =:g_1(\underbar{x}, \underbar{y}, \eps), \label{LvB}
\end{eqnarray}
which turns out to be a standard singular perturbation problem,
up to the factor of $\dot{\underbar{y}}$ which is $\eps^{3/2}$ rather than $\eps$.
To achieve the latter, we transform
\[ x(t)=\underbar{x}(\sqrt{\eps}t),\quad y(t)=\underbar{y}(\sqrt{\eps}t), \]
and arrive at the system
\begin{equation}\label{stasky}
   \dot{x}=\sqrt{\eps}f_1(x, y, \eps)=:f(x, y, \eps),\quad
   \eps\dot{y}=g_1(x, y, \eps)=:g(x, y, \eps),
\end{equation}
which is of standard form and can be shown to allow for the existence
of a (locally) invariant manifold if $\eps>0$ is small enough.

To carry out this program, we first have to investigate the solvability
of (\ref{2-N-eqs}).

\begin{lemma} There exist $\eps_0>0$ and a smooth function
$\underbar{U}_{\,2N}: \R^{3N}\times\R^{3N}\times\R^3\times [0, \eps_0]\to\R^{3N-3}$,
$(\dot{\underbar{u}}_2, \ldots, \dot{\underbar{u}}_N)
=\underbar{U}_{\,2N}(\underbar{r}, \underbar{u}, \dot{\underbar{u}}_1, \eps)$,
such that (\ref{augst}) is satisfied.
\end{lemma}
{\bf Proof\,:} With $z=M^{{\rm reg}}(A\underbar{u}, A\dot{\underbar{u}},
\eps)-G^{{\rm reg}}(A\underbar{r}, A\underbar{u}, A\dot{\underbar{u}}, \eps)
\in\R^{3N}$ we need to solve
\begin{equation}\label{lazut}
   0=e_2 z_1-e_1 z_2,\quad 0=e_3 z_2-e_2 z_3,\quad 0=e_4 z_3-e_3 z_4,\quad\ldots\,,
   \quad 0=e_N z_{N-1}-e_{N-1}z_N,
\end{equation}
for $(\dot{\underbar{u}}_2, \ldots, \dot{\underbar{u}}_N)$, cf.~(\ref{At-def}).
According to (\ref{brien}) and (\ref{gabb}) we have
\[ z_\alpha=M_{\alpha}^{{\rm reg}}\Big((A\underbar{u})_\alpha, \eps\Big)
   (A\dot{\underbar{u}})_\alpha-G_{\alpha}^{{\rm reg}}\Big(A\underbar{r},
   A\underbar{u}, A\dot{\underbar{u}}, \eps\Big)\in\R^3,\quad
   1\le\alpha\le N. \]
To decompose $z_\alpha$ appropriately, we introduce the notation
$\dot{\underbar{u}}=(\dot{\underbar{u}}_1, \eta)\in\R^3\times\R^{3N-3}$
with $\eta=(\dot{\underbar{u}}_2, \ldots, \dot{\underbar{u}}_N)$,
and accordingly we split $A\dot{\underbar{u}}=A_1\dot{\underbar{u}}_1
+A_{2N}\eta$ for suitable linear $A_1: \R^3\to\R^{3N}$
and $A_{2N}: \R^{3N-3}\to\R^{3N}$, as given in (\ref{Amart-def}).
{}From the definition of $M_{\alpha}^{{\rm reg}}$
and $G_{\alpha}^{{\rm reg}}$, cf.~(\ref{Malph-def2}), (\ref{Galph-def2}),
and (\ref{reguluxi}), we then obtain
\begin{eqnarray*}
   M_{\alpha}^{{\rm reg}}\Big((A\underbar{u})_\alpha, \eps\Big)
   (A\dot{\underbar{u}})_\alpha & = & m_\alpha (A\dot{\underbar{u}})_\alpha
   +\eps\bigg(\frac{1}{2}\,m_\alpha^{\ast}
   \Big((A\underbar{u})_\alpha^{{\rm reg}}\Big)^2
   +m_\alpha^{\ast}\Big((A\underbar{u})_\alpha^{{\rm reg}}
   \cdot (A\dot{\underbar{u}})_\alpha\Big)(A\underbar{u})_\alpha^{{\rm reg}}\bigg)
   \\ & = & m_\alpha (A_{2N}\eta)_\alpha+\bigg[m_\alpha (A_1\dot{\underbar{u}}_1)_\alpha
   +\frac{\eps}{2}\,m_\alpha^{\ast}\Big((A\underbar{u})_\alpha^{{\rm reg}}\Big)^2
   \\ & & +\eps m_\alpha^{\ast}\Big((A\underbar{u})_\alpha^{{\rm reg}}
   \cdot (A_1\dot{\underbar{u}}_1)_\alpha\Big)(A\underbar{u})_\alpha^{{\rm reg}}\bigg]
   +\eps m_\alpha^{\ast}\Big((A\underline{u})_\alpha^{{\rm reg}}
   \cdot (A_{2N}\eta)_\alpha\Big)(A\underbar{u})_\alpha^{{\rm reg}}
   \\ & =: & m_\alpha (A_{2N}\eta)_\alpha
   +M_{\alpha}^{(1)}\Big((A\underbar{u})_\alpha^{{\rm reg}},
   (A_1\dot{\underbar{u}}_1)_\alpha, \eps\Big)
   +\eps M_{\alpha}^{(2)}\Big((A\underbar{u})_\alpha^{{\rm reg}},
   (A_{2N}\eta)_\alpha\Big), \\
   G_{\alpha}^{{\rm reg}}\Big(A\underbar{r},
   A\underbar{u}, A\dot{\underbar{u}}, \eps\Big) & = &
   \frac{e_\alpha}{4\pi}\sum_{\stackrel{\beta=1}{\beta\neq\alpha}}^N
   e_\beta\frac{\zeta_{\alpha\beta}}{|\zeta_{\alpha\beta}|^3} \\ & &
   +\eps\frac{e_\alpha}{4\pi}\sum_{\stackrel{\beta=1}{\beta\neq\alpha}}^N
   e_\beta\Bigg(-\frac{1}{2|\zeta_{\alpha\beta}|}\,(A\dot{\underbar{u}})_\beta
   -\frac{((A\dot{\underbar{u}})_\beta\cdot\zeta_{\alpha\beta})}
   {2|\zeta_{\alpha\beta}|^3}\,\zeta_{\alpha\beta} \\ & & \hspace{5em}
   +\,\frac{\Big((A\underbar{u})_\beta^{{\rm reg}}\Big)^2}
   {2|\zeta_{\alpha\beta}|^3}\,\zeta_{\alpha\beta}
   -\frac{3{\Big((A\underbar{u})_\beta^{{\rm reg}}\cdot\zeta_{\alpha\beta}\Big)}^2}
   {2|\zeta_{\alpha\beta}|^5}\,\zeta_{\alpha\beta} \\[1ex] & & \hspace{5em}
   -\,\frac{((A\underbar{u})_\alpha^{{\rm reg}}\cdot
   (A\underbar{u})_\beta^{{\rm reg}})}{|\zeta_{\alpha\beta}|^3}\,
   \zeta_{\alpha\beta}+\frac{((A\underbar{u})_\alpha^{{\rm reg}}\cdot\zeta_{\alpha\beta})}
   {|\zeta_{\alpha\beta}|^3}\,(A\underbar{u})_\beta^{{\rm reg}}\Bigg)
   \\ & =: & G_{\alpha}^{(0)}(A\underbar{r})
   +\eps G_{\alpha}^{(2)}\Big(A\underbar{r}, A_{2N}\eta\Big)
   +\eps G_{\alpha}^{(1)}\Big(A\underbar{r}, (A\underbar{u})^{{\rm reg}},
   A_1\dot{\underbar{u}}_1\Big),
\end{eqnarray*}
with
\begin{equation}\label{scatle}
   \zeta_{\alpha\beta}=\chi_2\Big(|(A\underbar{r})_\alpha-(A\underbar{r})_\beta|\Big)
   \Big[(A\underbar{r})_\alpha-(A\underbar{r})_\beta\Big],
\end{equation}
cf.~(\ref{reguluxi}), and
\begin{eqnarray*}
   G_{\alpha}^{(0)}(A\underbar{r}) & = & \frac{e_\alpha}{4\pi}\sum_{\stackrel{\beta=1}
   {\beta\neq\alpha}}^N e_\beta\frac{\zeta_{\alpha\beta}}{|\zeta_{\alpha\beta}|^3},
   \\ G_{\alpha}^{(2)}\Big(A\underbar{r}, A_{2N}\eta\Big)
   & = & \frac{e_\alpha}{4\pi}\sum_{\stackrel{\beta=1}{\beta\neq\alpha}}^N
   e_\beta\Bigg(-\frac{1}{2|\zeta_{\alpha\beta}|}\,(A_{2N}\eta)_\beta
   -\frac{((A_{2N}\eta)_\beta\cdot\zeta_{\alpha\beta})}
   {2|\zeta_{\alpha\beta}|^3}\,\zeta_{\alpha\beta}\Bigg).
\end{eqnarray*}
In view of (\ref{regul-bd}) we note the bounds
\begin{equation}\label{fmino}
   \Big|M_{\alpha}^{(2)}\Big((A\underbar{u})_\alpha^{{\rm reg}},
   (A_{2N}\eta)_\alpha\Big)\Big|+\Big|G_{\alpha}^{(2)}\Big(A\underbar{r},
   A_{2N}\eta\Big)\Big|\le C|\eta|,\quad 1\le\alpha\le N,
\end{equation}
which are valid for all $(\underbar{r}, \underbar{u}, \eta)\in\R^{3N}\times\R^{3N}
\times\R^{3N-3}$. Omitting the arguments and recalling (\ref{Amart-def}),
the equations from (\ref{lazut}) can be rewritten as
\begin{eqnarray*}
   0 & = & \Big(e_2^2m_1+e_1^2m_2\Big)\dot{\underbar{u}}_2
   -e_1e_3m_2\,\dot{\underbar{u}}_3
   +\eps\Big(e_2 M_1^{(2)}-e_1 M_2^{(2)}\Big)
   +\eps\Big(e_1G_2^{(2)}-e_2G_1^{(2)}\Big)
   \\ & & +\,\Big(e_1 G_2^{(0)}-e_2 G_1^{(0)}\Big)
   +\Big(e_2 M_1^{(1)}-e_1 M_2^{(1)}\Big)
   +\eps\Big(e_1G_2^{(1)}-e_2G_1^{(1)}\Big), \\
   0 & = & -e_1e_3m_2\,\dot{\underbar{u}}_2
   +\Big(e_3^2m_2+e_2^2m_3\Big)\dot{\underbar{u}}_3
   -e_2e_4m_3\,\dot{\underbar{u}}_4+\eps\Big(e_3 M_2^{(2)}-e_2 M_3^{(2)}\Big)
   +\eps\Big(e_2G_3^{(2)}-e_3G_2^{(2)}\Big)
   \\ & & +\,\Big(e_2 G_3^{(0)}-e_3 G_2^{(0)}\Big)
   +\Big(e_3 M_2^{(1)}-e_2 M_3^{(1)}\Big)
   +\eps\Big(e_2G_3^{(1)}-e_3G_2^{(1)}\Big), \\
   0 & = & -e_2e_4m_3\,\dot{\underbar{u}}_3
   +\Big(e_4^2m_3+e_3^2m_4\Big)\dot{\underbar{u}}_4
   -e_3e_5m_4\,\dot{\underbar{u}}_5+\eps\Big(e_4 M_3^{(2)}-e_3 M_4^{(2)}\Big)
   +\eps\Big(e_3G_4^{(2)}-e_4G_3^{(2)}\Big)
   \\ & & +\,\Big(e_3 G_4^{(0)}-e_4 G_3^{(0)}\Big)
   +\Big(e_4 M_3^{(1)}-e_3 M_4^{(1)}\Big)
   +\eps\Big(e_3G_4^{(1)}-e_4G_3^{(1)}\Big), \\
   \vdots & & \hspace{13em} \vdots \\
   0 & = & -e_{N-2}e_N m_{N-1}\,\dot{\underbar{u}}_{N-1}
   +\Big(e_N^2m_{N-1}+e_{N-1}^2m_N\Big)\dot{\underbar{u}}_N
   \\ & & +\,\eps\Big(e_N M_{N-1}^{(2)}-e_{N-1} M_N^{(2)}\Big)
   +\eps\Big(e_{N-1}G_N^{(2)}-e_N G_{N-1}^{(2)}\Big)
   \\ & & +\,\Big(e_{N-1} G_N^{(0)}-e_N G_{N-1}^{(0)}\Big)
   +\Big(e_N M_{N-1}^{(1)}-e_{N-1} M_N^{(1)}\Big)
   +\eps\Big(e_{N-1}G_N^{(1)}-e_N G_{N-1}^{(1)}\Big).
\end{eqnarray*}
Note that this is a linear system for
$\eta=(\dot{\underbar{u}}_2, \ldots, \dot{\underbar{u}}_N)$, since both
$M_\alpha^{(2)}$ and $G_\alpha^{(2)}$ do depend on $\eta$ linearly,
whereas $M_\alpha^{(1)}$, $G_\alpha^{(0)}$, and $G_\alpha^{(1)}$ are
independent of $\eta$. Accordingly, the system has the form
\begin{equation}\label{kpt-gl}
   0={\cal M}^{(0)}\eta+\eps {\cal M}^{(2)}(\underbar{r}, \underbar{u})\eta
   -{\cal R}(\underbar{r}, \underbar{u}, \dot{\underbar{u}}_1, \eps),
\end{equation}
with the $(3N-3)\times (3N-3)$-matrix ${\cal M}^{(0)}={\cal M}^{(0)}(e_1, \ldots, e_N,
m_1, \ldots, m_N)$ being defined as
\begin{eqnarray}\label{M0def}
   {\cal M}^{(0)}\eta & = & \bigg(\Big(e_2^2m_1+e_1^2m_2\Big)\eta_2
   -e_1e_3m_2\eta_3,\,-e_1e_3m_2\eta_2+\Big(e_3^2m_2+e_2^2m_3\Big)\eta_3-e_2e_4m_3\eta_4,
   \nonumber\\ & & \hspace{2em} -e_2e_4m_3\eta_3+\Big(e_4^2m_3+e_3^2m_4\Big)\eta_4
   -e_3e_5m_4\eta_5\,,\ldots, \nonumber\\ & & \hspace{2em} -e_{N-2}e_N m_{N-1}\eta_{N-1}
   +\Big(e_N^2m_{N-1}+e_{N-1}^2m_N\Big)\eta_N\bigg),
\end{eqnarray}
and moreover
\begin{eqnarray*}
   {\cal M}^{(2)}(\underbar{r}, \underbar{u})\eta
   & = & \Big(e_\alpha M_{\alpha-1}^{(2)}-e_{\alpha-1} M_\alpha^{(2)}
   +e_{\alpha-1} G_{\alpha}^{(2)}-e_\alpha G_{\alpha-1}^{(2)}\Big)_{2\le\alpha\le N},
   \\ {\cal R}(\underbar{r}, \underbar{u}, \dot{\underbar{u}}_1, \eps)
   & = & \Big(e_\alpha G_{\alpha-1}^{(0)}-e_{\alpha-1} G_\alpha^{(0)}
   +e_{\alpha-1} M_\alpha^{(1)}-e_\alpha M_{\alpha-1}^{(1)}
   +\eps e_\alpha G_{\alpha-1}^{(1)}-\eps e_{\alpha-1} G_\alpha^{(1)}\Big)_{2\le\alpha\le N}.
\end{eqnarray*}
By (\ref{fmino}) we have
\[ \Big|{\cal M}^{(2)}(\underbar{r}, \underbar{u})\eta\Big|
   \le C|\eta|,\quad (\underbar{r}, \underbar{u}, \eta)\in\R^{3N}
   \times\R^{3N}\times\R^{3N-3}, \]
i.e., $|{\cal M}^{(2)}(\underbar{r}, \underbar{u})|\le C$
as a linear map, uniformly in $(\underbar{r}, \underbar{u})$.
Choosing $\eps>0$ small enough, hence (\ref{kpt-gl}) can be solved
for $\eta$ (even explicitly by means of a von Neumann-series),
as soon as we know that ${\cal M}^{(0)}$ is invertible;
this is verified in Lemma \ref{M0invert} below.
The associated solution function $\underbar{U}_{2N}$ is smooth w.r.t.~all variables since,
due to regularizing, $\zeta_{\alpha\beta}$ is a smooth function
of $\underbar{r}$. {\hfill$\Box$}\bigskip

\begin{lemma}\label{M0invert}
Denote ${\cal M}^{(0)}={\cal M}^{(0)}_N\in\R^{(3N-3)\times (3N-3)}$ the matrix
defined by (\ref{M0def}). Then
\[ \det {\cal M}^{(0)}=\bigg[\bigg(\prod_{j=2}^{N-1} e_j^2\bigg)
   \bigg(\sum_{j=1}^N e_j^2\prod_{\stackrel{i=1}{i\neq j}}^N m_i\bigg)\bigg]^3>0. \]
\end{lemma}
{\bf Proof\,:} The first observation to make is that
$\det {\cal M}^{(0)}_N=\Big[\det {\cal A}_N\Big]^3$, where
\[ {\cal A}_N=\left(\begin{array}{ccccc} e_2^2m_1+e_1^2m_2
   & -e_1e_3m_2 & 0 & \ldots & 0 \\
   -e_1e_3m_2 & e_3^2m_2+e_2^2m_3 & -e_2e_4m_3 & \ldots & 0 \\
   0 & -e_2e_4m_3 & e_4^2m_3+e_3^2m_4 & \ddots & 0 \\
   \vdots & \vdots & \ddots & \ddots & \vdots \\
   0 & 0 & 0 & \ldots & e_N^2m_{N-1}+e_{N-1}^2m_N
   \end{array}\right)\in\R^{N-1}; \]
this can be verified by induction. Hence we need to prove that
\begin{equation}\label{AN-det}
   \det {\cal A}_N=\bigg(\prod_{j=2}^{N-1} e_j^2\bigg)
   \bigg(\sum_{j=1}^N e_j^2\prod_{\stackrel{i=1}{i\neq j}}^N m_i\bigg).
\end{equation}
For $N=2$ we have $\det {\cal A}_2={\cal A}_2=e_2^2m_1+e_1^2m_2$,
whereas for $N=3$ we calculate $\det {\cal A}_3=
e_2^2(e_1^2m_2m_3+e_2^2m_1m_3+e_3^2m_1m_2)$, thus (\ref{AN-det})
is satisfied. If this is already known for some $N$, then
$\det {\cal A}_{N+1}=(e_{N+1}^2m_N+e_N^2m_{N+1})\det {\cal A}_N
-e_{N-1}^2e_{N+1}^2m_N^2 \det {\cal A}_{N-1}$ and the induction
hypothesis lead to
\begin{eqnarray*}
   \det {\cal A}_{N+1} & = & (e_{N+1}^2m_N+e_N^2m_{N+1})\bigg(\prod_{j=2}^{N-1} e_j^2\bigg)
   \bigg(\sum_{j=1}^N e_j^2\prod_{\stackrel{i=1}{i\neq j}}^N m_i\bigg)
   \\ & & -\,e_{N-1}^2e_{N+1}^2m_N^2\bigg(\prod_{j=2}^{N-2} e_j^2\bigg)
   \bigg(\sum_{j=1}^{N-1} e_j^2\prod_{\stackrel{i=1}{i\neq j}}^{N-1} m_i\bigg)
   \\ & = & \bigg(\prod_{j=2}^{N-1}e_j^2\bigg)
   \bigg\{e_{N+1}^2m_Ne_N^2\bigg(\prod_{\stackrel{i=1}{i\neq N}}^N m_i\bigg)
   +e_{N+1}^2m_N\bigg(\sum_{j=1}^{N-1} e_j^2\prod_{\stackrel{i=1}{i\neq j}}^N m_i\bigg)
   \\ & & \hspace{5em}
   +\,e_N^2m_{N+1}\bigg(\sum_{j=1}^N e_j^2\prod_{\stackrel{i=1}{i\neq j}}^N m_i\bigg)
   -e_{N+1}^2m_N^2\bigg(\sum_{j=1}^{N-1} e_j^2
   \prod_{\stackrel{i=1}{i\neq j}}^{N-1} m_i\bigg)\bigg\}
   \\ & = & \bigg(\prod_{j=2}^N e_j^2\bigg)
   \bigg\{e_{N+1}^2m_N\bigg(\prod_{\stackrel{i=1}{i\neq N}}^N m_i\bigg)
   +m_{N+1}\bigg(\sum_{j=1}^N e_j^2\prod_{\stackrel{i=1}{i\neq j}}^N m_i\bigg)
   \bigg\}=\bigg(\prod_{j=2}^N e_j^2\bigg)
   \bigg(\sum_{j=1}^{N+1} e_j^2\prod_{\stackrel{i=1}{i\neq j}}^{N+1} m_i\bigg),
\end{eqnarray*}
completing the proof of (\ref{AN-det}). {\hfill$\Box$}\bigskip

The next step is to study the second equation in (\ref{stasky})
for $\eps=0$ in greater detail.

\begin{lemma}\label{sing-vor} For $\eps=0$ the equation
$g(\underbar{x}, \underbar{y}, 0)=0$ is solved for $\underbar{y}=\dot{\underbar{u}}_1$
by the function
\[ \underbar{y}=\underbar{h}_0(\underbar{x})=\underbar{h}_0(\underbar{r}, \underbar{u})
   =\underbar{h}_0(\underbar{r})=\frac{1}{4\pi e^2}\sum_{1\le\alpha<\beta\le N}
   e_\alpha e_\beta\bigg(\frac{e_\alpha}{m_\alpha}-\frac{e_\beta}{m_\beta}\bigg)
   \frac{\zeta_{\alpha\beta}}{|\zeta_{\alpha\beta}|^3},\quad\mbox{with}\quad
   e^2=\sum_{\alpha=1}^N e_\alpha^2, \]
cf.~also (\ref{scatle}). Moreover, the eigenvalues
of $D_{\underbar{y}}\,g(\underbar{x}, \underbar{y}, 0)\in\R^{3\times 3}$
are $\frac{6\pi}{e^4}\sum_{\alpha=1}^N e_\alpha^2 m_{\alpha}>0$ ($3$ times),
and hence bounded away from the imaginary axis.
\end{lemma}
{\bf Proof\,:} The simplest way to find $\underbar{y}=\underbar{h}_0(\underbar{x})$
is to use the original variables and consider (\ref{eff-regul-P})
for $\eps=0$. Since
\begin{equation}\label{lita}
   M^{{\rm reg}}(u, \dot{u}, 0)=\Big(m_{\alpha}\dot{u}_\alpha\Big)_{1\le\alpha\le N}
   \quad\mbox{and}\quad G^{{\rm reg}}(r, u, \dot{u}, 0)
   =\bigg(\frac{e_\alpha}{4\pi}\sum_{\stackrel{\beta=1}{\beta\neq\alpha}}^N
   e_\beta\,\frac{\xi_{\alpha\beta}^{{\rm reg}}}{|\xi_{\alpha\beta}^{{\rm reg}}|^3}
   \bigg)_{1\le\alpha\le N},
\end{equation}
cf.~(\ref{Malph-def2}) and (\ref{Galph-def2}), eq.~(\ref{eff-regul-P}) reads as
\begin{equation}\label{psycch}
   m_{\alpha}\dot{u}_\alpha=\frac{e_\alpha}{4\pi}
   \sum_{\stackrel{\beta=1}{\beta\neq\alpha}}^N
   e_\beta\,\frac{\xi_{\alpha\beta}^{{\rm reg}}}{|\xi_{\alpha\beta}^{{\rm reg}}|^3},
   \quad 1\le\alpha\le N.
\end{equation}
The relevant transformations are $r=A\underbar{r}$ and $\dot{u}=A\dot{\underbar{u}}$,
whence $\dot{\underbar{u}}_1=(A^{-1}\dot{u})_1$. It may be verified that
$A^{-1}$ has the general form
\begin{equation}\label{Ainv-form}
   A^{-1}z=e^{-2}(\sum_{\alpha=1}^N e_\alpha z_\alpha,
   \ast, \ldots, \ast),\quad z=(z_1, \ldots, z_N)\in (\R^3)^N,
\end{equation}
and therefore (\ref{psycch}) implies
\[ \underbar{y}=\dot{\underbar{u}}_1
   =\frac{1}{4\pi e^2}\sum_{\alpha=1}^N\frac{e_\alpha^2}{m_\alpha}
   \sum_{\stackrel{\beta=1}{\beta\neq\alpha}}^N
   e_\beta\,\frac{\zeta_{\alpha\beta}}{|\zeta_{\alpha\beta}|^3}
   =\frac{1}{4\pi e^2}\sum_{1\le\alpha<\beta\le N}
   e_\alpha e_\beta\bigg(\frac{e_\alpha}{m_\alpha}-\frac{e_\beta}{m_\beta}\bigg)
   \frac{\zeta_{\alpha\beta}}{|\zeta_{\alpha\beta}|^3}, \]
the latter due to $\zeta_{\beta\alpha}=-\zeta_{\alpha\beta}$.
For the eigenvalues of $D_{\underbar{y}}\,g(\underbar{x}, \underbar{y}, 0)$,
we note that due to (\ref{stasky}), (\ref{LvB}), (\ref{Phi-defi}), and (\ref{At-def})
\begin{eqnarray*}
   g(\underbar{x}, \underbar{y}, 0) & =& \Phi_1\Big(\underbar{r}, \underbar{u},
   \dot{\underbar{u}}_1, \underbar{U}_{\,2N}(\underbar{r}, \underbar{u},
   \dot{\underbar{u}}_1, 0), 0\Big) \\
    & = & 6\pi e^{-4}\,\Big[A^t\Big(M^{{\rm reg}}(A\underbar{u}, A\dot{\underbar{u}}, 0)
   -G^{{\rm reg}}(A\underbar{r}, A\underbar{u}, A\dot{\underbar{u}}, 0)\Big)\Big]_1
   \\ & = & 6\pi e^{-4}\sum_{\alpha=1}^N e_\alpha
   \Big([M^{{\rm reg}}(A\underbar{u}, A\dot{\underbar{u}}, 0)]_\alpha
   -[G^{{\rm reg}}(A\underbar{r}, A\underbar{u}, A\dot{\underbar{u}}, 0)]_\alpha\Big)
   \\ & = & 6\pi e^{-4}\sum_{\alpha=1}^N e_\alpha\bigg(m_{\alpha}\dot{u}_\alpha
   -\frac{e_\alpha}{4\pi}\sum_{\stackrel{\beta=1}{\beta\neq\alpha}}^N
   e_\beta\,\frac{\xi_{\alpha\beta}^{{\rm reg}}}{|\xi_{\alpha\beta}^{{\rm reg}}|^3}\bigg),
\end{eqnarray*}
where we have used (\ref{lita}) and passed to the original variables.
Observing $\underbar{y}=\dot{\underbar{u}}_1$,
$\dot{u}_\alpha=[A\dot{\underbar{u}}]_\alpha$, and (\ref{Amart-def}), we hence obtain
\[ D_{\underbar{y}}\,g(\underbar{x}, \underbar{y}, 0)
   =6\pi e^{-4}\sum_{\alpha=1}^N e_\alpha m_{\alpha}
   \bigg(\frac{\partial\dot{u}_\alpha}{\partial\dot{\underbar{u}}_1}\bigg)
   =6\pi e^{-4}\sum_{\alpha=1}^N e_\alpha m_{\alpha} e_\alpha {\rm id}_{\R^3}
   =\bigg(6\pi e^{-4}\sum_{\alpha=1}^N e_\alpha^2 m_{\alpha}\bigg)\,{\rm id}_{\R^3}, \]
and this yields the claim.
{\hfill$\Box$}\bigskip

According to Lemma \ref{sing-vor} we see that the assumptions $(H1)$-$(H3)$ of
\cite[Sect.~1.1\,\& 1.2]{jones} are satisfied, and therefore we find
a (locally) invariant manifold for (\ref{stasky}); cf.~\cite[Thm.~2]{jones}.
Transferred back to (\ref{WAM}) and (\ref{LvB}) this result can be applied as follows.
We define
\[ \underbar{K}_0=\Big\{\underbar{x}=(\underbar{r}, \underbar{u})\in\R^{3N}\times\R^{3N}:
       |A\underbar{r}|\le 4C_q,\,|A\underbar{u}|\le 4C_v\Big\}  \]
with $C_q$ and $C_v$ from (\ref{qbd}) and (\ref{v-bound}), respectively,
and we let
\[ \underbar{I}_0=\Big\{(\underbar{x}, \underbar{y}):
   \underbar{y}=\underbar{h}_0(\underbar{x}),\,\underbar{x}\in\underbar{K}_0\Big\}, \]
with $\underbar{h}_0(\underbar{x})$ from Lemma \ref{sing-vor}.

\begin{lemma}\label{prel} For every $k\in\N$ with $k\ge 4$ there exist $\eps_1>0$
and a $C^k$-function $\underbar{h}: [0, \eps_1]\times \underbar{K}_0\to\R^3$ such that
\begin{equation}\label{barI-eps}
   \underbar{I}_\eps=\Big\{(\underbar{x}, \underbar{y}):
   \underbar{y}=\underbar{h}_\eps(\underbar{x}),\,\underbar{x}\in\underbar{K}_0\Big\}
\end{equation}
is locally invariant w.r.t.~(\ref{WAM}) and (\ref{LvB}), where
$\underbar{h}_\eps(\underbar{x})=\underbar{h}(\eps, \underbar{x})$.
In particular this means the following.
Consider $(\underbar{x}(\tau_0), \underbar{y}(\tau_0))
=(\underbar{r}(\tau_0), \underbar{u}(\tau_0), \dot{\underbar{u}}_1(\tau_0))
\in\R^{3N}\times\R^{3N}\times\R^3$ such that
$|A\underbar{r}(\tau_0)|\le 2C_q$ as well as $|A\underbar{u}(\tau_0)|\le 2C_v$ and
\[ \dot{\underbar{u}}_1(\tau_0)
   =\underbar{h}_\eps\Big(\underbar{r}(\tau_0), \underbar{u}(\tau_0)\Big) \]
hold. Then the corresponding solution $(\underbar{x}(t), \underbar{y}(t))
=(\underbar{r}(t), \underbar{u}(t), \dot{\underbar{u}}_1(t))$
of (\ref{WAM}) and (\ref{LvB}) with this data at $t=\tau_0$ exists at least until
$\tau_1>\tau_0$ and satisfies
\begin{equation}\label{wldsch}
   \dot{\underbar{u}}_1(t)=\underbar{h}_\eps\Big(\underbar{r}(t), \underbar{u}(t)\Big),
   \quad t\in [\tau_0, \tau_1],
\end{equation}
where $\tau_1>\tau_0$ denotes the longest time such that $|A\underbar{r}(t)|\le 3C_q$
and $|A\underbar{u}(t)|\le 3C_v$ for $t\in [\tau_0, \tau_1]$. Moreover we have
\begin{equation}\label{C1-diff}
   {|\underbar{h}_\eps-\underbar{h}_0|}_{C^1_b(\underline{K}_0)}\le C\eps
\end{equation}
for a constant $C>0$ depending only on the input constants $C_q$ and $C_v$.
The dynamics on $\underbar{I}_\eps$ is governed by $\dot{\underbar{x}}
=f_1(\underbar{x}, \underbar{h}_\eps(\underbar{x}), \eps)$, i.e.,
\begin{equation}\label{effgl-bar}
   \dot{\underbar{r}}=\underbar{u},\quad
   \dot{\underbar{u}}_1=\underbar{h}_\eps(\underbar{r}, \underbar{u}),\quad
   (\dot{\underbar{u}}_2, \ldots, \dot{\underbar{u}}_N)
   =\underbar{U}_{\,2N}\Big(\underbar{r}, \underbar{u},
   \underbar{h}_\eps(\underbar{r}, \underbar{u}), \eps\Big),
\end{equation}
cf.~(\ref{WAM}).
\end{lemma}

The $C^1_b$-estimate (\ref{C1-diff}) is not given explicitly in \cite{jones},
but may be validated along the lines of \cite{saka,ryba} where the analogous
statement is shown for $C^0_b$. With this preparation we can finally
come to the \medskip

\noindent
{\bf Proof of Theorem \ref{invmfthm}\,:} We only have to undo
the transformations to find $\hat{h}_\eps$, and this way we arrive at
\[ \hat{h}_\eps(r, u)=A\Big(\underbar{h}_\eps(A^{-1}r, A^{-1}u),\,
   \underbar{U}_{2N}(A^{-1}r, A^{-1}u, \underbar{h}_\eps(A^{-1}r, A^{-1}u),
   \eps)\Big),\quad (r, u)\in K_0. \]
To verify that then ${\cal I}_\eps$ from (\ref{Ieps-def}) is locally invariant
in the sense stated, we note that (\ref{samk}) and the definition
of $\chi_1$ and $\chi_2$, cf.~(\ref{reguluxi}), imply that
(\ref{effect-equ}) and (\ref{eff-regul}) do agree for $t\in [\tau_0, \tau_1]$.
Hence (\ref{effect-equ}) is equivalent to (\ref{WAM}) and (\ref{LvB})
via the transformation from (\ref{r-barr-trafo}) for $t\in [\tau_0, \tau_1]$,
and therefore the local invariance of ${\cal I}_\eps$ is a direct consequence
of the local invariance of $\underbar{I}_\eps$ from (\ref{barI-eps})
in Lemma \ref{prel}, for (\ref{uh-gl}) cf.~(\ref{wldsch}).
Concerning (\ref{sumdotu-form}), in view of (\ref{C1-diff})
we can write $\underbar{h}_\eps(\underbar{r}, \underbar{u})
=\underbar{h}_0(\underbar{r})+\Delta_\eps$
with ${|\Delta_\eps|}_{C^1_b(\underline{K}_0)}\le C\eps$.
Hence (\ref{Ainv-form}), (\ref{effgl-bar}), and Lemma \ref{sing-vor} imply
\begin{eqnarray}\label{fox2}
    e^{-2}\sum_{\beta=1}^N e_\beta\dot{u}_\beta & = & {(A^{-1}\dot{u})}_1
    =\dot{\underbar{u}}_1=\underbar{h}_\eps(\underbar{r}, \underbar{u})
    =\underbar{h}_0(\underbar{r})+\Delta_\eps \nonumber
    \\ & = & \frac{1}{4\pi e^2}\sum_{1\le\beta<\beta'\le N}
    e_\beta e_{\beta'}\bigg(\frac{e_\beta}{m_\beta}-\frac{e_{\beta'}}{m_{\beta'}}\bigg)
    \frac{\zeta_{\beta\beta'}}{|\zeta_{\beta\beta'}|^3}+\Delta_\eps \nonumber
    \\ & = & \frac{1}{4\pi e^2}\sum_{1\le\beta<\beta'\le N}
    e_\beta e_{\beta'}\bigg(\frac{e_\beta}{m_\beta}-\frac{e_{\beta'}}{m_{\beta'}}\bigg)
    \frac{\xi_{\beta\beta'}}{|\xi_{\beta\beta'}|^3}+\Delta_\eps,
\end{eqnarray}
the latter since here $\zeta_{\beta\beta'}=(A\underline{r})_\beta
-(A\underline{r})_{\beta'}=r_\beta-r_{\beta'}=\xi_{\beta\beta'}$
according to (\ref{samk}), cf.~(\ref{scatle}). {}From (\ref{fox2})
we obtain (\ref{sumdotu-form}) by differentiation and observing that
${|\dot{\Delta}_\eps|}_{C^0_b(\underline{K}_0)}\le C\eps$.
{\hfill$\Box$}\bigskip


\setcounter{equation}{0}

\section{Comparison of the full and the effective system}

We prove Theorem \ref{main-thm} and assume that the stage is set
as is described in the theorem. Then we define
\begin{equation}\label{hepsi-def}
   h_\eps(r, u)=\eps^2\hat{h}_\eps\Big(\eps r, \eps^{-1/2}u\Big),
   \quad (r, u)\in K_\eps,
\end{equation}
with the function $\hat{h}$ from Theorem \ref{invmfthm};
note that $(r, u)\in K_\eps$ is equivalent to $(\eps r, \eps^{-1/2}u)\in K_0$,
cf.~(\ref{K0-def}). We let $\tau_0=\eps^{3/2}t_0$ and
\begin{equation}\label{trafo2}
   \bar{r}_\alpha(t)=\eps r_\alpha(\eps^{-3/2}t),\quad
   \bar{u}_\alpha(t)=\eps^{-1/2} u_\alpha(\eps^{-3/2}t),\quad 1\le\alpha\le N.
\end{equation}
Hence (\ref{ZM-data}) implies that
\begin{equation}\label{dorgem}
   \dot{\bar{u}}_\alpha(\tau_0)=\eps^{-2}\dot{u}_\alpha(t_0)
   =\eps^{-2} h_\eps\Big(r_\alpha(t_0), u_\alpha(t_0)\Big)
   =\hat{h}_\eps\Big(\eps r_\alpha(t_0), \eps^{-1/2}u_\alpha(t_0)\Big)
   =\hat{h}_\eps\Big(\bar{r}_\alpha(\tau_0), \bar{u}_\alpha(\tau_0)\Big)
\end{equation}
for $1\le\alpha\le N$. In addition we deduce from (\ref{qbd}), (\ref{v-bound}),
and (\ref{diff-bound}) that
\begin{eqnarray*}
   & & |\bar{r}(\tau_0)|=\eps |r(t_0)|=\eps |q(t_0)|
   :=\eps\max_{1\le\alpha\le N}|q_\alpha(t_0)|\le C_q, \\
   & & |\bar{u}(\tau_0)|=\eps^{-1/2} |u(t_0)|=\eps^{-1/2} |v(t_0)|\le C_v,\quad
   \mbox{and}\quad \\ & &
   C_\ast\le \eps |q_\alpha(t_0)-q_\beta(t_0)|
   =|\bar{r}_\alpha(\tau_0)-\bar{r}_\beta(\tau_0)|
   \le C^\ast\quad (\alpha\neq\beta),
\end{eqnarray*}
whence taking into account (\ref{dorgem}) we see that the assumptions
of Theorem \ref{invmfthm} are satisfied. We denote by $t_1\in ]t_0, T\eps^{-3/2}]$
the longest time such that
\begin{equation}\label{long-tim}
  |r(t)|\le 3C_q\eps^{-1},\quad |u(t)|\le 3C_v\sqrt{\eps},\quad\mbox{and}\quad
   (C_\ast/3)\eps^{-1}\le |r_\alpha(t)-r_\beta(t)|\le 3C^\ast\eps^{-1}
   \,\,\,(\alpha\neq\beta),
\end{equation}
are valid simultaneously for $t\in [t_0, t_1]$, corresponding to the longest time
$\tau_1\in ]\tau_0, T]$ such that
\[ |\bar{r}(t)|\le 3C_q,\quad |\bar{u}(t)|\le 3C_v,\quad\mbox{and}\quad
   (C_\ast/3)\le |\bar{r}_\alpha(t)-\bar{r}_\beta(t)|\le 3C^\ast
   \,\,\,(\alpha\neq\beta), \]
are verified for $t\in [\tau_0, \tau_1]$, cf.~(\ref{samk}). We infer
from Theorem \ref{invmfthm} that the solution $(r(t), u(t), \dot{u}(t))$
of the effective equation (\ref{LD-dyn-eff}) with data given by (\ref{ZM-data})
exists at least for $t\in [t_0, t_1]$ and satisfies
\begin{equation}\label{uh-gl-2}
   \dot{u}(t)=h_\eps\Big(r(t), u(t)\Big),\quad t\in [t_0, t_1],
\end{equation}
due to (\ref{uh-gl}). Moreover,
\begin{eqnarray}\label{sumdotu-form-2}
   \sum_{\beta=1}^N e_\beta\ddot{u}_\beta
   & = & \frac{1}{2}\sum_{\stackrel{\beta, \beta'=1}{\beta\neq\beta'}}^N
   \frac{e_\beta e_{\beta'}}{4\pi}
   \bigg(\frac{e_\beta}{m_\beta}-\frac{e_{\beta'}}{m_{\beta'}}\bigg)
   \bigg[\frac{1}{|\xi_{\beta\beta'}|^3}(u_\beta-u_{\beta'})
   \nonumber \\ & & \hspace{12em}
   -\frac{3}{|\xi_{\beta\beta'}|^5}\,\xi_{\beta\beta'}\cdot (u_\beta-u_{\beta'})
   \,\xi_{\beta\beta'}\bigg]+{\cal O}(\eps^{9/2})
\end{eqnarray}
for $t\in [t_0, t_1]$, by transforming (\ref{sumdotu-form}) back
utilizing (\ref{trafo2}).

We need to prove that $t_1=T\eps^{-3/2}$ holds,
and for this purpose we compare the true solution and the solution
of the effective equation (\ref{LD-dyn-eff}) for times $t\in [t_0, t_1]$.
We recall from (\ref{bjack}) in Lemma \ref{force-esti} that
\begin{equation}\label{cazen}
   M_\alpha(v_\alpha)\dot{v}_\alpha=G_\alpha(q, v, \dot{v})
   +\frac{e_\alpha}{6\pi}\sum_{\beta=1}^N e_\beta\ddot{v}_\beta
   +{\cal O}(\eps^4)=G_\alpha(q, v, \dot{v})+{\cal O}(\eps^{7/2}),
   \quad\alpha=1, \ldots, N,
\end{equation}
for $t\in [t_0, T\eps^{-3/2}]$, where in the last step we have used Lemma \ref{q3esti},
noting that $t_0\ge\tau_{\ast\ast}$. In addition, direct calculation reveals
that (\ref{LD-dyn-eff}) may be reformulated as
\begin{equation}\label{jdde}
   M_\alpha(u_\alpha)\dot{u}_\alpha=G_\alpha(r, u, \dot{u})
   +\frac{e_\alpha}{6\pi}\sum^N_{\beta=1}
   e_{\beta}\ddot{u}_{\beta},\quad\alpha=1, \ldots, N.
\end{equation}
{}From (\ref{sumdotu-form-2}) and (\ref{long-tim}) we deduce that
\[ \bigg|\sum_{\beta=1}^N e_\beta\ddot{u}_\beta\bigg|\le C\eps^{7/2},
   \quad t\in [t_0, t_1], \]
with $C>0$ depending only on the input constants, and accordingly we obtain
from (\ref{jdde}) that
 \begin{equation}\label{jdde2}
   M_\alpha(u_\alpha)\dot{u}_\alpha=G_\alpha(r, u, \dot{u})
   +{\cal O}(\eps^{7/2}),\quad\alpha=1, \ldots, N,\quad t\in [t_0, t_1].
\end{equation}
Next we observe that (\ref{uh-gl-2}) and (\ref{hepsi-def}) yield
\begin{equation}\label{woale}
   |\dot{u}_\alpha(t)|\le C\eps^2,\quad 1\le\alpha\le N,\quad
   t\in [t_0, t_1],
\end{equation}
since in particular $\hat{h}_\eps: [0, \eps_1]\times K_0\to\R^{3N}$ is bounded.
As we can use both the bounds from Lemma \ref{esti} and the bounds from
(\ref{long-tim}) and (\ref{woale}) for $t\in [t_0, t_1]$,
it is therefore possible to proceed exactly as in \cite[p.~449/450]{Nteil}
and to deduce, comparing (\ref{cazen}) and (\ref{jdde2}), that
\begin{equation}\label{diff3}
   |q_\alpha(t)-r_\alpha(t)|\le C\sqrt{\eps},\quad |v_\alpha(t)-u_\alpha(t)|\le C\eps^2,
   \quad 1\le\alpha\le N,\quad t\in [t_0, t_1];
\end{equation}
the only property of $t_1$ which enters here is $t_1\le C\eps^{-3/2}$.
However, we know from Lemma \ref{esti} that
\begin{equation}\label{long-tim-3}
   |q(t)|\le C_q\,\eps^{-1},\quad |v(t)|\le C_v\sqrt{\eps},\quad\mbox{and}\quad
   C_\ast\eps^{-1}\le |q_\alpha(t)-q_\beta(t)|\le C^\ast\eps^{-1}
   \,\,\,(\alpha\neq\beta),
\end{equation}
in particular for $t\in [t_0, T\eps^{-3/2}]$. Since all constants thus far
do depend only on the input constants, by choosing $\eps>0$ small enough
and observing (\ref{long-tim-3}) and (\ref{diff3}), we therefore arrive at
\[ |r(t)|\le 2C_q\eps^{-1},\quad |u(t)|\le 2C_v\sqrt{\eps},\quad\mbox{and}\quad
   (C_\ast/2)\eps^{-1}\le |r_\alpha(t)-r_\beta(t)|\le 2C^\ast\eps^{-1}
   \,\,\,(\alpha\neq\beta), \]
being valid for $t\in [t_0, t_1]$. In view of the definition of $t_1$
this leads to a contradiction, unless $t_1=T\eps^{-3/2}$. Hence
(\ref{diff3}) shows that (\ref{zollv}) holds as well,
since we can use (\ref{cazen}) and (\ref{jdde2}) to obtain from (\ref{diff3})
the further estimate $|\dot{v}_\alpha(t)-\dot{u}_\alpha(t)|\le C\eps^{7/2}$,
cf.~\cite[(4.6)]{Nteil}. In addition, (\ref{LD-dyn-eff-3})
is a consequence of (\ref{LD-dyn-eff}) and (\ref{sumdotu-form-2}).
Finally to verify (\ref{H-diff}), it follows by means of direct calculation
from Lemma \ref{force-esti} and from (\ref{v-bound}) that
\begin{equation}\label{haubes}
   \frac{d}{dt}\,{\cal H}_{{\rm RR}}(q(t), v(t), \dot{v}(t))
   =-\frac{1}{6\pi}\bigg(\sum_{\alpha=1}^N e_\alpha\dot{v}_\alpha(t)\bigg)^2
   +{\cal O}(\eps^{9/2}),\quad t\in [t_0, T\eps^{-3/2}].
\end{equation}
Therefore (\ref{zollv}), (\ref{haubes}), (\ref{energ-dec}), (\ref{ZM-data}),
(\ref{wend}), and (\ref{woale}) yield
\begin{eqnarray*}
   \lefteqn{{\cal H}_{{\rm D}}(q(t), v(t))-{\cal H}_{{\rm D}}(r(t), u(t))} \\
   & = & {\cal H}_{{\rm RR}}(q(t), v(t), \dot{v}(t))
   -{\cal H}_{{\rm RR}}(r(t), u(t), \dot{u}(t))
   +\sum_{\alpha, \beta=1}^N \frac{e_\alpha e_\beta}{6\pi}
   \,\Big[v_\alpha(t)\cdot\dot{v}_\beta(t)-u_\alpha(t)\cdot\dot{u}_\beta(t)\Big]
   \\ & = & {\cal H}_{{\rm RR}}(q(t), v(t), \dot{v}(t))
   -{\cal H}_{{\rm RR}}(r(t), u(t), \dot{u}(t))+{\cal O}(\eps^4)
   \\ & = & \int_{t_0}^t \frac{d}{dt'}\Big({\cal H}_{{\rm RR}}(q, v, \dot{v})
   -{\cal H}_{{\rm RR}}(r, u, \dot{u})\Big)\,dt'
   \\ & & +\,{\cal H}_{{\rm RR}}(q(t_0), v(t_0), \dot{v}(t_0))
   -{\cal H}_{{\rm RR}}(r(t_0), u(t_0), \dot{u}(t_0))+{\cal O}(\eps^4)
   \\ & = & \int_{t_0}^t\bigg[\frac{1}{6\pi}\bigg(\sum_{\alpha=1}^N
   e_\alpha\dot{u}_\alpha\bigg)^2-\frac{1}{6\pi}\bigg(\sum_{\alpha=1}^N
   e_\alpha\dot{v}_\alpha\bigg)^2+{\cal O}(\eps^{9/2})\,\bigg]\,dt'
   \\ & & +\,\sum_{\alpha, \beta=1}^N \frac{e_\alpha e_\beta}{6\pi}
   \,\Big[u_\alpha(t_0)\cdot\dot{u}_\beta(t_0)-v_\alpha(t_0)\cdot\dot{v}_\beta(t_0)\Big]
   +{\cal O}(\eps^4) \\ & = & \int_{t_0}^t\bigg[{\cal O}(\eps^{11/2})
   +{\cal O}(\eps^{9/2})\,\bigg]\,dt'+{\cal O}(\eps^4)={\cal O}(\eps^3),
\end{eqnarray*}
for $t\le C\eps^{-3/2}$. This completes the proof of Theorem \ref{main-thm}.
{\hfill$\Box$}\bigskip


\setcounter{equation}{0}

\section{Appendix A: Proof of Lemma \ref{q4esti}}
\label{append-sect}

This appendix is devoted to the proof of Lemma \ref{q4esti}.
To calculate $\stackrel{...}{v}(t)$ from (\ref{system2}), we first note that
$\frac{d}{dt}(m_{{\rm b}\alpha}\gamma_\alpha v_\alpha(t))
=m_{0\alpha}(v_\alpha(t))\dot{v}_\alpha(t)$, where the $(3\times 3)$-matrices
$m_{0\alpha}(v_\alpha)$ are defined as
\begin{equation}\label{m0form}
   m_{0\alpha}(v_\alpha)(z)=m_{{\rm b}\alpha}(\gamma_\alpha z
   +\gamma_\alpha^3 (v_\alpha\cdot z)v_\alpha),\quad z\in\R^3.
\end{equation}
Thus differentiating (\ref{system2}) once, it follows that
for $\alpha=1, \ldots, N$ we have
\begin{eqnarray}\label{dotv-1}
   \dot{v}_\alpha & = & {m_{0\alpha}(v_\alpha)}^{-1}
   \int d^3x\,\rho_\alpha(x-q_\alpha)\Big([E(x)-E_{v_\alpha}(x-q_\alpha)]
   +v_\alpha\wedge [B(x)-B_{v_\alpha}(x-q_\alpha)]\Big) \nonumber\\
   & = & {m_{0\alpha}(v_\alpha)}^{-1}
   \int d^3x\,\rho_\alpha(x)\Big(Z_1(x+q_\alpha, t)
   +v_\alpha\wedge Z_2(x+q_\alpha, t)\Big) + R_\alpha (t),
\end{eqnarray}
with
\begin{equation}\label{m0-1form}
   {m_{0\alpha}(v_\alpha)}^{-1}z={m_{{\rm b}\alpha}}^{-1}
   \gamma_\alpha^{-1}(z-(v_\alpha\cdot z)v_\alpha),\quad z\in\R^3,
\end{equation}
denoting the matrix inverse of $m_{0\alpha}(v_\alpha)$; here and henceforth
we often omit the argument $t$ of $q_\alpha$, $v_\alpha$, $\dot{v}_\alpha$, etc.
Moreover,
\begin{equation}\label{R-def}
   R_\alpha(t)={m_{0\alpha}(v_\alpha)}^{-1}
   \bigg(\sum_{\stackrel{\beta=1}{\beta\neq\alpha}}^N
   \int d^3x\,\rho_\alpha(x-q_\alpha)
   \Big[E_{v_\beta}(x-q_\beta)+v_\alpha\wedge
   B_{v_\beta}(x-q_\beta)\Big]\bigg),
\end{equation}
as well as
\begin{equation}\label{Z-def}
   Z(x, t)=\left(\begin{array}{c} Z_1(x, t) \\ Z_2(x, t)
   \end{array}\right)
   =\displaystyle\left(\begin{array}{c} E(x, t)-\sum_{\beta=1}^N
   E_{v_\beta(t)}(x-q_\beta(t)) \\[1ex]
   B(x, t)-\sum_{\beta=1}^N B_{v_\beta(t)}(x-q_\beta(t))\end{array}\right)
\end{equation}
in (\ref{dotv-1}). An important observation is that
\begin{equation}\label{Z-gleich}
   \dot{Z}(t)={\cal A}Z(t)-f(t)\,,\quad\mbox{with}\quad
   {\cal A}=\left(\begin{array}{cc} 0 & \nabla\wedge
   \\ -\nabla\wedge & 0 \end{array}\right)
\end{equation}
the Maxwell operator, and
\begin{equation}\label{f-def}
   f(x, t)=\left(\begin{array}{c} f_1(x, t) \\ f_2(x, t)
   \end{array}\right)=\sum_{\beta=1}^N\left(\begin{array}{c}
   (\dot{v}_\beta(t)\cdot \nabla_v) E_{v_\beta(t)}(x-q_\beta(t))
   \\[1ex] (\dot{v}_\beta(t)\cdot \nabla_v)
   B_{v_\beta(t)}(x-q_\beta(t))\end{array}\right);
\end{equation}
see \cite[Section 5.2]{Nteil}. We also note that $\nabla\cdot f_1=0=\nabla\cdot f_2$,
since $\nabla\cdot E_{v_\beta}=e_\beta\varphi$ and $\nabla\cdot B_{v_\beta}=0$
are calculated from (\ref{EBv-def}). Differentiating (\ref{dotv-1}) once more,
we obtain
\begin{eqnarray}\label{dotv-2}
   \ddot{v}_\alpha & = & \bigg(\frac{d}{dt}\,{m_{0\alpha}(v_\alpha)}^{-1}\bigg)
    m_{0\alpha}(v_\alpha)\Big[\dot{v}_\alpha-R_\alpha(t)\Big]
    +{m_{0\alpha}(v_\alpha)}^{-1}\,M_\alpha(t)
    \nonumber \\ & & +\,{m_{0\alpha}(v_\alpha)}^{-1}
    \int d^3x\,\rho_\alpha(x)\Big(\dot{v}_\alpha\wedge Z_2(x+q_\alpha, t)\Big)
    +\dot{R}_\alpha (t),
\end{eqnarray}
with the main term
\begin{equation}\label{Malph-def}
   M_\alpha(t)=\int d^3x\,\rho_\alpha(x)\Big[(L_\alpha(t)Z_1)
   (x+q_\alpha(t), t)+v_\alpha(t)\wedge
   (L_\alpha(t)Z_2)(x+q_\alpha(t), t)\Big],
\end{equation}
where $L_\alpha(t)\phi=(v_\alpha(t)\cdot\nabla)\phi+\dot{\phi}$ for a
general function $\phi=\phi(x, t)$. Finally, upon differentiating
(\ref{dotv-2}) it follows that
\begin{eqnarray}\label{dotv-3}
   \stackrel{...}{v}_\alpha & = & \bigg(\frac{d^2}{dt^2}\,{m_{0\alpha}(v_\alpha)}^{-1}
   \bigg)m_{0\alpha}(v_\alpha)\Big[\dot{v}_\alpha-R_\alpha(t)\Big]
   +\bigg(\frac{d}{dt}\,{m_{0\alpha}(v_\alpha)}^{-1}
   \bigg)\bigg(\frac{d}{dt}\,m_{0\alpha}(v_\alpha)\bigg)
   \Big[\dot{v}_\alpha-R_\alpha(t)\Big]
   \nonumber\\ & & +\,\bigg(\frac{d}{dt}\,{m_{0\alpha}(v_\alpha)}^{-1}
   \bigg)m_{0\alpha}(v_\alpha)\Big[\ddot{v}_\alpha-\dot{R}_\alpha(t)\Big]
   +\bigg(\frac{d}{dt}\,{m_{0\alpha}(v_\alpha)}^{-1}
   \bigg)M_\alpha(t)+{m_{0\alpha}(v_\alpha)}^{-1}\dot{M}_\alpha(t)
   \nonumber\\ & & +\,\ddot{R}_\alpha(t)+\bigg(\frac{d}{dt}
   \,{m_{0\alpha}(v_\alpha)}^{-1}\bigg)
   \int d^3x\,\rho_\alpha(x)\Big(\dot{v}_\alpha\wedge Z_2(x+q_\alpha, t)\Big)
   \nonumber\\ & & +\,{m_{0\alpha}(v_\alpha)}^{-1}
   \int d^3x\,\rho_\alpha(x)\Big[\ddot{v}_\alpha\wedge Z_2(x+q_\alpha, t)
   +\dot{v}_\alpha\wedge (L_\alpha(t)Z_2)(x+q_\alpha, t)\Big].
\end{eqnarray}
Most of these terms are directly seen to be at least of the desired
order ${\cal O}(\eps^5)$. Indeed, using (\ref{m0form}), (\ref{m0-1form}),
Lemma \ref{esti}, and Lemma \ref{q3esti}, one derives that
\begin{eqnarray}
   & \displaystyle\Big|m_{0\alpha}(v_\alpha)\Big|
   +\Big|{m_{0\alpha}(v_\alpha)}^{-1}\Big|\le C,\quad
   \bigg|\frac{d}{dt}\,{m_{0\alpha}(v_\alpha)}\bigg|
   +\bigg|\frac{d}{dt}\,{m_{0\alpha}(v_\alpha)}^{-1}\bigg|\le C\eps^{5/2},
   \quad t\in [0, T\eps^{-3/2}],\qquad\label{cava1} & \\
   & \displaystyle\bigg|\frac{d^2}{dt^2}\,{m_{0\alpha}(v_\alpha)}^{-1}\bigg|
   \le C\eps^4, \quad t\in [\tau_{\ast\ast}, T\eps^{-3/2}].
   \label{cava2} &
\end{eqnarray}
Moreover, following \cite[Section 5.2\,\& 5.3]{Nteil} we have
\begin{equation}\label{haberl1}
   |R_\alpha(t)|+\bigg|\int d^3x\,\rho_\alpha(x)Z(x+q_\alpha(t), t)\bigg|
   \le C\eps^2,\quad
   |\dot{R}_\alpha(t)|\le C\eps^{7/2},\quad t\in [0, T\eps^{-3/2}],
\end{equation}
and also
\begin{equation}\label{haberl2}
   |M_\alpha(t)|+\bigg|\int d^3x\,\rho_\alpha(x)(L_\alpha(t)Z)(x+q_\alpha(t), t)\bigg|
   \le C\eps^{7/2},\quad t\in [\tau_{\ast\ast}, T\eps^{-3/2}],
\end{equation}
for $1\le\alpha\le N$. Thus we find from (\ref{dotv-3}) and
(\ref{cava1})--(\ref{haberl2}) that
\begin{equation}\label{kaz}
   |\stackrel{...}{v}_\alpha(t)|\le C|\dot{M}_\alpha(t)|
   +|\ddot{R}_\alpha(t)|+C\eps^{11/2},\quad\alpha=1, \ldots, N,
   \quad t\in [\tau_{\ast\ast}, T\eps^{-3/2}].
\end{equation}
In Section \ref{ddotR-sect} below we will show
\begin{equation}\label{ddotR-bd-2}
   |\ddot{R}_\alpha(t)|\le C\eps^5,
   \quad\alpha=1, \ldots, N,\quad t\in [\tau_{\ast\ast}, T\eps^{-3/2}],
\end{equation}
cf.~(\ref{ddotR-bd}). Introducing
\begin{equation}\label{calL-form}
   {\cal L}_\alpha(t)\phi = (\dot{v}_\alpha(t)\cdot\nabla) \phi
   +(v_\alpha(t)\cdot\nabla)^2\phi+2(v_\alpha(t)\cdot\nabla)\dot{\phi}
   +\ddot{\phi}
\end{equation}
for a general $\phi=\phi(x, t)$ and observing
$\frac{d}{dt}[(L_\alpha(t)\phi)(x+q_\alpha(t))]=({\cal L}_\alpha(t)\phi)
(x+q_\alpha(t))$, we moreover deduce from (\ref{Malph-def}) that
\begin{eqnarray*}
   \dot{M}_\alpha(t) & = & \int d^3x\,\rho_\alpha(x)\Big[({\cal L}_\alpha(t)Z_1)
   (x+q_\alpha, t)+v_\alpha\wedge ({\cal L}_\alpha(t)Z_2)(x+q_\alpha, t)\Big]
   \nonumber \\ & & +\,\int d^3x\,\rho_\alpha(x)\Big(\dot{v}_\alpha\wedge
   (L_\alpha(t)Z_2)(x+q_\alpha, t)\Big).
\end{eqnarray*}
In view of Lemma \ref{esti} and (\ref{haberl2}) we hence obtain
\begin{eqnarray*}
   |\dot{M}_\alpha(t)| & \le &
   \bigg|\int d^3x\,\rho_\alpha(x)\Big[({\cal L}_\alpha(t)Z_1)
   (x+q_\alpha(t), t)+v_\alpha\wedge ({\cal L}_\alpha(t)Z_2)(x+q_\alpha(t), t)\Big]\bigg|
   +C\eps^{11/2} \\ & \le &
   C\,\bigg|\int d^3x\,\rho_\alpha(x)({\cal L}_\alpha(t)Z)
   (x+q_\alpha(t), t)\bigg|+C\eps^{11/2},
   \quad t\in [\tau_{\ast\ast}, T\eps^{-3/2}].
\end{eqnarray*}
Putting this together with (\ref{kaz}) and (\ref{ddotR-bd-2}),
we have seen that
\begin{equation}\label{tengl}
   |\stackrel{...}{v}_\alpha(t)|\le C\eps^5
   +C\,\bigg|\int d^3x\,\rho_\alpha(x)({\cal L}_\alpha(t)Z)
   (x+q_\alpha(t), t)\bigg|,\quad\alpha=1, \ldots, N,
   \quad t\in [\tau_{\ast\ast}, T\eps^{-3/2}].
\end{equation}
In order to bound the main term on the right-hand side of (\ref{tengl}),
we calculate
\begin{eqnarray*}
   \frac{d}{dt}({\cal L}_\alpha(t)Z(\cdot, t))
   & = & {\cal A}({\cal L}_\alpha(t)Z(\cdot, t))
   -{\cal L}_\alpha(t)f(\cdot, t)+(\ddot{v}_\alpha\cdot\nabla)Z(\cdot, t)
   \\ & & +\,2(v_\alpha\cdot\nabla)(\dot{v}_\alpha\cdot\nabla)Z(\cdot, t)
   +2(\dot{v}_\alpha\cdot\nabla)\dot{Z}(\cdot, t),
\end{eqnarray*}
where we have used (\ref{Z-gleich}). Denoting $U(t)$,
$t\in\R$, the group of isometries in ${L^2(\R^3)}^3\oplus {L^2(\R^3)}^3$
generated by the Maxwell operator ${\cal A}$ from (\ref{Z-gleich}),
we therefore find for any $t_1\in [0, t]$
\begin{eqnarray}\label{gbma}
   \int d^3x\,\rho_\alpha(x)({\cal L}_\alpha(t)Z)(x+q_\alpha(t), t) & = &
   \int d^3x\,\rho_\alpha(x)\Big[U(t-t_1)\Big({\cal L}_\alpha(t_1)Z(\cdot, t_1)\Big)
   \Big](x+q_\alpha(t)) \nonumber \\ & & +\,\int d^3x\,\rho_\alpha(x)\int_{t_1}^t ds\,
   \Big[U(t-s)\Big(-{\cal L}_\alpha(s)f(\cdot, s) \nonumber \\
   & & \hspace{10em} +\,(\ddot{v}_\alpha(s)\cdot\nabla) Z(\cdot, s) \nonumber \\
   & & \hspace{10em} +\,2(v_\alpha(s)\cdot\nabla)(\dot{v}_\alpha(s)\cdot\nabla)
   Z(\cdot, s) \nonumber \\ & & \hspace{10em}
   +\,2(\dot{v}_\alpha(s)\cdot\nabla)\dot{Z}(\cdot, s)\Big)\Big](x+q_\alpha(t))
   \nonumber \\ &=:&  T_{{\rm data}}+T_1+T_2+T_3+T_4.
\end{eqnarray}
This expression will be bounded term by term, for different choices
of $t_1\ge\tau_{\ast\ast}$; note that we need to select
$t_1\ge\tau_{\ast\ast}$ in order to use the bound on $\ddot{v}_\alpha(s)$
from Lemma \ref{q3esti} for $s\ge t_1$. First some estimates are derived below
in Sections \ref{T1-sect}-\ref{ddotR-sect} which are then used
in Section \ref{conclu-sect} to complete the proof of Lemma \ref{q4esti}.
Since our treatment of data terms in similar situations
in the previous papers \cite{Nteil,MK-S-2,MK-S-1} was not completely accurate,
we will give a more detailed account here.
To obtain the necessary estimates, we will frequently use the following
three technical lemmas.

\begin{lemma}\label{darst} For given $f=(f_1, f_2)$ with $\nabla\cdot f_1=0$
and $\nabla\cdot f_2=0$ we have for $W(t, s, x)=(W_1(t, s, x), W_2(t, s, x))
=[U(t-s)f(\cdot, s)](x)$ that
\begin{eqnarray*}
   W_1(t, s, x) & = & \frac{1}{4\pi (t-s)^2}\,\int_{|y-x|=(t-s)}\hspace{-1em}
   d^2y\,\Big[ (t-s)\nabla\wedge f_2(y, s)+f_1(y, s)+((y-x)\cdot\nabla)
   f_1(y, s)\Big], \\
   W_2(t, s, x) & = & \frac{1}{4\pi (t-s)^2}\,\int_{|y-x|=(t-s)}\hspace{-1em}
   d^2y\,\Big[ -(t-s)\nabla\wedge f_1(y, s)+f_2(y, s)+((y-x)\cdot\nabla)
   f_2(y, s)\Big].
\end{eqnarray*}
\end{lemma}
{\bf Proof\,:} See \cite[Lemma 5.1]{Nteil}. {\hfill$\Box$}\bigskip

\begin{lemma}\label{zetav-esti} Defining
\[ \zeta_v(x)=\frac{1}{{[(1-v^2)x^2+{(x\cdot v)}^2]}^{1/2}},\quad\mbox{we have}\quad
   \hat{\zeta}_v(k)=\sqrt{\frac{2}{\pi}}\,\frac{1}{k^2-(k\cdot v)^2},
   \quad |v|<1. \]
In addition, $|\nabla^j\zeta_v(x)|\le C{|x|}^{-(1+j)}$ for $0\le j\le 4$,
$x\in\R^3$, and $|v|\le\bar{v}<1$. More generally, even
$|\nabla_v^l\nabla^j\zeta_v(x)|\le C{|x|}^{-(j+1)}$ for $0\le l, j\le 4$.
\end{lemma}
{\bf Proof\,:} Through direct calculation. {\hfill$\Box$}\bigskip

\begin{lemma}\label{werdenf} Assume $|x|, |y|\le R_\varphi$
and $1\le\alpha, \beta\le N$, and recall
\[ \tau_{\ast\ast}=(C_\ast/8)\eps^{-1}, \]
with the constant $C_\ast$ from Lemma \ref{esti}, cf.~(\ref{tauastast-def}).
Moreover, let $\tilde{x}=x-y+q_\alpha(t)-q_\beta(t-\tau)-z$
with some $0\le\tau\le t\le T\eps^{-3/2}$ and $z\in\R^3$.
Then the following assertions hold for $\eps>0$ small enough.
\begin{itemize}
\item[(a)] If $\alpha\neq\beta$, $\tau\le\tau_{\ast\ast}$, and $|\tilde{x}|=\tau$,
then $|z|\ge (C_\ast/4)\eps^{-1}$.
\item[(b)] If $\alpha\neq\beta$, $\tau\ge\tau_{\ast\ast}$, and $|\tilde{x}|=\tau$,
then $|\tilde{x}|\ge (C_\ast/8)\eps^{-1}$.
\item[(c)] If $\alpha=\beta$, $\tau\ge 8R_\varphi$, and $|\tilde{x}|=\tau$,
then $|z|\ge\tau/4$.
\end{itemize}
\end{lemma}
{\bf Proof\,:} (a) By (\ref{diff-bound}) and (\ref{v-bound})
we have $|z|=|\tilde{x}-[x-y+q_\alpha(t)-q_\beta(t-\tau)]|
\ge |q_\alpha(t)-q_\beta(t)|-|x|-|y|-|q_\beta(t)-q_\beta(t-\tau)|-|\tilde{x}|
\ge C_\ast\eps^{-1}-2R_\varphi-C_v\sqrt{\eps}\tau-\tau\ge (C_\ast/2)\eps^{-1}
-2\tau\ge (C_\ast/4)\eps^{-1}$, if $\eps>0$ is chosen sufficiently small.
(b) Obvious. (c) Similarly as in (a) we find $|z|
=|\tilde{x}-[x-y+q_\alpha(t)-q_\alpha(t-\tau)]|\ge |\tilde{x}|-2R_\varphi
-|q_\alpha(t)-q_\alpha(t-\tau)|\ge\tau-2R_\varphi-C_v\sqrt{\eps}\tau
\ge\tau/2-2R_\varphi\ge\tau/4$. {\hfill$\Box$}\bigskip

\subsection{Bounding $T_1$}
\label{T1-sect}

To deal with
\begin{equation}\label{T1-form}
   T_1(t, t_1)=-\int d^3x\,\rho_\alpha(x)\int_{t_1}^t ds\,
   \Big[U(t-s)\Big({\cal L}_\alpha(s)f(\cdot, s)\Big)\Big](x+q_\alpha(t))
\end{equation}
from the right-hand side of (\ref{gbma}), we introduce the notation
\begin{equation}\label{fmitPhi}
   \Phi_v=\left(\begin{array}{c} E_v \\ B_v\end{array}\right),\quad
   \mbox{thus}\quad f(x, t)=\sum_{\beta=1}^N
   (\dot{v}_\beta(t)\cdot \nabla_v) \Phi_{v_\beta(t)}(x-q_\beta(t)).
\end{equation}
Utilizing (\ref{calL-form}), a somewhat lengthy calculation reveals that
\begin{eqnarray}\label{huls}
   ({\cal L}_\alpha(s)f)(x, s) & = & \sum_{\beta=1}^N\bigg\{
   (\stackrel{...}{v}_\beta\cdot\nabla_v)+3(\ddot{v}_\beta\cdot\nabla_v)
   (\dot{v}_\beta\cdot\nabla_v)+2([v_\alpha-v_\beta]\cdot\nabla)
   (\ddot{v}_\beta\cdot\nabla_v)
   \nonumber\\ & & \hspace{2.5em} +\,(\dot{v}_\beta\cdot\nabla_v)^3
   +([\dot{v}_\alpha-\dot{v}_\beta]\cdot\nabla)(\dot{v}_\beta\cdot\nabla_v)
   +2([v_\alpha-v_\beta]\cdot\nabla)(\dot{v}_\beta\cdot\nabla_v)^2
   \nonumber\\ & & \hspace{2.5em}
   +\,([v_\alpha-v_\beta]\cdot\nabla)^2(\dot{v}_\beta\cdot\nabla_v)\bigg\}
   \,\Phi_{v_\beta}(x-q_\beta),
\end{eqnarray}
where all $v_\alpha, v_\beta, \ldots$, etc., are evaluated at time $s$.
We will not go through the estimate of all these terms when
substituted back to (\ref{T1-form}), but only the first and the last one
will be dealt with in some detail. Since the last term contains the maximal number two
of $\nabla$'s and the minimal a priori power $\eps^3$ (cf.~Lemma \ref{esti}
and Lemma \ref{q3esti}), and as the first term contains a third derivative
$\stackrel{...}{v}_\beta$ which we are about to estimate, it is clear
that all other constituents of (\ref{huls}) will be easier to handle;
note that due to $T\eps^{-3/2}\ge t\ge s\ge t_1\ge\tau_{\ast\ast}$ in the integral,
we may as well use the a priori estimate on $\ddot{v}_\alpha$ from Lemma \ref{q3esti}.

To begin with, we consider the last expression $([v_\alpha-v_\beta]\cdot\nabla)^2
(\dot{v}_\beta\cdot\nabla_v)\Phi_{v_\beta}(x-q_\beta)$.
Due to the difference, we can restrict to $\alpha\neq\beta$, as will be supposed
in the sequel. Using this expression in (\ref{T1-form}), we see that it suffices to verify
\begin{equation}\label{schost}
   \bigg|\int d^3x\,\rho_\alpha(x)\int_{t_1}^t ds\,
   \Big[U(t-s)\Big(([v_\alpha(s)-v_\beta(s)]\cdot\nabla)^2
   (\dot{v}_\beta(s)\cdot\nabla_v)\,\Phi_{v_\beta(s)}(\cdot-q_\beta(s))
   \Big)\Big](x+q_\alpha(t))\bigg|\le C\eps^5
\end{equation}
for $t\in [t_1, T\eps^{-3/2}]$. By means of the solution formulas
from Lemma \ref{darst} we can rewrite this in Fourier transformed form.
Recalling $\Phi_v=(E_v, B_v)$ as well as $\rho_\alpha=e_\alpha\varphi$,
we hence need to show
\begin{eqnarray}\label{lilo}
   & & \Bigg|e_\alpha\int d^3k\,\overline{\hat{\varphi}(k)}\int_{t_1}^t ds\,
   e^{ik\cdot[q_\beta(s)-q_\alpha(t)]}\,
   ([v_\alpha(s)-v_\beta(s)]\cdot k)^2(\dot{v}_\beta(s)\cdot\nabla_v)
   \nonumber \\ & & \hspace{3em}\times\bigg\{\frac{\sin |k|(t-s)}{|k|}
   \,{\cal F}(\nabla\wedge B_{v_\beta(s)})(k)
   +\cos |k|(t-s){\cal F}(E_{v_\beta(s)})(k)\bigg\}\,\Bigg|\le C\eps^5
\end{eqnarray}
for $t\in [t_1, T\eps^{-3/2}]$, with ${\cal F}$ denoting Fourier transform.
For simplicity, we will concentrate only on the first expression
containing $\frac{\sin |k|(t-s)}{|k|}$,
the other term with $\cos |k|(t-s)$ can be bounded in a similar way.
{}From (\ref{EBv-def}) we deduce the relation ${\cal F}(\nabla\wedge B_v)(k)
=[k^2v-(v\cdot k)k]\hat{\phi}_v(k)$, and (\ref{phiv-def}) yields
\[ \dot{v}\cdot\nabla_v\hat{\phi}_v(k)=2e\,\frac{(v\cdot k)(\dot{v}\cdot k)}
   {k^2-(v\cdot k)^2}\,\hat{\phi}_v(k), \]
where $e=e_\beta$ for $v=v_\beta(s)$. Moreover, Lemma \ref{zetav-esti}
and (\ref{phiv-def}) imply $\hat{\phi}_v(k)=\sqrt{\frac{\pi}{2}}e\,\hat{\varphi}(k)
\hat{\zeta}_v(k)$. Thus calculating $(\dot{v}_\beta(s)\cdot\nabla_v)
{\cal F}(\nabla\wedge B_{v_\beta(s)})(k)$ explicitly,
it follows that it is enough to prove that
\begin{eqnarray}\label{jedodeu}
   & & \Bigg|\int d^3k\,|\hat{\varphi}(k)|^2\int_{t_1}^t ds\,
   e^{ik\cdot[q_\beta(s)-q_\alpha(t)]}\,
   ([v_\alpha(s)-v_\beta(s)]\cdot k)^2\,\frac{\sin |k|(t-s)}{|k|}
   \,\hat{\zeta}_{v_\beta(s)}(k) \nonumber \\
   & & \,\,\,\times\bigg[k^2\dot{v}_\beta(s)-(\dot{v}_\beta(s)\cdot k)k
   +2k^2\frac{(v_\beta(s)\cdot k)(\dot{v}_\beta(s)\cdot k)}
   {k^2-(v_\beta(s)\cdot k)^2}\,v_\beta(s)-2\,\frac{(v_\beta(s)\cdot k)^2
   (\dot{v}_\beta(s)\cdot k)}{k^2-(v_\beta(s)\cdot k)^2}\,k\,\bigg]\Bigg|
   \le C\eps^5 \nonumber \\ & &
\end{eqnarray}
for $t\in [t_1, T\eps^{-3/2}]$. Counting the powers of $\eps$
in view of Lemma \ref{esti}, we see that
both last terms have an additional $\eps$ compared to the first two terms,
the order in $k$ being $k^2$ for all four terms. Therefore the last
two terms are easier to handle, and thus dropped, since the same
method can be used as will be explained for the first two terms.
Hence we are going to show that $|A^{(1)}(t, t_1)|\le C\eps^5$
for $t\in [t_1, T\eps^{-3/2}]$, with
\begin{eqnarray}\label{A1-form}
   & & A^{(1)}(t, t_1)=\int d^3k\,|\hat{\varphi}(k)|^2\int_0^{t-t_1} d\tau\,
   e^{ik\cdot[q_\beta(t-\tau)-q_\alpha(t)]}\,
   ([v_\alpha(t-\tau)-v_\beta(t-\tau)]\cdot k)^2\nonumber\\ & & \hspace{12.5em}
   \times\,\frac{\sin |k|\tau}{|k|}\,\hat{\zeta}_{v_\beta(t-\tau)}(k)
   \Big[k^2\dot{v}_\beta(t-\tau)-(\dot{v}_\beta(t-\tau)\cdot k)k\Big].
\end{eqnarray}
Recalling the definition of $\tau_{\ast\ast}=(C_\ast/8)\eps^{-1}$
from Lemma \ref{werdenf}, we split the integral
\begin{equation}\label{split-int}
   \int_0^{t-t_1} d\tau=\int_0^{\tau_{\ast\ast}}d\tau+\int_{\tau_{\ast\ast}}^{t-t_1} d\tau,
\end{equation}
and accordingly we decompose $A^{(1)}(t, t_1)=A^{(1)}_{[0, \tau_{\ast\ast}]}(t)
+A^{(1)}_{[\tau_{\ast\ast}, t-t_1]}(t, t_1)$. In case that $t-t_1\le\tau_{\ast\ast}
={\cal O}(\eps^{-1})$, the whole term $A^{(1)}(t, t_1)$ can be bounded as is
$A^{(1)}_{[0, \tau_{\ast\ast}]}(t)$.

To begin with $A^{(1)}_{[0, \tau_{\ast\ast}]}(t)$, we write this term as a
double convolution in space variables as
\begin{eqnarray}\label{hobet}
   \Big|A^{(1)}_{[0, \tau_{\ast\ast}]}(t)\Big|
   & = & C\,\bigg|\int_0^{\tau_{\ast\ast}}d\tau\int\int d^3x\,d^3y\,
   \varphi(x)\varphi(y) \nonumber \\ & & \hspace{-9em}
   \times\int d^3z\,(\eta_{\alpha\beta}(\tau)\cdot\nabla)^2
   \Big(\nabla\wedge (\dot{v}_\beta(t-\tau)\wedge\nabla)\Big)
   {\zeta}_{v_\beta(t-\tau)}(z)
   \frac{1}{|\tilde{x}|}\,\delta(|\tilde{x}|-\tau)
   \Big|_{\tilde{x}=x-y+q_\alpha(t)-q_\beta(t-\tau)-z}\,\bigg|\,,
\end{eqnarray}
where we have set $\eta_{\alpha\beta}(\tau)=v_\alpha(t-\tau)-v_\beta(t-\tau)$
for brevity; recall that ${\cal F}(\frac{\sin |k|\tau}{|k|})
=\frac{1}{4\pi |x|}\delta(|x|-\tau)$, and note also $k\wedge (\dot{v}\wedge k)
=k^2\dot{v}-(\dot{v}\cdot k)k$. Using Lemma \ref{esti}, Lemma \ref{zetav-esti}
with $j=4$, and Lemma \ref{werdenf}(a), we deduce that for $t\in [0, T\eps^{-3/2}]$
\begin{eqnarray}\label{A11}
   \Big|A^{(1)}_{[0, \tau_{\ast\ast}]}(t)\Big| & \le & C\eps^3\int_0^{\tau_{\ast\ast}}d\tau
   \int\int d^3x\,d^3y\,\varphi(x)\varphi(y)\int d^3z\,\frac{1}{|z|^5}\,
   \frac{1}{|\tilde{x}|}\,\delta(|\tilde{x}|-\tau)
   \nonumber\\ & \le & C\eps^8\int_0^{\tau_{\ast\ast}}\frac{d\tau}{\tau}
   \int\int d^3x\,d^3y\,\varphi(x)\varphi(y)
   \int_{|z-[x-y+q_\alpha(t)-q_\beta(t-\tau)]|=\tau} d^2 z
   \nonumber \\ & \le & C\eps^8\tau_{\ast\ast}^2\le C\eps^6.
\end{eqnarray}

Next we turn to bound $A^{(1)}_{[\tau_{\ast\ast}, t]}(t, t_1)$,
corresponding to $\int_{\tau_{\ast\ast}}^{t-t_1} d\tau (\ldots)$ in (\ref{A1-form}).
Again we write this term as a double convolution in space variables, but this time as
\begin{eqnarray*}
   \Big|A^{(1)}_{[\tau_{\ast\ast}, t]}(t, t_1)\Big|
   & = & C\,\bigg|\int_{\tau_{\ast\ast}}^{t-t_1} d\tau\int\int d^3x\,d^3y\,
   \varphi(x)\varphi(y)\int d^3z\,{\zeta}_{v_\beta(t-\tau)}(z)
   \\ & & \hspace{3em} \times (\eta_{\alpha\beta}(\tau)\cdot\nabla)^2
   \Big(\nabla\wedge (\dot{v}_\beta(t-\tau)\wedge\nabla)\Big)\frac{1}{|\tilde{x}|}
   \,\delta(|\tilde{x}|-\tau)\Big|_{\tilde{x}=x-y+q_\alpha(t)-q_\beta(t-\tau)-z}\,\bigg|\,,
\end{eqnarray*}
where $\nabla=\nabla_{\tilde{x}}$. Hence it follows from Lemma \ref{esti},
Lemma \ref{zetav-esti} with $j=0$, and Lemma \ref{werdenf}(b)
that for $t\in [0, T\eps^{-3/2}]$
\begin{eqnarray}
   \Big|A^{(1)}_{[\tau_{\ast\ast}, t]}(t, t_1)\Big| & \le & C\eps^3\int\int d^3x\,d^3y\,
   \varphi(x)\varphi(y)\int_{\tau_{\ast\ast}}^{t-t_1} d\tau\int d^3z\,\frac{1}{|z|}
   \,\frac{1}{|\tilde{x}|^5}\,\delta(|\tilde{x}|-\tau)
   \label{wustra} \\ & \le & C\eps^8\int\int d^3x\,d^3y\,
   \varphi(x)\varphi(y)\int_{\tau_{\ast\ast}}^{t-t_1} d\tau
   \int_{|z-[x-y+q_\alpha(t)-q_\beta(t-\tau)]|=\tau} d^2z\,\frac{1}{|z|}.
   \label{relev}
\end{eqnarray}
Now observe that for $a\in\R^3$ and $\tau>0$
\begin{equation}\label{surf-int}
   \int_{|z-a|=\tau} d^2z\,\frac{1}{|z|}=\frac{2\pi\tau}{|a|}
   \Big(|a|+\tau-||a|-\tau|\Big)\le 4\pi\tau,
\end{equation}
in both cases $|a|\ge\tau$ and $|a|\le\tau$. We thus deduce from
(\ref{relev}) that
\begin{eqnarray}\label{A12}
   \Big|A^{(1)}_{[\tau_{\ast\ast}, t]}(t, t_1)\Big|
   \le C\eps^8 \int_{\tau_{\ast\ast}}^{t-t_1} d\tau\tau
   \le C\eps^8 t^2\le C\eps^5
\end{eqnarray}
for $t\in [0, T\eps^{-3/2}]$. Summarizing (\ref{A11}) and (\ref{A12}),
we have proved that $|A^{(1)}(t, t_1)|\le C\eps^5$ for $t\in [t_1, T\eps^{-3/2}]$,
and bounding the remaining terms similarly, we conclude that (\ref{schost}) is satisfied.

Hence we can return from (\ref{huls}) to (\ref{T1-form}) and carry through
this argument just elaborated for each of the terms besides the one containing
$\stackrel{...}{v}_\beta$. In this way we arrive at
\begin{equation}\label{T1-esti-1}
   |T_1(t, t_1)|\le \bigg|\sum_{\beta=1}^N\,\int d^3x\,\rho_\alpha(x)\int_{t_1}^t ds\,
   \Big[U(t-s)\Big((\stackrel{...}{v}_\beta(s)\cdot\nabla_v)\Phi_{v_\beta(s)}
   (\cdot-q_\beta(s))\Big)\Big](x+q_\alpha(t))\bigg|+C\eps^5
\end{equation}
being satisfied for $t_1\ge\tau_{\ast\ast}$ and $t\in [t_1, T\eps^{-3/2}]$.
In order to deal with the $\stackrel{...}{v}_\beta$--term in (\ref{T1-esti-1}),
we will consider this expression under different circumstances,
elaborated in the next three sections.

\subsubsection{A standard estimate}
\label{stand-esti-sect}

We are going to show first that
\begin{eqnarray}\label{wamoz}
   & & \bigg|\int d^3x\,\rho_\alpha(x)\int_{t_2}^t ds\,
   \Big[U(t-s)\Big((\stackrel{...}{v}_\beta(s)\cdot\nabla_v)
   \,\Phi_{v_\beta(s)}(\cdot-q_\beta(s))\Big)\Big](x+q_\alpha(t))\bigg|\nonumber
   \\ & & \le C\Big(\max_{1\le\kappa\le N}|e_\kappa|^2\Big)
   \Big(\sup_{s\in [t_2, T\eps^{-3/2}]}\max_{1\le\kappa\le N}|\stackrel{...}{v}_\kappa(s)|\Big)
\end{eqnarray}
for $1\le\alpha, \beta\le N$ and $t\in [t_2, T\eps^{-3/2}]$,
with $t_2\in [0, t]$ remaining to be specified only later.
The argument for $\alpha\neq\beta$ is similar to the one leading to the previous estimates.
Once more we can employ the solution formulas from Lemma \ref{darst}
and rewrite the term in question in Fourier transformed form.
Dropping again the term with $\cos |k|(t-s)$, we may proceed as before and calculate
$(\stackrel{...}{v}_\beta(s)\cdot\nabla_v){\cal F}(\nabla\wedge B_{v_\beta(s)})(k)$
explicitly. This results in
\begin{eqnarray*}
   & & \Bigg|\,e_\alpha e_\beta\int d^3k\,|\hat{\varphi}(k)|^2\int_{t_2}^t ds\,
   e^{ik\cdot[q_\beta(s)-q_\alpha(t)]}\,\frac{\sin |k|(t-s)}{|k|}
   \,\hat{\zeta}_{v_\beta(s)}(k)\,\bigg[k^2\stackrel{...}{v}_\beta(s)
   -\,(\stackrel{...}{v}_\beta(s)\cdot k)k \\ & & \hspace{7em}
   +\,2k^2\frac{(v_\beta(s)\cdot k)(\stackrel{...}{v}_\beta(s)\cdot k)}
   {k^2-(v_\beta(s)\cdot k)^2}\,v_\beta(s)-2\,\frac{(v_\beta(s)\cdot k)^2
   (\stackrel{...}{v}_\beta(s)\cdot k)}{k^2-(v_\beta(s)\cdot k)^2}\,k\,\bigg]\Bigg|
   \\ & & \le C\Big(\max_{1\le\kappa\le N}|e_\kappa|^2\Big)
   \Big(\sup_{s\in [t_2, T\eps^{-3/2}]}\max_{1\le\kappa\le N}|
   \stackrel{...}{v}_\kappa(s)|\Big)
\end{eqnarray*}
to be verified for $t\in [t_2, T\eps^{-3/2}]$. Note that here we kept track
of $e_\alpha$ and $e_\beta$, since they being small will be important at a later point.
Again the last two terms on the left-hand side have an additional $\eps$,
and hence are omitted. Thus we need to show $|A^{(2)}(t, t_2)|
\le C(\sup_{s\in [t_2, T\eps^{-3/2}]}\max_{1\le\kappa\le N}|\stackrel{...}{v}_\kappa(s)|)$
for $t\in [t_2, T\eps^{-3/2}]$, with
\begin{eqnarray}\label{A2-form}
   & & A^{(2)}(t, t_2)=\int d^3k\,|\hat{\varphi}(k)|^2\int_0^{t-t_2} d\tau\,
   e^{ik\cdot[q_\beta(t-\tau)-q_\alpha(t)]}\,\frac{\sin |k|\tau}{|k|}
   \,\hat{\zeta}_{v_\beta(t-\tau)}(k) \nonumber \\ & & \hspace{15em}
   \times\,\Big[k^2\stackrel{...}{v}_\beta(t-\tau)
   -(\stackrel{...}{v}_\beta(t-\tau)\cdot k)k\Big].
\end{eqnarray}

First we consider the case $\alpha\neq\beta$, and for this purpose we recall the
definition of $\tau_{\ast\ast}=(C_\ast/8)\eps^{-1}$ from Lemma \ref{werdenf}.
Analogously to (\ref{split-int}) we write $\int_0^{t-t_2} d\tau
=\int_0^{\tau_{\ast\ast}}d\tau+\int_{\tau_{\ast\ast}}^{t-t_2} d\tau$,
and correspondingly split $A^{(2)}(t, t_2)=A^{(2)}_{[0, \tau_{\ast\ast}]}(t)
+A^{(2)}_{[\tau_{\ast\ast}, t-t_2]}(t, t_2)$;
again the case $t-t_2\le\tau_{\ast\ast}$ is simpler so that we are going to assume
$t-t_2\ge\tau_{\ast\ast}$. To start with, from Lemma \ref{zetav-esti}
and Lemma \ref{werdenf}(a) we find
\begin{eqnarray}\label{A21}
   \Big|A^{(2)}_{[0, \tau_{\ast\ast}]}(t)\Big|
   & = & C\,\bigg|\int_0^{\tau_{\ast\ast}}d\tau\int\int d^3x\,d^3y\,
   \varphi(x)\varphi(y)\int d^3z\,
   \Big(\nabla\wedge (\stackrel{...}{v}_\beta(t-\tau)\wedge\nabla)\Big)
   {\zeta}_{v_\beta(t-\tau)}(z) \nonumber \\ & & \hspace{17em} \times\frac{1}{|\tilde{x}|}
   \,\delta(|\tilde{x}|-\tau)\Big|_{\tilde{x}=x-y+q_\alpha(t)-q_\beta(t-\tau)-z}\,\bigg|
   \nonumber \\ & \le & C\Big(\sup_{s\in [t-\tau_{\ast\ast}, t]}\,\max_{1\le\kappa\le N}|
   \stackrel{...}{v}_\kappa(s)|\Big)\int\int d^3x\,d^3y\,
   \varphi(x)\varphi(y)\int_0^{\tau_{\ast\ast}} d\tau\int d^3z\,\frac{1}{|z|^3}
   \,\frac{1}{|\tilde{x}|}\,\delta(|\tilde{x}|-\tau)
   \nonumber \\ & \le &
   C\eps^3\Big(\sup_{s\in [t-\tau_{\ast\ast}, t]}\,\max_{1\le\kappa\le N}|
   \stackrel{...}{v}_\kappa(s)|\Big)\int\int d^3x\,d^3y\,
   \varphi(x)\varphi(y)\int_0^{\tau_{\ast\ast}} \frac{d\tau}{\tau}
   \int_{|z-[x-y+q_\alpha(t)-q_\beta(t-\tau)]|=\tau}\hspace{-2em} d^2z
   \nonumber \\ & \le & C\eps^3\tau_{\ast\ast}^2\Big(\sup_{s\in [t-\tau_{\ast\ast}, t]}
   \,\max_{1\le\kappa\le N}|\stackrel{...}{v}_\kappa(s)|\Big)
   \le C\eps\Big(\sup_{s\in [t_2, T\eps^{-3/2}]}
   \,\max_{1\le\kappa\le N}|\stackrel{...}{v}_\kappa(s)|\Big).
\end{eqnarray}
Concerning $A^{(2)}_{[\tau_{\ast\ast}, t-t_2]}(t, t_2)$, in view of Lemma \ref{zetav-esti},
Lemma \ref{werdenf}(b), and (\ref{surf-int}) we can estimate
\begin{eqnarray}\label{A22}
   \Big|A^{(2)}_{[\tau_{\ast\ast}, t-t_2]}(t, t_2)\Big|
   & = & C\,\bigg|\int_{\tau_{\ast\ast}}^{t-t_2} d\tau\int\int d^3x\,d^3y\,
   \varphi(x)\varphi(y)\int d^3z\,{\zeta}_{v_\beta(t-\tau)}(z)
   \nonumber \\ & & \hspace{5em}
   \times\Big(\nabla\wedge (\stackrel{...}{v}_\beta(t-\tau)\wedge\nabla)\Big)
   \frac{1}{|\tilde{x}|}\,\delta(|\tilde{x}|-\tau)
   \Big|_{\tilde{x}=x-y+q_\alpha(t)-q_\beta(t-\tau)-z}\,\bigg|
   \nonumber \\ & \le & C\Big(\sup_{s\in [t_2, t-\tau_{\ast\ast}]}\,\max_{1\le\kappa\le N}|
   \stackrel{...}{v}_\kappa(s)|\Big)\int\int d^3x\,d^3y\,
   \varphi(x)\varphi(y) \nonumber \\ & & \hspace{15em}
   \times\,\int_{\tau_{\ast\ast}}^{t-t_2} d\tau\int d^3z\,\frac{1}{|z|}
   \,\frac{1}{|\tilde{x}|^3}\,\delta(|\tilde{x}|-\tau) \nonumber \\ & \le &
   C\eps^3\Big(\sup_{s\in [t_2, t-\tau_{\ast\ast}]}\,\max_{1\le\kappa\le N}|
   \stackrel{...}{v}_\kappa(s)|\Big)\int\int d^3x\,d^3y\,
   \varphi(x)\varphi(y) \nonumber \\ & &
   \hspace{15em}\times\,\int_{\tau_{\ast\ast}}^{t-t_2} d\tau
   \int_{|z-[x-y+q_\alpha(t)-q_\beta(t-\tau)]|=\tau} \frac{d^2 z}{|z|}
   \nonumber \\ & \le & C\eps^3 t^2\,\Big(\sup_{s\in [t_2, t-\tau_{\ast\ast}]}
   \,\max_{1\le\kappa\le N}|\stackrel{...}{v}_\kappa(s)|\Big)
   \le C\Big(\sup_{s\in [t_2, T\eps^{-3/2}]}\,\max_{1\le\kappa\le N}|
   \stackrel{...}{v}_\kappa(s)|\Big)
\end{eqnarray}
for $t\le T\eps^{-3/2}$. Summarizing (\ref{A21}) and (\ref{A22}),
we have seen that
\begin{equation}\label{A2}
   |A^{(2)}(t, t_2)|\le C\Big(\sup_{s\in [t_2, T\eps^{-3/2}]}\,\max_{1\le\kappa\le N}|
   \stackrel{...}{v}_\kappa(s)|\Big),\quad t\in [t_2, T\eps^{-3/2}],
   \quad\alpha\neq\beta.
\end{equation}

To handle the case $\alpha=\beta$, we split $\int_0^t d\tau
=\int_0^{8R_\varphi}d\tau+\int_{8R_\varphi}^{t-t_2} d\tau$ in (\ref{A2-form}),
and accordingly $A^{(2)}(t, t_2)=A^{(2)}_{[0, 8R_\varphi]}(t)
+A^{(2)}_{[8R_\varphi, t-t_2]}(t, t_2)$; again w.l.o.g.~we may assume
that $t-t_2\ge 8R_\varphi$. Then Lemma \ref{zetav-esti} and Lemma \ref{werdenf}(c) imply
\begin{eqnarray}\label{ecke}
   \Big|A^{(2)}_{[8R_\varphi, t-t_2]}(t, t_2)\Big|
   & = & C\,\bigg|\int_{8R_\varphi}^{t-t_2} d\tau
   \int\int d^3x\,d^3y\,\varphi(x)\varphi(y)\int d^3z\,
   \Big(\nabla\wedge (\stackrel{...}{v}_\alpha(t-\tau)\wedge\nabla)\Big)
   {\zeta}_{v_\alpha(t-\tau)}(z) \nonumber \\ & & \hspace{16em}
   \times\,\frac{1}{|\tilde{x}|}\,\delta(|\tilde{x}|-\tau)
   \Big|_{\tilde{x}=x-y+q_\alpha(t)-q_\alpha(t-\tau)-z}\,\bigg|
   \nonumber \\ & \le & C\Big(\sup_{s\in [t_2, t-8R_\varphi]}\,\max_{1\le\kappa\le N}|
   \stackrel{...}{v}_\kappa(s)|\Big)\int\int d^3x\,d^3y\,
   \varphi(x)\varphi(y) \nonumber \\ & & \hspace{16em}
   \times\,\int_{8R_\varphi}^{t-t_2} d\tau\int d^3z\,\frac{1}{|z|^3}
   \,\frac{1}{|\tilde{x}|}\,\delta(|\tilde{x}|-\tau)
   \nonumber \\ & \le & C\Big(\sup_{s\in [t_2, t-8R_\varphi]}\,\max_{1\le\kappa\le N}|
   \stackrel{...}{v}_\kappa(s)|\Big)\int\int d^3x\,d^3y\,
   \varphi(x)\varphi(y)\nonumber \\ & & \hspace{16em}
   \times\,\int_{8R_\varphi}^{t-t_2}\frac{d\tau}{\tau^4}
   \int_{|z-[x-y+q_\alpha(t)-q_\alpha(t-\tau)]|=\tau} \hspace{-2em} d^2z
   \nonumber \\ & \le & C\Big(\sup_{s\in [t_2, t-8R_\varphi]}\,\max_{1\le\kappa\le N}|
   \stackrel{...}{v}_\kappa(s)|\Big)\int\int d^3x\,d^3y\,
   \varphi(x)\varphi(y)\int_{8R_\varphi}^{t-t_2}\frac{d\tau}{\tau^2}
   \nonumber \\ & \le & C\Big(\sup_{s\in [t_2, T\eps^{-3/2}]}\,\max_{1\le\kappa\le N}|
   \stackrel{...}{v}_\kappa(s)|\Big).
\end{eqnarray}
On the other hand, we have the simple estimate
\begin{eqnarray}\label{clift}
   \Big|A^{(2)}_{[0, 8R_\varphi]}(t)\Big| & = & C\,\bigg|\int_0^{8R_\varphi} d\tau
   \int\int d^3x\,d^3y\,\Big(\nabla\wedge (\stackrel{...}{v}_\alpha(t-\tau)
   \wedge\nabla)\Big)\varphi(x)\varphi(y) \nonumber \\ & & \hspace{7em}
   \times\,\int d^3z\,{\zeta}_{v_\alpha(t-\tau)}(z)
   \frac{1}{|\tilde{x}|}\,\delta(|\tilde{x}|-\tau)
   \Big|_{\tilde{x}=x-y+q_\alpha(t)-q_\alpha(t-\tau)-z}\,\bigg|
   \nonumber \\ & \le & C\Big(\sup_{s\in [t-8R_\varphi, t]}\,\max_{1\le\kappa\le N}
   |\stackrel{...}{v}_\kappa(s)|\Big)\int\int d^3x\,d^3y\,
   |\nabla^2\varphi(x)|\varphi(y) \nonumber \\ & & \hspace{13em}
   \times\,\int_0^{8R_\varphi} d\tau\int d^3z\,\frac{1}{|z|}
   \,\frac{1}{|\tilde{x}|}\,\delta(|\tilde{x}|-\tau)
   \nonumber \\ & \le & C\Big(\sup_{s\in [t-8R_\varphi, t]}\,\max_{1\le\kappa\le N}
   |\stackrel{...}{v}_\kappa(s)|\Big)\int\int d^3x\,d^3y\,
   |\nabla^2\varphi(x)|\varphi(y) \nonumber \\ & &
   \hspace{13em}\times\,\int_0^{8R_\varphi}\frac{d\tau}{\tau}
   \int_{|z-[x-y+q_\alpha(t)-q_\alpha(t-\tau)]|=\tau}\frac{d^2z}{|z|}
   \nonumber \\ & \le & C\Big(\sup_{s\in [t_2, T\eps^{-3/2}]}\,\max_{1\le\kappa\le N}
   |\stackrel{...}{v}_\kappa(s)|\Big),
\end{eqnarray}
cf.~(\ref{surf-int}). In view of (\ref{A2}), (\ref{ecke}), and (\ref{clift}),
and bounding the remaining terms in the same manner,
it is verified that (\ref{wamoz}) holds. We will further use this in the form

\begin{eqnarray}\label{wamoz-sum}
   & & \bigg|\sum_{\beta=1}^N\,\int d^3x\,\rho_\alpha(x)\int_{t_2}^t ds\,
   \Big[U(t-s)\Big((\stackrel{...}{v}_\beta(s)\cdot\nabla_v)
   \,\Phi_{v_\beta(s)}(\cdot-q_\beta(s))\Big)\Big](x+q_\alpha(t))\bigg|
   \nonumber \\ & & \le C\Big(\max_{1\le\kappa\le N}|e_\kappa|^2\Big)
   \Big(\sup_{s\in [t_2, T\eps^{-3/2}]}\max_{1\le\kappa\le N}|
   \stackrel{...}{v}_\kappa(s)|\Big)
\end{eqnarray}
for $1\le\alpha\le N$ and $t\in [t_2, T\eps^{-3/2}]$, with $t_2\in [0, t]$
still being free to be chosen; the constant $C>0$ is independent of $t_2$.

\subsubsection{An a priori estimate to bound $|\stackrel{...}{v}_\alpha(t)|$ by $\eps^4$}
\label{v3-eps4}

The purpose of this section is to prove that
\begin{eqnarray}\label{lampa}
   |T_5(t, t_2)| & := & \bigg|\sum_{\beta=1}^N\,\int d^3x\,\rho_\alpha(x)
   \int_{t_2}^{t_2+\tau_{\ast\ast}} ds\,
   \Big[U(t-s)\Big((\stackrel{...}{v}_\beta(s)\cdot\nabla_v)
   \,\Phi_{v_\beta(s)}(\cdot-q_\beta(s))\Big)\Big](x+q_\alpha(t))\bigg|
   \nonumber \\ & \le & C\eps^4
\end{eqnarray}
for $1\le\alpha\le N$ and
\begin{equation}\label{tassum}
   t\in [t_2+\tau_{\ast\ast}, T\eps^{-3/2}],
\end{equation}
with $\tau_{\ast\ast}$ from (\ref{tauastast-def}).
This estimate will later play the key role in showing that at least
$|\stackrel{...}{v}|\cong\eps^4$.

Since $U(t)$ is the group with generator ${\cal A}$, we calculate
\begin{eqnarray}\label{grostrak}
   \lefteqn{\frac{d}{ds}\Big[U(t-s)\Big((\ddot{v}_\beta(s)\cdot\nabla_v)
   \,\Phi_{v_\beta(s)}(\cdot-q_\beta(s))\Big)\Big]} \nonumber \\ & = &
   -\,U(t-s)\Big((\ddot{v}_\beta(s)\cdot\nabla_v)
   \,{\cal A}\Phi_{v_\beta(s)}(\cdot-q_\beta(s))\Big)
   +U(t-s)\Big((\stackrel{...}{v}_\beta(s)\cdot\nabla_v)
   \,\Phi_{v_\beta(s)}(\cdot-q_\beta(s))\Big)
   \nonumber \\ & & +\,U(t-s)\Big((\ddot{v}_\beta(s)\cdot\nabla_v)
   (\dot{v}_\beta(s)\cdot\nabla_v)\,\Phi_{v_\beta(s)}(\cdot-q_\beta(s))\Big)
   \nonumber \\ & & -\,U(t-s)\Big((\ddot{v}_\beta(s)\cdot\nabla_v)
   (v_\beta(s)\cdot\nabla)\,\Phi_{v_\beta(s)}(\cdot-q_\beta(s))\Big).
\end{eqnarray}
Next, calculating
\begin{equation}\label{calAPhi}
   {\cal A}\Phi_v=(\nabla\wedge B_v, -\nabla\wedge E_v)
   =\Big((v\cdot\nabla)\nabla\phi_v-v\Delta\phi_v,
   (v\cdot\nabla)(v\wedge\nabla\phi_v)\Big)
\end{equation}
explicitly from (\ref{EBv-def}), we see that taking ${\cal A}\Phi_v$
has a similar effect as taking $(v\cdot\nabla)\Phi_v$, since both operations result
in an additional $v$ and an additional $\nabla$-derivative. For simplicity
we hence drop the term with ${\cal A}\Phi_v$ in (\ref{grostrak}).
Substituting the remainder of (\ref{grostrak}) back in the definition of $T_5(t, t_2)$
then shows that
\begin{eqnarray}\label{T5-split}
   \lefteqn{|T_5(t, t_2)|} \nonumber \\
   & \le & \sum_{\beta=1}^N\,\bigg|\int d^3x\,\rho_\alpha(x)
   \Big\{U(t-[t_2+\tau_{\ast\ast}])
   \Big((\ddot{v}_\beta(t_2+\tau_{\ast\ast})\cdot\nabla_v)
   \,\Phi_{v_\beta(t_2+\tau_{\ast\ast})}(\cdot-q_\beta(t_2+\tau_{\ast\ast}))
   \Big)\Big\} \nonumber \\ & & \hspace{33em} (x+q_\alpha(t))\bigg|
   \nonumber \\ & & +\,\sum_{\beta=1}^N\,\bigg|\int d^3x\,\rho_\alpha(x)
   \Big\{U(t-t_2)\Big((\ddot{v}_\beta(t_2)\cdot\nabla_v)
   \,\Phi_{v_\beta(t_2)}(\cdot-q_\beta(t_2))\Big)\Big\}(x+q_\alpha(t))\bigg|
   \nonumber \\ & & +\,\sum_{\beta=1}^N\,\bigg|\int d^3x\,\rho_\alpha(x)
   \int_{t_2}^{t_2+\tau_{\ast\ast}} ds\,\Big\{U(t-s)\Big((\ddot{v}_\beta(s)\cdot\nabla_v)
   (\dot{v}_\beta(s)\cdot\nabla_v)\,\Phi_{v_\beta(s)}(\cdot-q_\beta(s))\Big)\Big\}
   \nonumber \\ & & \hspace{33em}
   (x+q_\alpha(t))\bigg| \nonumber \\ & & +\,\sum_{\beta=1}^N\,
   \bigg|\int d^3x\,\rho_\alpha(x)\int_{t_2}^{t_2+\tau_{\ast\ast}} ds\,\Big\{
   U(t-s)\Big((\ddot{v}_\beta(s)\cdot\nabla_v)
   (v_\beta(s)\cdot\nabla)\,\Phi_{v_\beta(s)}(\cdot-q_\beta(s))\Big)\Big\}
   \nonumber \\ & & \hspace{33em} (x+q_\alpha(t))\bigg|
   \nonumber \\ & & =: \sum_{\beta=1}^N\bigg(|T_{5, 1}^\beta(t, t_2)|
   +|T_{5, 2}^\beta(t, t_2)|+|T_{5, 3}^\beta(t, t_2)|+|T_{5, 4}^\beta(t, t_2)|\bigg).
\end{eqnarray}
To bound these terms individually, we will make frequent use of the method
developed before in Sections \ref{T1-sect} and \ref{stand-esti-sect},
without expanding all the details again. The recipe is always the same:
(1.) Substitute the solution formulas from Lemma \ref{darst} and pass to Fourier
transformed form in the $\int d^3x\,\rho_\alpha(x)(\ldots)$; (2.) Drop the
term with factor $\cos |k|(t-s)$, since it can be handled the same way; (3.) Evaluate
explicitly the $\nabla_v$-derivatives and drop the easier terms which thereby
have gained additional $v$'s; (4.) If there is an $\int ds (\ldots)$, then change this
through $t-s=\tau$ to an $\int d\tau (\ldots)$. Afterwards split the latter
into two parts, by inserting $\tau_{\ast\ast}$ from Lemma \ref{werdenf} if $\alpha\neq\beta$,
and by inserting $8R_\varphi$ for $\alpha=\beta$; (5.) Finally rewrite
the whole expression from Fourier transformed form to double convolution
form $\int\int d^3x\,d^3y\,\varphi(x)\varphi(y)(\ldots)$. If $\alpha\neq\beta$
and $\tau\le\tau_{\ast\ast}$, then place the $k$'s as derivatives
on the $\zeta_{v_\beta(t-\tau)}(z)$ and use Lemma \ref{werdenf}(a),
but for $\tau\ge\tau_{\ast\ast}$ place the $k$'s as derivatives
on the $\frac{1}{|\tilde{x}|}$ and use Lemma \ref{werdenf}(b).
In case that $\alpha=\beta$, some extra care has to be taken,
as will be explained case by case later, nevertheless it should be kept in mind
that then Lemma \ref{werdenf}(c) applies for $\tau\ge 8R_\varphi$.

\medskip
\noindent
{\bf Part A: Bounding $|T_{5, 3}^\beta(t, t_2)|+|T_{5, 4}^\beta(t, t_2)|$.}
Let us start with the case $\alpha\neq\beta$. Noting that here (and everywhere else)
we have an additional $\nabla$-derivative resulting from the $(\nabla\wedge B_v)$-term
of the solution formula from Lemma \ref{darst}, cf.~the beginning
of Section \ref{T1-sect}, and recalling that $\Phi_v\cong\nabla\phi_v$,
we find that here
\begin{eqnarray*}
   & & |T_{5, 3}^\beta(t, t_2)|+|T_{5, 4}^\beta(t, t_2)| \\
   & & \le C\int_{t_2}^{t_2+\tau_{\ast\ast}} ds\,
   \bigg|\int\int d^3x\,d^3y\,\varphi(x)\varphi(y)
   \\ & & \hspace{9em}\times\,[\eps^{11/2}\,\nabla^2+\eps^4\,\nabla^3]
   \int d^3z\,\zeta_{v_\beta(s)}(z)\frac{1}{|\tilde{x}|}
   \,\delta(|\tilde{x}|-(t-s))\Big|_{\tilde{x}=x-y+q_\alpha(t)-q_\beta(s)-z}\,\bigg|\,,
\end{eqnarray*}
where we have also used the bounds from Lemma \ref{esti} and Lemma \ref{q3esti},
which is possible due to $s\ge t_2\ge\tau_{\ast\ast}$. Transforming
$t-s=\tau$, the integration interval becomes $\tau\in [t-(t_2+\tau_{\ast\ast}),
t-t_2]$. By hypothesis, $\tau_{\ast\ast}\le t-t_2$.
We may as well assume that $t-(t_2+\tau_{\ast\ast})\le\tau_{\ast\ast}$, since otherwise
the first of the integrals below simply drops out. Following the above recipe
we hence deduce from Lemma \ref{zetav-esti}, Lemma \ref{werdenf}(a), (b),
and (\ref{surf-int}) that
\begin{eqnarray}\label{maur}
   & & |T_{5, 3}^\beta(t, t_2)|+|T_{5, 4}^\beta(t, t_2)| \nonumber \\
   & & \le C\int\int d^3x\,d^3y\,\varphi(x)\varphi(y)\,
   \int_{t-(t_2+\tau_{\ast\ast})}^{\tau_{\ast\ast}} d\tau\,
   \bigg(\eps^{11/2}\,\eps^3\,\frac{1}{\tau}
   \int_{|z-[x-y+q_\alpha(t)-q_\beta(t-\tau)]|=\tau} d^2z
   \nonumber \\ & & \hspace{18em} +\,\eps^4\,\eps^4\,\frac{1}{\tau}
   \int_{|z-[x-y+q_\alpha(t)-q_\beta(t-\tau)]|=\tau} d^2z\bigg)
   \nonumber \\ & & \quad +\,C\int\int d^3x\,d^3y\,\varphi(x)\varphi(y)\,
   \int_{\tau_{\ast\ast}}^{t-t_2} d\tau\,
   \bigg(\eps^{11/2}\,\eps^3\,
   \int_{|z-[x-y+q_\alpha(t)-q_\beta(t-\tau)]|=\tau} d^2z\,\frac{1}{|z|}
   \nonumber \\ & & \hspace{17em} +\,\eps^4\,\eps^4\,
   \int_{|z-[x-y+q_\alpha(t)-q_\beta(t-\tau)]|=\tau} d^2z\,\frac{1}{|z|}\bigg)
   \nonumber \\ & & \le C\Big(\eps^{11/2}\,\eps^3+\eps^4\,\eps^4\Big)\tau_{\ast\ast}^2
   +C\Big(\eps^{11/2}\,\eps^3+\eps^4\,\eps^4\Big)t^2\le C\eps^5,
\end{eqnarray}
according to $\tau_{\ast\ast}={\cal O}(\eps^{-1})$ and $t\le T\eps^{-3/2}$.

Next we consider the case $\alpha=\beta$. Here we have $t-t_2\ge\tau_{\ast\ast}
={\cal O}(\eps^{-1})>8R_\varphi$, and again we can suppose w.l.o.g.~that
$t-(t_2+\tau_{\ast\ast})\le 8R_\varphi$. Partitioning the $d\tau$-integral this way,
it follows from Lemma \ref{zetav-esti}, Lemma \ref{werdenf}(c),
and (\ref{surf-int}) that
\begin{eqnarray}\label{backr}
   & & |T_{5, 3}^\alpha(t, t_2)|+|T_{5, 4}^\alpha(t, t_2)| \nonumber \\
   & & \le C\int_{t-(t_2+\tau_{\ast\ast})}^{8R_\varphi} d\tau\,
   \bigg|\int\int d^3x\,d^3y\,\Big([\eps^{11/2}\,\nabla^2+\eps^4\,\nabla^3]\varphi(x)\Big)
   \varphi(y) \nonumber \\ & & \hspace{9em}\times\,
   \int d^3z\,\zeta_{v_\alpha(t-\tau)}(z)\frac{1}{|\tilde{x}|}
   \,\delta(|\tilde{x}|-\tau)\Big|_{\tilde{x}=x-y+q_\alpha(t)-q_\alpha(t-\tau)-z}\,\bigg|
   \nonumber \\ & & \quad +\,C\int^{t-t_2}_{8R_\varphi} d\tau\,
   \bigg|\int\int d^3x\,d^3y\,\varphi(x)\varphi(y) \nonumber \\ & & \hspace{5em}\times\,
   \int d^3z\,\Big([\eps^{11/2}\,\nabla^2+\eps^4\,\nabla^3]
   \,\zeta_{v_\alpha(t-\tau)}(z)\Big)\,\frac{1}{|\tilde{x}|}
   \,\delta(|\tilde{x}|-\tau)\Big|_{\tilde{x}=x-y+q_\alpha(t)-q_\alpha(t-\tau)-z}\,\bigg|
   \nonumber \\ & & \le C\int\int d^3x\,d^3y\,\Big(\eps^{11/2}\,|\nabla^2\varphi(x)|
   +\eps^4\,|\nabla^3\varphi(x)|\Big)\varphi(y)
   \,\int_{t-(t_2+\tau_{\ast\ast})}^{8R_\varphi} d\tau\,\frac{1}{\tau}
   \nonumber \\ & & \hspace{25em} \times\,
   \int_{|z-[x-y+q_\alpha(t)-q_\alpha(t-\tau)]|=\tau} d^2z\,\frac{1}{|z|}
   \nonumber \\ & & \quad +\,C\int\int d^3x\,d^3y\,\varphi(x)\varphi(y)\,
   \int^{t-t_2}_{8R_\varphi} d\tau\,\Big(\eps^{11/2}\frac{1}{\tau^3}
   +\eps^4\frac{1}{\tau^4}\Big)\,\frac{1}{\tau}
   \,\int_{|z-[x-y+q_\alpha(t)-q_\alpha(t-\tau)]|=\tau} d^2z
   \nonumber \\ & & \le C\eps^4\int\int d^3x\,d^3y\,\Big(|\nabla^2\varphi(x)|
   +|\nabla^3\varphi(x)|\Big)\varphi(y)\,\int_0^{8R_\varphi} d\tau
   \nonumber \\ & & \quad +\,C\int\int d^3x\,d^3y\,\varphi(x)\varphi(y)\,
   \int^{t-t_2}_{8R_\varphi} d\tau\,\Big(\eps^{11/2}\frac{1}{\tau^2}
   +\eps^4\frac{1}{\tau^3}\Big)\le C\eps^4.
\end{eqnarray}
Summarizing (\ref{maur}) and (\ref{backr}), we have verified
\begin{equation}\label{T5354}
   |T_{5, 3}^\beta(t, t_2)|+|T_{5, 4}^\beta(t, t_2)|\le C\eps^4
\end{equation}
for all $1\le\beta\le N$ and all $t, t_2$ satisfying the assumptions
(\ref{tassum}) of this section.

\medskip
\noindent
{\bf Part B: Bounding $|T_{5, 1}^\beta(t, t_2)|$.} We go back to (\ref{T5-split})
and deal with the contribution of the first term on the right-hand side.
Again we begin with the case $\alpha\neq\beta$. Following the method used so far,
we obtain from Lemma \ref{q3esti} that
\begin{eqnarray}\label{geste}
   |T_{5, 1}^\beta(t, t_2)| & \le & C\eps^{7/2}\,\bigg|\int\int d^3x\,d^3y\,\varphi(x)
   \varphi(y) \nonumber \\ & & \hspace{3em}\times\,
   \nabla^2\int d^3z\,\zeta_{v_\beta(t_2+\tau_{\ast\ast})}(z)
   \frac{1}{|\tilde{x}|}\,\delta\Big(|\tilde{x}|-(t-[t_2+\tau_{\ast\ast}])\Big)
   \Big|_{\tilde{x}=x-y+q_\alpha(t)-q_\beta(t_2+\tau_{\ast\ast})-z}\,\bigg|\,.
   \nonumber \\ & &
\end{eqnarray}
First we assume that $\tau:=t-[t_2+\tau_{\ast\ast}]\le\tau_{\ast\ast}$.
Then we pass the derivatives to $\zeta_{v_\beta(t_2+\tau_{\ast\ast})}(z)$
and invoke Lemma \ref{werdenf}(a) and (\ref{surf-int}) to deduce
\begin{eqnarray*}
   |T_{5, 1}^\beta(t, t_2)| & \le & C\eps^{7/2}\eps^3\,\int\int d^3x\,d^3y\,\varphi(x)
   \varphi(y) \\ & & \hspace{5em}\times\,
   \int_{|z-[x-y+q_\alpha(t)-q_\beta(t_2+\tau_{\ast\ast})]|=\tau} d^2z
   \,\frac{1}{|x-y+q_\alpha(t)-q_\beta(t_2+\tau_{\ast\ast})-z|}
   \\ & \le & C\eps^{7/2}\eps^3\tau\le C\eps^{7/2}\eps^3\tau_{\ast\ast}
   \le C\eps^{11/2}.
\end{eqnarray*}
On the other hand, if $\tau\ge\tau_{\ast\ast}$ holds, then $\frac{1}{|\tilde{x}|}$
in (\ref{geste}) gets the $\nabla^2$, and Lemma \ref{werdenf}(b) together with
(\ref{surf-int}) implies
\begin{eqnarray*}
   |T_{5, 1}^\beta(t, t_2)| & \le & C\eps^{7/2}\eps^3\,\int\int d^3x\,d^3y\,\varphi(x)
   \varphi(y)\int_{|z-[x-y+q_\alpha(t)-q_\beta(t_2+\tau_{\ast\ast})]|=\tau} d^2z
   \,\frac{1}{|z|} \\ & \le & C\eps^{7/2}\eps^3\tau\le C\eps^{7/2}\eps^3 t
   \le C\eps^5
\end{eqnarray*}
for $t\le T\eps^{-3/2}$. Thus we have found
\begin{equation}\label{saschmi}
   |T_{5, 1}^\beta(t, t_2)| \le C\eps^5,\quad\alpha\neq\beta.
\end{equation}
The most difficult term to estimate in this section is
\begin{eqnarray}\label{T51alphalph}
   \lefteqn{T_{5, 1}^\alpha(t, t_2)} \nonumber \\ & &
   =\,\int d^3x\,\rho_\alpha(x)\Big\{U(t-[t_2+\tau_{\ast\ast}])
   \Big((\ddot{v}_\alpha(t_2+\tau_{\ast\ast})\cdot\nabla_v)
   \,\Phi_{v_\alpha(t_2+\tau_{\ast\ast})}(\cdot-q_\alpha(t_2+\tau_{\ast\ast}))
   \Big)\Big\}(x+q_\alpha(t)), \nonumber \\ & &
\end{eqnarray}
since for $\alpha=\beta$ there is no way to gain $\eps$'s through the particles
being far apart, and as here $t=t_2+\tau_{\ast\ast}$ is possible, thus
preventing us from using the decay induced by the Maxwell group $U(t)$
close to $t=t_2+\tau_{\ast\ast}$. What saves the argument is the observation
that the integrand vanishes at $t=t_2+\tau_{\ast\ast}$, as is most easily seen
by changing to Fourier transformed form. Indeed, e.g.
\begin{equation}\label{flori}
   \int d^3x\,\rho_\alpha(x) \Big\{E_v(\cdot-q_\alpha(t))\Big\}(x+q_\alpha(t))
   =\int d^3x\,\rho_\alpha(x) E_v(x)
   =C\int d^3k\,|\hat{\varphi}(k)|^2\,i[k-(v\cdot k)v]=0
\end{equation}
due to the rotational symmetry of $\varphi$, recall $(C)$. To exploit this,
we calculate
\begin{eqnarray}\label{horn}
   & & \frac{d}{ds}\Big[U(s-[t_2+\tau_{\ast\ast}])
   \,\Phi_{v_\alpha(t_2+\tau_{\ast\ast})}(\cdot+q_\alpha(s)
   -q_\alpha(t_2+\tau_{\ast\ast}))\Big]
   \nonumber \\ & & \quad =\,U(s-[t_2+\tau_{\ast\ast}])
   \,{\cal A}\Phi_{v_\alpha(t_2+\tau_{\ast\ast})}(\cdot+q_\alpha(s)
   -q_\alpha(t_2+\tau_{\ast\ast})) \nonumber \\ & & \qquad +\,U(s-[t_2+\tau_{\ast\ast}])
   (v_\alpha(s)\cdot\nabla)\,\Phi_{v_\alpha(t_2+\tau_{\ast\ast})}(\cdot+q_\alpha(s)
   -q_\alpha(t_2+\tau_{\ast\ast})).
\end{eqnarray}
Again we may rely on (\ref{calAPhi}) to see that ${\cal A}\Phi_v$ is, from the point
of view of estimates, exactly as $(v\cdot\nabla)\Phi_v$, so we will again drop
the term containing ${\cal A}\Phi_v$. Integrating (\ref{horn}) over
$s\in [t_2+\tau_{\ast\ast}, t]$, substituting the remainder
back into (\ref{T51alphalph}), and cancelling the zero term at $t=t_2+\tau_{\ast\ast}$,
we then obtain
\begin{eqnarray*}
   |T_{5, 1}^\alpha(t, t_2)| & \le & C\bigg|\int d^3x\,\rho_\alpha(x)
   \int_{t_2+\tau_{\ast\ast}}^t ds\,\Big\{U(s-[t_2+\tau_{\ast\ast}])
   \Big( \\ & & \hspace{4em} (\ddot{v}_\alpha(t_2+\tau_{\ast\ast})\cdot\nabla_v)
   (v_\alpha(s)\cdot\nabla)\,\Phi_{v_\alpha(t_2+\tau_{\ast\ast})}
   (\cdot-q_\alpha(t_2+\tau_{\ast\ast}))\Big)\Big\}(x+q_\alpha(s))\,\bigg|\,.
\end{eqnarray*}
Thus we have gained one $v_\alpha(s)\cong\sqrt{\eps}$ and one $\nabla$-derivative,
at the expense that now $\int_{t_2+\tau_{\ast\ast}}^t ds$ appears.
Applying our standard method yields in view of Lemma \ref{q3esti} and Lemma \ref{esti}
\begin{eqnarray*}
   |T_{5, 1}^\alpha(t, t_2)| & \le & C\eps^4\,\bigg|\int_0^{t-[t_2+\tau_{\ast\ast}]}
   d\tau\,\int\int d^3x\,d^3y\,\varphi(x)
   \varphi(y) \nonumber \\ & & \hspace{3em}\times\,
   \nabla^3\int d^3z\,\zeta_{v_\alpha(t_2+\tau_{\ast\ast})}(z)
   \frac{1}{|\tilde{x}|}\,\delta(|\tilde{x}|-\tau)
   \Big|_{\tilde{x}=x-y+q_\alpha(\tau+[t_2+\tau_{\ast\ast}])
   -q_\alpha(t_2+\tau_{\ast\ast})-z}\,\bigg|\,.
\end{eqnarray*}
We split the integral at $\tau=8R_\varphi$. If $t-[t_2+\tau_{\ast\ast}]\le 8R_\varphi$,
then the second term can be omitted and the estimate is simpler.
Passing the derivatives to $\zeta_{v_\alpha(t_2+\tau_{\ast\ast})}(z)$ in
$\int_{8R_\varphi}^{t-[t_2+\tau_{\ast\ast}]}d\tau (\ldots)$,
and then utilizing Lemma \ref{werdenf}(c) to find $|z|\ge C\tau$ there,
it follows that
\begin{eqnarray*}
   |T_{5, 1}^\alpha(t, t_2)| & \le & C\eps^4\,\int_0^{8R_\varphi}
   d\tau\,\int\int d^3x\,d^3y\,|\nabla^3\varphi(x)|
   \varphi(y) \\ & & \hspace{3em}\times\,\frac{1}{\tau}
   \int_{|z-[x-y+q_\alpha(\tau+[t_2+\tau_{\ast\ast}])
   -q_\alpha(t_2+\tau_{\ast\ast})]|=\tau} d^2z\,\frac{1}{|z|}
   \\ & & +\,C\eps^4\,\int_{8R_\varphi}^{t-[t_2+\tau_{\ast\ast}]}
   d\tau\,\int\int d^3x\,d^3y\,\varphi(x)\varphi(y)
   \\ & & \hspace{3em}\times\,\frac{1}{\tau^5}
   \int_{|z-[x-y+q_\alpha(\tau+[t_2+\tau_{\ast\ast}])
   -q_\alpha(t_2+\tau_{\ast\ast})]|=\tau} d^2z
   \\ & \le & C\eps^4.
\end{eqnarray*}
Therefore (\ref{saschmi}) allows us to summarize our findings as
\begin{equation}\label{T51}
   |T_{5, 1}^\beta(t, t_2)|\le C\eps^4
\end{equation}
for all $1\le\beta\le N$ and all $t, t_2$ obeying the assumptions (\ref{tassum})
of this section.

\medskip
\noindent
{\bf Part C: Bounding $|T_{5, 2}^\beta(t, t_2)|$.} Again we go back to (\ref{T5-split})
and investigate the second term on the right-hand side. This expression
is easier to handle, due to $t-t_2\ge\tau_{\ast\ast}={\cal O}(\eps^{-1})$.
Here the standard method yields, no matter whether $\alpha\neq\beta$
or $\alpha=\beta$, that
\begin{eqnarray*}
   |T_{5, 2}^\beta(t, t_2)| & \le & C\eps^{7/2}\,\bigg|\int\int d^3x\,d^3y\,\varphi(x)
   \varphi(y) \nonumber \\ & & \hspace{3em}\times\,
   \int d^3z\,\zeta_{v_\beta(t_2)}(z)
   \Big(\nabla^2\frac{1}{|\tilde{x}|}\Big)\,\delta(|\tilde{x}|-(t-t_2))
   \Big|_{\tilde{x}=x-y+q_\alpha(t)-q_\beta(t_2)-z}\,\bigg|
   \\ & \le & C\eps^{7/2}{(t-t_2)}^{-3}\int\int d^3x\,d^3y\,\varphi(x)\varphi(y)
   \,\int_{|z-[x-y+q_\alpha(t)-q_\beta(t_2)]|=(t-t_2)} d^2z\,\frac{1}{|z|}
   \\ & \le & C\eps^{7/2}{(t-t_2)}^{-2}\le C\eps^{11/2}.
\end{eqnarray*}
Together with (\ref{T5354}) and (\ref{T51}) this proves that indeed
(\ref{lampa}) is verified, under the assumptions (\ref{tassum}).

\subsubsection{Improving $\stackrel{...}{v}$-estimates by $\eps^{1/4}$}

Here we are going to show that
\begin{eqnarray}\label{lamperl}
   |T_6(t, t_3)| & := & \bigg|\sum_{\beta=1}^N\,\int d^3x\,\rho_\alpha(x)
   \int_{t_3}^{t_3+\tau_{\ast\ast}} ds\,
   \Big[U(t-s)\Big((\stackrel{...}{v}_\beta(s)\cdot\nabla_v)
   \,\Phi_{v_\beta(s)}(\cdot-q_\beta(s))\Big)\Big](x+q_\alpha(t))\bigg|
   \nonumber \\ & \le & C\Big(\sup_{s\in [t_3, t_3+\tau_{\ast\ast}]}
   \max_{1\le\kappa\le N}|\stackrel{...}{v}_\kappa(s)|\Big)\eps^{1/4}
\end{eqnarray}
for $1\le\alpha\le N$ and
\begin{equation}\label{tassum-2}
   t\in [t_3+\tau_{\ast\ast}, T\eps^{-3/2}].
\end{equation}

To verify (\ref{lamperl}), we define $T_6(t, t_3)=\sum_{\beta=1}^N T_6^\beta(t, t_3)$.
In case that $\alpha\neq\beta$, our standard method implies
\begin{eqnarray*}
   |T_6^\beta(t, t_3)| & \le & C\Big(\sup_{s\in [t_3, t_3+\tau_{\ast\ast}]}
   \max_{1\le\kappa\le N}|\stackrel{...}{v}_\kappa(s)|\Big)
   \,\bigg|\int_{t-[t_3+\tau_{\ast\ast}]}^{t-t_3}
   d\tau\,\int\int d^3x\,d^3y\,\varphi(x)
   \varphi(y) \\ & & \hspace{3em}\times\,
   \nabla^2\int d^3z\,\zeta_{v_\beta(t-\tau)}(z)
   \frac{1}{|\tilde{x}|}\,\delta(|\tilde{x}|-\tau)
   \Big|_{\tilde{x}=x-y+q_\alpha(t)-q_\beta(t-\tau)-z}\,\bigg|\,.
\end{eqnarray*}
Again we split the $d\tau$-integral by introducing $\tau_{\ast\ast}\le t-t_3$,
cf.~(\ref{tassum-2}), from Lemma \ref{werdenf}. W.l.o.g.~we can suppose that
$\tau_{\ast\ast}\ge t-[t_3+\tau_{\ast\ast}]$, as otherwise simply the first integral
below drops out. In this manner we find from Lemma \ref{werdenf}(a)
and (\ref{surf-int}) that
\begin{eqnarray*}
   |T_6^\beta(t, t_3)| & \le & C\Big(\sup_{s\in [t_3, t_3+\tau_{\ast\ast}]}
   \max_{1\le\kappa\le N}|\stackrel{...}{v}_\kappa(s)|\Big)
   \,\bigg(\,\bigg|\,\int_{t-[t_3+\tau_{\ast\ast}]}^{\tau_{\ast\ast}}
   d\tau\,\int\int d^3x\,d^3y\,\varphi(x)
   \varphi(y) \\ & & \hspace{3em}\times\,
   \int d^3z\,\Big(\nabla^2\zeta_{v_\beta(t-\tau)}(z)\Big)
   \frac{1}{|\tilde{x}|}\,\delta(|\tilde{x}|-\tau)
   \Big|_{\tilde{x}=x-y+q_\alpha(t)-q_\beta(t-\tau)-z}\,\bigg|
   \\ & & \hspace{1em} +\,\int_{\tau_{\ast\ast}}^{t-t_3}
   d\tau\,\int\int d^3x\,d^3y\,\varphi(x)
   \varphi(y) \\ & & \hspace{3em}\times\,
   \int d^3z\,\zeta_{v_\beta(t-\tau)}(z)
   \Big(\nabla^2\frac{1}{|\tilde{x}|}\Big)\,\delta(|\tilde{x}|-\tau)
   \Big|_{\tilde{x}=x-y+q_\alpha(t)-q_\beta(t-\tau)-z}\,\bigg|\,\bigg)
   \\ & \le & C\Big(\sup_{s\in [t_3, t_3+\tau_{\ast\ast}]}
   \max_{1\le\kappa\le N}|\stackrel{...}{v}_\kappa(s)|\Big)
   \,\bigg(\,\int\int d^3x\,d^3y\,\varphi(x)\varphi(y)\,
   \\ & & \hspace{3em}\times\,
   \eps^3\int_{t-[t_3+\tau_{\ast\ast}]}^{\tau_{\ast\ast}}d\tau\,\frac{1}{\tau}\,
   \int_{|z-[x-y+q_\alpha(t)-q_\beta(t-\tau)]|=\tau} d^2z
   \\ & & \hspace{1em} +\,\int\int d^3x\,d^3y\,\varphi(x)\varphi(y)\,
   \int_{\tau_{\ast\ast}}^{t-t_3}d\tau\,\frac{1}{\tau^3}\,
   \int_{|z-[x-y+q_\alpha(t)-q_\beta(t-\tau)]|=\tau} d^2z\,\frac{1}{|z|}\,\bigg)
   \\ & \le & C\Big(\sup_{s\in [t_3, t_3+\tau_{\ast\ast}]}
   \max_{1\le\kappa\le N}|\stackrel{...}{v}_\kappa(s)|\Big)\,\bigg(\eps^3\tau_{\ast\ast}^2
   +\int_{\tau_{\ast\ast}}^{t-t_3}d\tau\,\frac{1}{\tau^2}\bigg).
\end{eqnarray*}
Since $\tau_{\ast\ast}={\cal O}(\eps^{-1})$ and $\int_{\tau_{\ast\ast}}^{t-t_3}\,
\frac{d\tau}{\tau^2}\le\int_{\tau_{\ast\ast}}^\infty\,\frac{d\tau}{\tau^2}
=\tau_{\ast\ast}^{-1}$, we deduce that even
\begin{equation}\label{T6-neq}
   |T_6^\beta(t, t_3)|\le C\Big(\sup_{s\in [t_3, t_3+\tau_{\ast\ast}]}
   \max_{1\le\kappa\le N}|\stackrel{...}{v}_\kappa(s)|\Big)\eps,
   \quad\alpha\neq\beta.
\end{equation}
As in the previous Section \ref{v3-eps4} the term where $\alpha=\beta$, i.e.,
\[ T_6^\alpha(t, t_3)=\int d^3x\,\rho_\alpha(x)
   \int_{t_3}^{t_3+\tau_{\ast\ast}} ds\,
   \Big[U(t-s)\Big((\stackrel{...}{v}_\alpha(s)\cdot\nabla_v)
   \,\Phi_{v_\alpha(s)}(\cdot-q_\alpha(s))\Big)\Big](x+q_\alpha(t)), \]
is critical. To derive the desired bound, we introduce
\begin{equation}\label{hat-tau}
   \hat{\tau}=t_3+\tau_{\ast\ast}-\eps^{-1/4}.
\end{equation}
Since $\tau_{\ast\ast}={\cal O}(\eps^{-1})$, we have $\hat{\tau}\ge t_3$.
According to
\[ \int_{t_3}^{t_3+\tau_{\ast\ast}} ds
   =\int_{t_3}^{\hat{\tau}} ds+\int_{\hat{\tau}}^{t_3+\tau_{\ast\ast}} ds, \]
we split
\begin{equation}\label{T6-split}
   T_6^\alpha(t, t_3)=T_{6, 1}^\alpha(t, t_3)+T_{6, 2}^\alpha(t, t_3).
\end{equation}
Following our standard method, we first see that in view of (\ref{surf-int})
\begin{eqnarray*}
   |T_{6, 1}^\alpha(t, t_3)| & \le & C\Big(\sup_{s\in [t_3, t_3+\tau_{\ast\ast}]}
   \max_{1\le\kappa\le N}|\stackrel{...}{v}_\kappa(s)|\Big)
   \,\bigg|\int_{t_3}^{\hat{\tau}} ds\,\int\int d^3x\,d^3y\,\varphi(x)
   \varphi(y) \\ & & \hspace{3em}\times\,
   \int d^3z\,\zeta_{v_\alpha(s)}(z)
   \Big(\nabla^2\frac{1}{|\tilde{x}|}\Big)\,\delta(|\tilde{x}|-(t-s))
   \Big|_{\tilde{x}=x-y+q_\alpha(t)-q_\alpha(s)-z}\,\bigg|
   \\ & \le & C\Big(\sup_{s\in [t_3, t_3+\tau_{\ast\ast}]}
   \max_{1\le\kappa\le N}|\stackrel{...}{v}_\kappa(s)|\Big)
   \,\int\int d^3x\,d^3y\,\varphi(x)\varphi(y)\, \\ & & \hspace{3em}\times\,
   \int_{t_3}^{\hat{\tau}} ds\,(t-s)^{-3}\,
   \int_{|z-[x-y+q_\alpha(t)-q_\alpha(s)]|=(t-s)} d^2z\,\frac{1}{|z|}
   \\ & \le & C\Big(\sup_{s\in [t_3, t_3+\tau_{\ast\ast}]}
   \max_{1\le\kappa\le N}|\stackrel{...}{v}_\kappa(s)|\Big)
   \,\int_{t_3}^{\hat{\tau}} ds\,(t-s)^{-2}
   \\ & = & C\Big(\sup_{s\in [t_3, t_3+\tau_{\ast\ast}]}
   \max_{1\le\kappa\le N}|\stackrel{...}{v}_\kappa(s)|\Big)
   \,\frac{\hat{\tau}-t_3}{(t-\hat{\tau})(t-t_3)}.
\end{eqnarray*}
Now (\ref{tassum-2}) and (\ref{hat-tau}) yield
\[ \frac{\hat{\tau}-t_3}{(t-\hat{\tau})(t-t_3)}
   =\frac{\tau_{\ast\ast}-\eps^{-1/4}}{(t-\hat{\tau})(t-t_3)}
   \le\frac{1}{t-\hat{\tau}}\le\eps^{1/4},  \]
whence
\begin{equation}\label{T61}
   |T_{6, 1}^\alpha(t, t_3)|\le C\Big(\sup_{s\in [t_3, t_3+\tau_{\ast\ast}]}
   \max_{1\le\kappa\le N}|\stackrel{...}{v}_\kappa(s)|\Big)\eps^{1/4}.
\end{equation}
Consequently it remains to derive the same bound for
\[ T_{6, 2}^\alpha(t, t_3)
   =\int_{\hat{\tau}}^{t_3+\tau_{\ast\ast}} ds\,
   (\stackrel{...}{v}_\alpha(s)\cdot\nabla_v)\int d^3x\,\rho_\alpha(x)
   \,\Big[U(t-s)\Big(\Phi_{v_\alpha(s)}(\cdot-q_\alpha(s))\Big)\Big](x+q_\alpha(t)). \]
Again the main observation is that $\int d^3x\,\rho_\alpha(x)\,
[U(t-s)(\Phi_{v_\alpha(s)}(\cdot-q_\alpha(s)))](x+q_\alpha(t))=0$
for $s=t$, cf.~(\ref{flori}). Similar to (\ref{horn}) we calculate
\begin{eqnarray}\label{gobles}
   & & \frac{d}{d\bar{s}}\Big[U(t-\bar{s})
   \,\Phi_{v_\alpha(\bar{s})}(\cdot+q_\alpha(t)-q_\alpha(\bar{s}))\Big]
   \nonumber \\ & & \quad = -\,U(t-\bar{s})
   \,{\cal A}\Phi_{v_\alpha(\bar{s})}(\cdot+q_\alpha(s)
   -q_\alpha(\bar{s})) \nonumber \\ & & \qquad +\,U(t-\bar{s})
   (\dot{v}_\alpha(\bar{s})\cdot\nabla_v)\,\Phi_{v_\alpha(\bar{s})}(\cdot+q_\alpha(s)
   -q_\alpha(\bar{s})) \nonumber \\ & & \qquad -\,U(t-\bar{s})
   (v_\alpha(\bar{s})\cdot\nabla)\,\Phi_{v_\alpha(\bar{s})}(\cdot+q_\alpha(s)
   -q_\alpha(\bar{s})).
\end{eqnarray}
Using ${\cal A}\Phi_v\cong (v\cdot\nabla)\Phi_v$, cf.~(\ref{calAPhi}),
we notice that the term containing ${\cal A}\Phi_v$ can be handled similarly
to the last one on the right-hand side of (\ref{gobles}), and hence it is dropped.
Integrating the remainder over $\bar{s}\in [s, t]$, we get
\begin{eqnarray}\label{waxens}
   |T_{6, 2}^\alpha(t, t_3)|
   & \le & \bigg|\int_{\hat{\tau}}^{t_3+\tau_{\ast\ast}} ds\,
   \int_s^t d\bar{s}\,
   (\stackrel{...}{v}_\alpha(s)\cdot\nabla_v)\int d^3x\,\rho_\alpha(x)
   \,\Big[U(t-\bar{s})\Big((v_\alpha(\bar{s})\cdot\nabla)
   \Phi_{v_\alpha(\bar{s})}(\cdot-q_\alpha(\bar{s}))\Big)
   \nonumber \\ & & \hspace{7em} -\,U(t-\bar{s})(\dot{v}_\alpha(\bar{s})\cdot\nabla)
   \Phi_{v_\alpha(\bar{s})}(\cdot-q_\alpha(\bar{s}))\Big)\Big](x+q_\alpha(t))\bigg|.
\end{eqnarray}
Let
\begin{equation}\label{eps14-zerle}
   I_1=\{s\in [\hat{\tau}, t_3+\tau_{\ast\ast}]: t-s\ge 1\},\quad
   I_2=[\hat{\tau}, t_3+\tau_{\ast\ast}]\setminus I_1.
\end{equation}
Then
\begin{equation}\label{kasberg}
   \int_{I_1} ds\,\int_s^t d\bar{s}=\int_{I_1} ds\,\bigg(\int_s^{t-1} d\bar{s}
   +\int_{t-1}^t d\bar{s}\bigg).
\end{equation}
For the first part we infer from the standard method and (\ref{surf-int}) that
\begin{eqnarray}\label{jue}
   \bigg|\int_{I_1} ds\,\int_s^{t-1} d\bar{s}\,\ldots\bigg|
   & \le & C\Big(\sup_{s\in [t_3, t_3+\tau_{\ast\ast}]}
   \max_{1\le\kappa\le N}|\stackrel{...}{v}_\kappa(s)|\Big)
   \,\int^{t-\hat{\tau}}_{t-[t_3+\tau_{\ast\ast}]} d\tau\,\int_1^\tau d\bar{\tau}\,
   \int\int d^3x\,d^3y\,\varphi(x)\varphi(y) \nonumber
   \\ & & \hspace{-3em} \times\,\bigg|\int d^3z\,\zeta_{v_\alpha(\bar{s})}(z)
   \Big([\sqrt{\eps}\,\nabla^3+\eps^2\nabla^2]
   \frac{1}{|\tilde{x}|}\Big)\,\delta(|\tilde{x}|-\bar{\tau})
   \Big|_{\tilde{x}=x-y+q_\alpha(t)-q_\alpha(t-\bar{\tau})-z}\,\bigg|
   \nonumber \\ & \le & C\Big(\sup_{s\in [t_3, t_3+\tau_{\ast\ast}]}
   \max_{1\le\kappa\le N}|\stackrel{...}{v}_\kappa(s)|\Big)
   \,\int^{t-\hat{\tau}}_{t-[t_3+\tau_{\ast\ast}]} d\tau\,\int_1^\tau d\bar{\tau}\,
   \int\int d^3x\,d^3y\,\varphi(x)\varphi(y)
   \nonumber \\ & & \hspace{5em} \times\,\Big[\sqrt{\eps}\,\frac{1}{\bar{\tau}^4}
   +\eps^2\frac{1}{\bar{\tau}^3}\Big]
   \,\int_{|z-[x-y+q_\alpha(t)-q_\alpha(t-\bar{\tau})]|=\bar{\tau}} d^2z\,\frac{1}{|z|}
   \nonumber \\ & \le & C\Big(\sup_{s\in [t_3, t_3+\tau_{\ast\ast}]}
   \max_{1\le\kappa\le N}|\stackrel{...}{v}_\kappa(s)|\Big)
   \,\int^{t-\hat{\tau}}_{t-[t_3+\tau_{\ast\ast}]} d\tau\,\int_1^\tau d\bar{\tau}\,
   \Big[\sqrt{\eps}\,\frac{1}{\bar{\tau}^3}+\eps^2\frac{1}{\bar{\tau}^2}\Big].
   \nonumber \\ & \le & C\Big(\sup_{s\in [t_3, t_3+\tau_{\ast\ast}]}
   \max_{1\le\kappa\le N}|\stackrel{...}{v}_\kappa(s)|\Big)
   \,\int^{t-\hat{\tau}}_{t-[t_3+\tau_{\ast\ast}]} d\tau\,[\sqrt{\eps}+\eps^2]
   \nonumber \\ & \le & C\Big(\sup_{s\in [t_3, t_3+\tau_{\ast\ast}]}
   \max_{1\le\kappa\le N}|\stackrel{...}{v}_\kappa(s)|\Big)\eps^{1/4},
\end{eqnarray}
the latter since $-\hat{\tau}+[t_3+\tau_{\ast\ast}]=\eps^{-1/4}$
due to (\ref{hat-tau}). For the second part of (\ref{kasberg}), we once more
apply the standard method, and passing all derivatives to $\varphi$,
it follows that
\begin{eqnarray}\label{abm}
   \bigg|\int_{I_1} ds\,\int_{t-1}^t d\bar{s}\,\ldots\bigg|
   & \le & C\Big(\sup_{s\in [t_3, t_3+\tau_{\ast\ast}]}
   \max_{1\le\kappa\le N}|\stackrel{...}{v}_\kappa(s)|\Big)
   \,\int^{t-\hat{\tau}}_{t-[t_3+\tau_{\ast\ast}]} d\tau\,\int_0^1 d\bar{\tau}\,
   \nonumber \\ & & \hspace{3em} \times\,
   \int\int d^3x\,d^3y\,\Big[\sqrt{\eps}\,|\nabla^3\varphi(x)|
   +\eps^2 |\nabla^2\varphi(x)|\Big]\varphi(y)
   \nonumber \\ & & \hspace{3em} \times\,
   \int d^3z\,|\zeta_{v_\alpha(\bar{s})}(z)|
   \,\frac{1}{|\tilde{x}|}\,\delta(|\tilde{x}|-\bar{\tau})
   \Big|_{\tilde{x}=x-y+q_\alpha(t)-q_\alpha(t-\bar{\tau})-z}
   \nonumber \\ & \le & C\Big(\sup_{s\in [t_3, t_3+\tau_{\ast\ast}]}
   \max_{1\le\kappa\le N}|\stackrel{...}{v}_\kappa(s)|\Big)\sqrt{\eps}
   \,\int^{t-\hat{\tau}}_{t-[t_3+\tau_{\ast\ast}]} d\tau
   \nonumber \\ & \le & C\Big(\sup_{s\in [t_3, t_3+\tau_{\ast\ast}]}
   \max_{1\le\kappa\le N}|\stackrel{...}{v}_\kappa(s)|\Big)\eps^{1/4}.
\end{eqnarray}
According to (\ref{eps14-zerle}), it thus remains to bound the contribution of
\[ \bigg|\int_{I_2} ds\,\int_s^t d\bar{s}\,\ldots\bigg| \]
to the right-hand side of (\ref{waxens}). However, since for $s\in I_2$ we have
$t-s\le 1$, the length of the $d\bar{s}$-integration interval is bounded by $1$.
Hence we may repeat the argument leading to (\ref{abm}), as we gain an $\eps^{1/2}$
from $v_\alpha(\bar{s})$, but we loose an $\eps^{-1/4}$ due to $|I_2|\le
t_3+\tau_{\ast\ast}-\hat{\tau}=\eps^{-1/4}$. Summarizing this observation
to (\ref{T6-neq}), (\ref{T6-split}), (\ref{T61}), (\ref{waxens}), (\ref{jue}),
and (\ref{abm}), we have completed the proof of (\ref{lamperl}).

\subsection{Bounding $T_{{\rm data}}$}
\label{data-sect}

The data contribution to (\ref{gbma}) is
\begin{equation}\label{Tdata}
   T_{{\rm data}}(t, t_1)=\int d^3x\,\rho_\alpha(x)
   \Big[U(t-t_1)\Big({\cal L}_\alpha(t_1)Z(\cdot, t_1)\Big)\Big](x+q_\alpha(t)).
\end{equation}
Our aim here is to prove that
\begin{equation}\label{data-bd}
   |T_{{\rm data}}(t, t_1)|\le C\eps^{11/2},\quad t_1\in [\tau_{\ast\ast}, t],
   \quad t\ge t_1+\tau_{\ast\ast}.
\end{equation}
First we recall from (\ref{calL-form}) that
\[ {\cal L}_\alpha(t_1)Z(\cdot, t_1) = (\dot{v}_\alpha(t_1)\cdot\nabla)Z(\cdot, t_1)
   +(v_\alpha(t_1)\cdot\nabla)^2 Z(\cdot, t_1)+2(v_\alpha(t_1)\cdot\nabla)\dot{Z}(\cdot, t_1)
   +\ddot{Z}(\cdot, t_1). \]
In view of $\dot{Z}={\cal A}Z-f$, cf.~(\ref{Z-gleich}), this can be rewritten as
\begin{eqnarray}\label{supas}
   {\cal L}_\alpha(t_1)Z(\cdot, t_1) & = & (\dot{v}_\alpha(t_1)\cdot\nabla)Z(\cdot, t_1)
   +(v_\alpha(t_1)\cdot\nabla)^2 Z(\cdot, t_1)+2(v_\alpha(t_1)\cdot\nabla)
   [{\cal A}Z(\cdot, t_1)-f(\cdot, t_1)] \nonumber \\ & & +\,{\cal A}^2 Z(\cdot, t_1)
   -{\cal A}f(\cdot, t_1)-\dot{f}(\cdot, t_1).
\end{eqnarray}
According to (\ref{Z-def}) and (\ref{ini-bed}) we have $Z(x, 0)\equiv 0$,
whence (\ref{Z-gleich}) yields
\begin{eqnarray}\label{thoma}
   Z(x, t_1) & = & -\int_0^{t_1} ds\,\Big[U(t_1-s)f(\cdot, s)\Big](x)
   \nonumber \\ & = & -\sum_{\beta=1}^N\int_0^{t_1} ds\,\Big[U(t_1-s)
   \Big((\dot{v}_\beta(s)\cdot\nabla_v)\Phi_{v_\beta(s)}(\cdot-q_\beta(s))\Big)\Big](x).
\end{eqnarray}
Since ${\cal A}\Phi_v\cong (v\cdot\nabla)\Phi_v$, cf.~(\ref{calAPhi}),
and ${\cal A}U(t)=U(t){\cal A}$, we take the liberty to treat ${\cal A}Z(\cdot, t_1)$
as $(v_\alpha(t_1)\cdot\nabla)Z(\cdot, t_1)$, ${\cal A}f(\cdot, t_1)$
as $(v_\alpha(t_1)\cdot\nabla)f(\cdot, t_1)$, and ${\cal A}^2 Z(\cdot, t_1)$
as $(v_\alpha(t_1)\cdot\nabla)^2 Z(\cdot, t_1)$, although in fact $v_\beta(s)$
do appear, but both $|v_\alpha(s)|$ and $|v_\beta(s)|$ will only be estimated
by $\sqrt{\eps}$ below, so this makes no difference. Hence from the point of view
of estimates (\ref{supas}) becomes
\begin{equation}\label{hyanni}
   {\cal L}_\alpha(t_1)Z(\cdot, t_1)\cong (\dot{v}_\alpha(t_1)\cdot\nabla)Z(\cdot, t_1)
   +(v_\alpha(t_1)\cdot\nabla)^2 Z(\cdot, t_1)+(v_\alpha(t_1)\cdot\nabla)
   f(\cdot, t_1)+\dot{f}(\cdot, t_1),
\end{equation}
where we also dropped factors and ``$-$''-signs. Since
\begin{eqnarray}\label{taima}
   \dot{f}(x, t_1) & = & \sum_{\beta=1}^N
   \Big\{(\ddot{v}_\beta(t_1)\cdot\nabla_v)\Phi_{v_\beta(t_1)}(x-q_\beta(t_1))
   +(\dot{v}_\beta(t_1)\cdot\nabla_v)^2\Phi_{v_\beta(t_1)}(x-q_\beta(t_1))
   \nonumber \\ & & \hspace{2.5em}
   -\,(v_\beta(t_1)\cdot\nabla)(\dot{v}_\beta(t_1)\cdot\nabla_v)
   \Phi_{v_\beta(t_1)}(x-q_\beta(t_1))\Big\},
\end{eqnarray}
we see that in fact also
\begin{equation}\label{wienna}
   {\cal L}_\alpha(t_1)Z(\cdot, t_1)\cong (\dot{v}_\alpha(t_1)\cdot\nabla)Z(\cdot, t_1)
   +(v_\alpha(t_1)\cdot\nabla)^2 Z(\cdot, t_1)+\dot{f}(\cdot, t_1).
\end{equation}
Substituting (\ref{wienna})
back into (\ref{Tdata}), and taking into account (\ref{thoma}) and (\ref{taima}),
we arrive at
\begin{eqnarray}\label{alpina}
   |T_{{\rm data}}(t, t_1)| & \le & C\sum_{\beta=1}^N\bigg|\int d^3x\,\rho_\alpha(x)\,
   \int_0^{t_1} ds\,\Big[U(t-s)\Big((\dot{v}_\alpha(t_1)\cdot\nabla)
   (\dot{v}_\beta(s)\cdot\nabla_v) \nonumber \\ & & \hspace{9em}
   +\,(v_\alpha(t_1)\cdot\nabla)^2 (\dot{v}_\beta(s)\cdot\nabla_v)\Big)
   \Phi_{v_\beta(s)}(\cdot-q_\beta(s))\Big](x+q_\alpha(t))\bigg|
   \nonumber \\ & & +\,C\sum_{\beta=1}^N\bigg|\int d^3x\,\rho_\alpha(x)\,
   \Big[U(t-t_1)\Big((\ddot{v}_\beta(t_1)\cdot\nabla_v)
   +(\dot{v}_\beta(t_1)\cdot\nabla_v)^2
   \nonumber \\ & & \hspace{9em} +\,(v_\beta(t_1)\cdot\nabla)
   (\dot{v}_\beta(t_1)\cdot\nabla_v)\Big)\Phi_{v_\beta(t_1)}(\cdot-q_\beta(t_1))\Big]
   (x+q_\alpha(t))\bigg| \nonumber \\ & =: & C\sum_{\beta=1}^N\Big
   (|T_{{\rm data}, 1}^{\beta}(t, t_1)|+|T_{{\rm data}, 2}^{\beta}(t, t_1)|\Big).
\end{eqnarray}
To bound these terms we apply the standard method from Section \ref{T1-sect},
and utilizing Lemma \ref{esti} and Lemma \ref{q3esti} we find
\begin{eqnarray*}
   \lefteqn{|T_{{\rm data}, 1}^{\beta}(t, t_1)|+|T_{{\rm data}, 2}^{\beta}(t, t_1)|}
   \\ & & \le C\,\bigg|\int_{t-t_1}^t d\tau\int\int d^3x\,d^3y\,\varphi(x)\varphi(y)
   \\ & & \hspace{4em}
   \times\,[\eps^4\nabla^3+\eps^3\nabla^4]\,\int d^3z\,\zeta_{v_\beta(t-\tau)}(z)
   \frac{1}{|\tilde{x}|}\,\delta(|\tilde{x}|-\tau)
   \Big|_{\tilde{x}=x-y+q_\alpha(t)-q_\beta(t-\tau)-z}\bigg|
   \\ & & +C\,\bigg|\int\int d^3x\,d^3y\,\varphi(x)\varphi(y)
   \\ & & \hspace{4em}\times\,[\eps^{7/2}\nabla^2+\eps^4\nabla^2+\eps^{5/2}\nabla^3]
   \,\int d^3z\,\zeta_{v_\beta(t_1)}(z)
   \frac{1}{|\tilde{x}|}\,\delta(|\tilde{x}|-(t-t_1))
   \Big|_{\tilde{x}=x-y+q_\alpha(t)-q_\beta(t_1)-z}\bigg|\,.
\end{eqnarray*}
Here all the terms can be handled in the same manner. Passing the derivatives
to $\frac{1}{|\tilde{x}|}$ and observing $\tau\ge t-t_1\ge\tau_{\ast\ast}
={\cal O}(\eps^{-1})$, cf.~(\ref{data-bd}), we deduce with (\ref{surf-int}) that
\begin{eqnarray*}
   \lefteqn{|T_{{\rm data}, 1}^{\beta}(t, t_1)|+|T_{{\rm data}, 2}^{\beta}(t, t_1)|}
   \\ & \le & C\int\int d^3x\,d^3y\,\varphi(x)\varphi(y)\int_{t-t_1}^t d\tau\,
   \Big[\eps^4\frac{1}{\tau^4}+\eps^3\frac{1}{\tau^5}\Big]
   \,\int_{|z-[x-y+q_\alpha(t)-q_\beta(t-\tau)]|=\tau} d^2z\,\frac{1}{|z|}
   \\ & & +\,C\int\int d^3x\,d^3y\,\varphi(x)\varphi(y)
   \,\Big[\eps^{7/2}(t-t_1)^{-3}+\eps^{5/2}(t-t_1)^{-4}\Big]
   \,\int_{|z-[x-y+q_\alpha(t)-q_\beta(t_1)]|=(t-t_1)} d^2z\,\frac{1}{|z|}
   \\ & \le & C\int_{t-t_1}^\infty d\tau\,
   \Big[\eps^4\frac{1}{\tau^3}+\eps^3\frac{1}{\tau^4}\Big]
   +C\Big[\eps^{7/2}(t-t_1)^{-2}+\eps^{5/2}(t-t_1)^{-3}\Big]\le C\eps^{11/2}.
\end{eqnarray*}
This completes the proof of (\ref{data-bd}).

\subsection{Bounding $T_2$}

The contribution $T_2$ to (\ref{gbma}) is
\begin{equation}\label{T2-form}
   T_2(t, t_1)=\int d^3x\,\rho_\alpha(x)\int_{t_1}^t d\tau\,
   \Big[U(t-\tau)\Big((\ddot{v}_\alpha(\tau)\cdot\nabla)
   Z(\cdot, \tau)\Big)\Big](x+q_\alpha(t)).
\end{equation}
We are going to show that
\begin{equation}\label{T2-bd}
   |T_2(t, t_1)|\le C\eps^5,\quad t_1\in [\tau_{\ast\ast}, t],\quad t\ge\tau_{\ast\ast}.
\end{equation}
In view of (\ref{Z-gleich}) we have $\frac{d}{dt}(\nabla Z)={\cal A}(\nabla Z)-\nabla f$.
Since $Z(x, 0)\equiv 0$ by (\ref{ini-bed}), also $\nabla Z(x, 0)\equiv 0$, whence
\begin{equation}\label{nablaZ}
   \nabla Z(\cdot, \tau)=-\int_0^\tau ds\,U(\tau-s)\nabla f(\cdot, s).
\end{equation}
Utilizing this and (\ref{fmitPhi}) in (\ref{T2-form}), we obtain
\[ T_2(t, t_1)=-\sum_{\beta=1}^N\int d^3x\,\rho_\alpha(x)
   \int_{t_1}^t d\tau\int_0^\tau ds\,\Big[U(t-s)\Big((\ddot{v}_\alpha(\tau)
   \cdot\nabla)(\dot{v}_\beta(s)\cdot\nabla_v)\,\Phi_{v_\beta(s)}(\cdot-q_\beta(s))
    \Big)\Big](x+q_\alpha(t)). \]
Defining the respective terms in the sum on the right-hand side as $T_2^\beta(t, t_1)$,
we first note that for $\alpha\neq\beta$ the standard method applies.
By assumption $t\ge\tau_{\ast\ast}$, and for $\bar{\tau}\in [0, t-t_1]$
it plays no role whether $\tau_{\ast\ast}\ge\bar{\tau}$ or $\tau_{\ast\ast}\le\bar{\tau}$,
since e.g.~in the former case it follows from Lemma \ref{werdenf}(a), (b) that
\begin{eqnarray*}
   & & \bigg|\int_{\bar{\tau}}^t d\bar{s}\,\int\int d^3x\,d^3y\,\varphi(x)\varphi(y)
   \nabla^3\int d^3z\,\zeta_{v_\beta(t-\bar{s})}(z)
   \frac{1}{|\tilde{x}|}\,\delta(|\tilde{x}|-\bar{s})
   \Big|_{\tilde{x}=x-y+q_\alpha(t)-q_\beta(t-\bar{s})-z}\bigg|
   \\ & & \quad\le C\int\int d^3x\,d^3y\,\varphi(x)\varphi(y)
   \,\bigg(\int_{\bar{\tau}}^{\tau_{\ast\ast}}
   d\bar{s}\,\eps^4\frac{1}{\bar{s}}
   \,\int_{|z-[x-y+q_\alpha(t)-q_\beta(t-\bar{s})]|=\bar{s}} d^2z
   \\ & & \hspace{14em}+\,\int_{\tau_{\ast\ast}}^t d\bar{s}\,\eps^4
   \,\int_{|z-[x-y+q_\alpha(t)-q_\beta(t-\bar{s})]|=\bar{s}}
   d^2z\,\frac{1}{|z|}\bigg)
   \\ & & \quad\le C\Big(\eps^4\tau_{\ast\ast}^2+\eps^4 t^2\Big)\le C\eps
\end{eqnarray*}
for $t\le T\eps^{-3/2}$, and in case that $\tau_{\ast\ast}\le\bar{\tau}$
the same result is obtained. Hence from Lemma \ref{esti}, Lemma \ref{q3esti},
and by means of the standard method we infer that
\begin{eqnarray}\label{jrop}
   |T_2^\beta(t, t_1)| & \le & C\eps^{11/2}
   \bigg|\int_0^{t-t_1} d\bar{\tau}\,\int_{\bar{\tau}}^t d\bar{s}
   \,\int\int d^3x\,d^3y\,\varphi(x)\varphi(y)
   \nonumber\\ & & \hspace{11em}\times\,\nabla^3\int d^3z\,\zeta_{v_\beta(t-\bar{s})}(z)
   \frac{1}{|\tilde{x}|}\,\delta(|\tilde{x}|-\bar{s})
   \Big|_{\tilde{x}=x-y+q_\alpha(t)-q_\beta(t-\bar{s})-z}\bigg|
   \nonumber\\ & \le & C\eps^{11/2} t\eps\le C\eps^5,\quad\alpha\neq\beta.
\end{eqnarray}
Hence it remains to investigate the case $\alpha=\beta$.
Then for e.g.~$\bar{\tau}\ge 1$ we have
\begin{eqnarray*}
   & & \bigg|\int_{\bar{\tau}}^t d\bar{s}\,\int\int d^3x\,d^3y\,\varphi(x)\varphi(y)
   \int d^3z\,\zeta_{v_\alpha(t-\bar{s})}(z)
   \Big(\nabla^3\frac{1}{|\tilde{x}|}\Big)\,\delta(|\tilde{x}|-\bar{s})
   \Big|_{\tilde{x}=x-y+q_\alpha(t)-q_\alpha(t-\bar{s})-z}\bigg|
   \\ & & \quad\le C\int\int d^3x\,d^3y\,\varphi(x)\varphi(y)
   \,\int_{\bar{\tau}}^{\infty} d\bar{s}\,\frac{1}{\bar{s}^4}
   \,\int_{|z-[x-y+q_\alpha(t)-q_\alpha(t-\bar{s})]|=\bar{s}} d^2z\,\frac{1}{|z|}
   \\ & & \quad\le C\int_{\bar{\tau}}^{\infty} d\bar{s}\,\frac{1}{\bar{s}^3}
   \le C\frac{1}{\bar{\tau}^2}.
\end{eqnarray*}
On the other hand, simply
\[ \bigg|\int_0^1 d\bar{s}\,\int\int d^3x\,d^3y\,(\nabla^3\varphi(x))\varphi(y)
   \int d^3z\,\zeta_{v_\alpha(t-\bar{s})}(z)
   \frac{1}{|\tilde{x}|}\,\delta(|\tilde{x}|-\bar{s})
   \Big|_{\tilde{x}=x-y+q_\alpha(t)-q_\alpha(t-\bar{s})-z}\bigg|\le C \]
holds. Consequently by the standard method
\begin{eqnarray}\label{maborn}
   |T_2^\alpha(t, t_1)| & \le & C\eps^{11/2}
   \bigg|\int_0^1 d\bar{\tau}\,\bigg(\int_{\bar{\tau}}^1+\int_1^t\bigg) d\bar{s}
   \,\int\int d^3x\,d^3y\,\varphi(x)\varphi(y)
   \nonumber\\ & & \hspace{11em}\times\,\nabla^3\int d^3z\,\zeta_{v_\alpha(t-\bar{s})}(z)
   \frac{1}{|\tilde{x}|}\,\delta(|\tilde{x}|-\bar{s})
   \Big|_{\tilde{x}=x-y+q_\alpha(t)-q_\alpha(t-\bar{s})-z}\bigg|
   \nonumber\\ & & +\,C\eps^{11/2}
   \bigg|\int_1^{t-t_1} d\bar{\tau}\,\int_{\bar{\tau}}^t d\bar{s}
   \,\int\int d^3x\,d^3y\,\varphi(x)\varphi(y)
   \nonumber\\ & & \hspace{11em}\times\,\nabla^3\int d^3z\,\zeta_{v_\alpha(t-\bar{s})}(z)
   \frac{1}{|\tilde{x}|}\,\delta(|\tilde{x}|-\bar{s})
   \Big|_{\tilde{x}=x-y+q_\alpha(t)-q_\alpha(t-\bar{s})-z}\bigg| \nonumber
   \\ & \le & C\eps^{11/2}+C\eps^{11/2}\int_1^{t-t_1} d\bar{\tau}\,\frac{1}{\bar{\tau}^2}
   \le C\eps^{11/2}.
\end{eqnarray}
Summarizing (\ref{jrop}) and (\ref{maborn}), we see that (\ref{T2-bd}) holds.

\subsection{Bounding $T_3+T_4$}

In this section we will be dealing with
\begin{eqnarray}\label{T34-form}
   T_3(t, t_1)+T_4(t, t_1) & = & 2\int d^3x\,\rho_\alpha(x)\int_{t_1}^t d\tau\,
   \Big[U(t-\tau)\Big((v_\alpha(\tau)\cdot\nabla)(\dot{v}_\alpha(\tau)\cdot\nabla)
   Z(\cdot, \tau) \nonumber \\ & & \hspace{11em}
   +\,(\dot{v}_\alpha(\tau)\cdot\nabla)\dot{Z}(\cdot, \tau)
   \Big)\Big](x+q_\alpha(t)),
\end{eqnarray}
and it will be verified that
\begin{equation}\label{T34-bd}
   |T_3(t, t_1)+T_4(t, t_1)|\le C\eps^5,\quad t_1\ge\tau_{\ast\ast},\quad
   t\in [t_1+\tau_{\ast\ast}, T\eps^{-3/2}].
\end{equation}
Introducing
\begin{equation}\label{Palhdef}
   P_\alpha(t)\phi=(v_\alpha(t)\cdot\nabla)\nabla\phi+\nabla\dot{\phi}
\end{equation}
for a general function $\phi=\phi(x, t)$, (\ref{T34-form}) can be rewritten as
\begin{equation}\label{T34-form-2}
   T_3(t, t_1)+T_4(t, t_1)
   =2\int d^3x\,\rho_\alpha(x)\int_{t_1}^t d\tau\,
   \Big[U(t-\tau)\Big(\dot{v}_\alpha(\tau)\cdot
   P_\alpha(\tau)Z(\cdot, \tau)\Big)\Big](x+q_\alpha(t)).
\end{equation}
Then
\[ \frac{d}{dt}\Big(P_\alpha(t)\phi\Big)=(\dot{v}_\alpha(t)\cdot\nabla)\nabla\phi
   +P_\alpha(t)\dot{\phi} \]
implies in view of (\ref{Z-gleich}) that
\[ \frac{d}{dt}\Big(P_\alpha(t)Z\Big)=(\dot{v}_\alpha(t)\cdot\nabla)\nabla Z
   +P_\alpha(t)[{\cal A}Z-f]={\cal A}\Big(P_\alpha(t)Z\Big)
   +(\dot{v}_\alpha(t)\cdot\nabla)\nabla Z-P_\alpha(t)f, \]
and this leads to
\begin{equation}\label{beda}
   P_\alpha(\tau)Z(\cdot, \tau)=U(\tau-t_1)[P_\alpha(t_1)Z(\cdot, t_1)]
   +\int_{t_1}^{\tau} ds\,U(\tau-s)\Big\{(\dot{v}_\alpha(s)\cdot\nabla)\nabla Z(\cdot, s)
   -P_\alpha(s)f(\cdot, s)\Big\}.
\end{equation}
Resubstituting (\ref{beda}) into (\ref{T34-form-2}) we get two terms,
one from the data, and one main term.

\subsubsection{Bounding the data term}

The data term part of (\ref{T34-form-2}) is
\begin{eqnarray*}
   T_{3+4, {\rm data}}(t, t_1) & = & 2\int d^3x\,\rho_\alpha(x)\int_{t_1}^t d\tau\,
   \Big[U(t-t_1)\Big(\dot{v}_\alpha(\tau)\cdot P_\alpha(t_1)Z(\cdot, t_1)
   \Big)\Big](x+q_\alpha(t)) \\ & = & 2\int d^3x\,\rho_\alpha(x)
   \Big[U(t-t_1)\Big([v_\alpha(t)-v_\alpha(t_1)]\cdot P_\alpha(t_1)Z(\cdot, t_1)
   \Big)\Big](x+q_\alpha(t)).
\end{eqnarray*}
Now (\ref{Palhdef}) and (\ref{Z-gleich}) imply
\begin{eqnarray*}
   P_\alpha(t_1)Z(\cdot, t_1) & = & (v_\alpha(t_1)\cdot\nabla)\nabla Z(\cdot, t_1)
   +\nabla\dot{Z}(\cdot, t_1) \\ & = & (v_\alpha(t_1)\cdot\nabla)\nabla Z(\cdot, t_1)
   +{\cal A}\nabla Z(\cdot, t_1)-\nabla f(\cdot, t_1)
   \\ & \cong & (v_\alpha(t_1)\cdot\nabla)\nabla Z(\cdot, t_1)+\nabla f(\cdot, t_1),
\end{eqnarray*}
again from the point of view of estimates, since ${\cal A}\Phi_v\cong (v\cdot\nabla)\Phi_v$,
cf.~the remarks before (\ref{hyanni}). Using (\ref{nablaZ}) and (\ref{fmitPhi})
we thus find
\begin{eqnarray*}
   |T_{3+4, {\rm data}}(t, t_1)| & \le & C\sum_{\beta=1}^N
   \bigg|\int d^3x\,\rho_\alpha(x)\int_0^{t_1} ds\,
   \Big[U(t-s)\Big(([v_\alpha(t)-v_\alpha(t_1)]\cdot\nabla)
   (v_\alpha(t_1)\cdot\nabla) \\ & & \hspace{13em}
   \times\,(\dot{v}_\beta(s)\cdot\nabla_v)
   \Phi_{v_\beta(s)}(\cdot -q_\beta(s))\Big)\Big](x+q_\alpha(t))\bigg|
   \\ & & +\,C\sum_{\beta=1}^N
   \bigg|\int d^3x\,\rho_\alpha(x)
   \Big[U(t-t_1)\Big(([v_\alpha(t)-v_\alpha(t_1)]\cdot\nabla)
    \\ & & \hspace{13em} \times\,(\dot{v}_\beta(t_1)\cdot\nabla_v)
    \Phi_{v_\beta(t_1)}(\cdot-q_\beta(t_1))\Big)\Big](x+q_\alpha(t))\bigg|\,.
\end{eqnarray*}
Concerning the right-hand side, counting powers of $\eps$ and $\nabla$-derivatives
we see that we have an $\eps^3\nabla^4$ for the first term and an $\eps^{5/2}\nabla^3$
for the second term (with $U(t)$ and $\Phi_v$ counting one $\nabla$ each).
Since $t-t_1\ge\tau_{\ast\ast}={\cal O}(\eps^{-1})$ we hence find
\begin{equation}\label{T34data}
   |T_{3+4, {\rm data}}(t, t_1)|\le C\eps^{11/2},
\end{equation}
exactly as the estimates on $T_{{\rm data}, 1}^{\beta}(t, t_1)$
and $T_{{\rm data}, 2}^{\beta}(t, t_1)$ from (\ref{alpina}) have been derived
in Section \ref{data-sect}. Note that here no $\ddot{v}$-term appears,
whence we do not need to assume $t_1\ge\tau_{\ast\ast}$ at this point.

\subsubsection{Bounding the main term}

By this we mean the contribution
\begin{eqnarray}\label{T5split}
   T_{3+4, {\rm main}}(t, t_1)& := &
   2\int d^3x\,\rho_\alpha(x)\int_{t_1}^t d\tau\int_{t_1}^{\tau} ds\,
   \Big[U(t-s)\Big(\dot{v}_\alpha(\tau)\cdot
   \Big\{(\dot{v}_\alpha(s)\cdot\nabla)\nabla Z(\cdot, s)
   \nonumber \\ & & \hspace{20em}
   -\,P_\alpha(s)f(\cdot, s)\Big\}\Big)\Big](x+q_\alpha(t))
   \nonumber \\ & =: & T_{3+4, {\rm main}, 1}(t, t_1)
   -T_{3+4, {\rm main}, 2}(t, t_1)
\end{eqnarray}
to (\ref{T34-form-2}).

To begin with $T_{3+4, {\rm main}, 1}(t, t_1)$, we can use (\ref{nablaZ}) to
rewrite this expression as
\begin{eqnarray*}
   \lefteqn{T_{3+4, {\rm main}, 1}(t, t_1)} \\ & & =-2\,\int d^3x\,\rho_\alpha(x)
   \int_{t_1}^t d\tau\int_{t_1}^{\tau} ds\int_0^s d\sigma
   \,\Big[U(t-\sigma)\Big((\dot{v}_\alpha(\tau)\cdot\nabla)
   (\dot{v}_\alpha(s)\cdot\nabla)f(\cdot, \sigma)\Big)\Big](x+q_\alpha(t)).
\end{eqnarray*}
Substituting (\ref{fmitPhi}) for $f$ and observing that no $\ddot{v}$-terms
are to be estimated, we hence arrive at
\begin{eqnarray*}
   \lefteqn{|T_{3+4, {\rm main}, 1}(t, t_1)|} \\ & \le &
   C\sum_{\beta=1}^N\int_0^t d\tau\int_0^{\tau} ds\int_0^s d\sigma
   \,\bigg|\int d^3x\,\rho_\alpha(x)
   \,\Big[U(t-\sigma)\Big((\dot{v}_\alpha(\tau)\cdot\nabla)
   (\dot{v}_\alpha(s)\cdot\nabla) \\ & & \hspace{18em} \times\,
   (\dot{v}_\beta(\sigma)\cdot\nabla_v)\,\Phi_{v_\beta(\sigma)}(\cdot-q_\beta(\sigma))
   \Big)\Big](x+q_\alpha(t))\bigg|\,.
\end{eqnarray*}
Observing $\int_0^t d\tau\int_0^\tau ds\int_0^s d\sigma
=\int_0^t d\sigma\int_{\sigma}^t ds\int_s^t d\tau$,
transforming $\tilde{\sigma}=t-\sigma$, $\tilde{s}=t-s$,
and $\tilde{\tau}=t-\tau$, and then omitting the tilde, we see that
\begin{eqnarray*}
   \lefteqn{|T_{3+4, {\rm main}, 1}(t, t_1)|} \\ & \le &
   C\sum_{\beta=1}^N\int_0^t d\sigma\int_0^{\sigma} ds\int_0^s d\tau
   \,\bigg|\int d^3x\,\rho_\alpha(x)
   \,\Big[U(\sigma)\Big((\dot{v}_\alpha(t-\tau)\cdot\nabla)
   (\dot{v}_\alpha(t-s)\cdot\nabla) \\ & & \hspace{13em} \times\,
   (\dot{v}_\beta(t-\sigma)\cdot\nabla_v)\,\Phi_{v_\beta(t-\sigma)}(\cdot-q_\beta(t-\sigma))
   \Big)\Big](x+q_\alpha(t))\bigg|\,.
\end{eqnarray*}
Invoking Lemma \ref{esti} and the standard method, it follows that
\begin{eqnarray}\label{juife}
   |T_{3+4, {\rm main}, 1}(t, t_1)| & \le &
   C\eps^6\sum_{\beta=1}^N\int_0^t d\sigma\int_0^{\sigma} ds\int_0^s d\tau
   \,\bigg|\int\int d^3x\,d^3y\,\varphi(x)\varphi(y) \nonumber
   \\ & & \hspace{7em} \times\,\nabla^4\int d^3z\,\zeta_{v_\beta(t-\sigma)}(z)
   \frac{1}{|\tilde{x}|}\,\delta(|\tilde{x}|-\sigma)
   \Big|_{\tilde{x}=x-y+q_\alpha(t)-q_\beta(t-\sigma)-z}\bigg|
   \nonumber \\ & \le &
   C\eps^6\sum_{\beta=1}^N\int_0^t d\sigma\,\sigma^2
   \,\bigg|\int\int d^3x\,d^3y\,\varphi(x)\varphi(y) \nonumber
   \\ & & \hspace{7em} \times\,\nabla^4\int d^3z\,\zeta_{v_\beta(t-\sigma)}(z)
   \frac{1}{|\tilde{x}|}\,\delta(|\tilde{x}|-\sigma)
   \Big|_{\tilde{x}=x-y+q_\alpha(t)-q_\beta(t-\sigma)-z}\bigg|.
   \nonumber \\ & &
\end{eqnarray}
First we consider $\alpha\neq\beta$. We may assume that $t\ge\tau_{\ast\ast}$,
with $\tau_{\ast\ast}$ from Lemma \ref{werdenf}, the case $t\le\tau_{\ast\ast}$
being simpler. Estimating $\sigma^2\le t^2\le C\eps^{-3}$ for $t\le T\eps^{-3/2}$ and
invoking Lemma \ref{werdenf}(a), (b), as well as (\ref{surf-int}), we deduce
\begin{eqnarray}\label{alpsp}
   & & \eps^6\bigg(\int_0^{\tau_{\ast\ast}}+\int_{\tau_{\ast\ast}}^t\bigg)d\sigma\,\sigma^2
   \,\bigg|\int\int d^3x\,d^3y\,\varphi(x)\varphi(y) \nonumber
   \\ & & \hspace{11em} \times\,\nabla^4\int d^3z\,\zeta_{v_\beta(t-\sigma)}(z)
   \frac{1}{|\tilde{x}|}\,\delta(|\tilde{x}|-\sigma)
   \Big|_{\tilde{x}=x-y+q_\alpha(t)-q_\beta(t-\sigma)-z}\bigg|
   \nonumber \\ & & \le C\eps^3\int_0^{\tau_{\ast\ast}} d\sigma\,
   \int\int d^3x\,d^3y\,\varphi(x)\varphi(y)\,\eps^5\,\frac{1}{\sigma}
   \int_{|z-[x-y+q_\alpha(t)-q_\beta(t-\sigma)]|=\sigma} d^2z
   \nonumber \\ & & \quad +\,C\eps^3\int_{\tau_{\ast\ast}}^t d\sigma\,
   \int\int d^3x\,d^3y\,\varphi(x)\varphi(y)\,\eps^5\,
   \int_{|z-[x-y+q_\alpha(t)-q_\beta(t-\sigma)]|=\sigma} d^2z\,\frac{1}{|z|}
   \nonumber \\ & & \le C\eps^3\Big(\eps^5\tau_{\ast\ast}^2+\eps^5 t^2\Big)\le C\eps^5
\end{eqnarray}
for $t\le T\eps^{-3/2}$. On the other hand, for $\alpha=\beta$ we split
at $\sigma=8R_\varphi$. Then Lemma \ref{werdenf}(c) yields
\begin{eqnarray}\label{zugsp}
   & & \eps^6\bigg(\int_0^{8R_\varphi}+\int_{8R_\varphi}^t\bigg)d\sigma\,\sigma^2
   \,\bigg|\int\int d^3x\,d^3y\,\varphi(x)\varphi(y)
   \nonumber \\ & & \hspace{11em}\times\,\nabla^4\int d^3z\,\zeta_{v_\alpha(t-\sigma)}(z)
   \frac{1}{|\tilde{x}|}\,\delta(|\tilde{x}|-\sigma)
   \Big|_{\tilde{x}=x-y+q_\alpha(t)-q_\alpha(t-\sigma)-z}\bigg|
   \nonumber \\ & & \le C\eps^6\int_0^{8R_\varphi} d\sigma\,\sigma^2
   \int\int d^3x\,d^3y\,|\nabla^4\varphi(x)|\varphi(y)\,\frac{1}{\sigma}
   \int_{|z-[x-y+q_\alpha(t)-q_\alpha(t-\sigma)]|=\sigma} d^2z\,\frac{1}{|z|}
   \nonumber \\ & & \quad +\,C\eps^6\int_{8R_\varphi}^t d\sigma\,\sigma^2
   \int\int d^3x\,d^3y\,\varphi(x)\varphi(y)\,\frac{1}{\sigma^6}\,
   \int_{|z-[x-y+q_\alpha(t)-q_\alpha(t-\sigma)]|=\sigma} d^2z
   \nonumber \\ & & \le C\eps^6.
\end{eqnarray}
Summarizing (\ref{alpsp}) and (\ref{zugsp}), and going back to (\ref{juife}),
we have shown that
\begin{equation}\label{T34main1}
   |T_{3+4, {\rm main}, 1}(t, t_1)|\le C\eps^5
\end{equation}
for the $t$, $t_1$ in question.

Finally we return to (\ref{T5split}) and bound
\begin{equation}\label{egyt}
   T_{3+4, {\rm main}, 2}(t, t_1)
   =2\int d^3x\,\rho_\alpha(x)\int_{t_1}^t d\tau\int_{t_1}^{\tau} ds\,
   \Big[U(t-s)\Big(\dot{v}_\alpha(\tau)\cdot
   P_\alpha(s)f(\cdot, s)\Big)\Big](x+q_\alpha(t)).
\end{equation}
By means of (\ref{Palhdef}) and (\ref{fmitPhi}) it is calculated that
\begin{equation}\label{algor}
   P_\alpha(s)f(\cdot, s)=\nabla\,\sum_{\beta=1}^N\bigg\{([v_\alpha-v_\beta]\cdot\nabla)
   (\dot{v}_\beta\cdot\nabla_v)+(\ddot{v}_\beta\cdot\nabla_v)
   +(\dot{v}_\beta\cdot\nabla_v)^2\bigg\}\,\Phi_{v_\beta}(\cdot-q_\beta),
\end{equation}
where all $v_\alpha, v_\beta, \ldots$, etc., are evaluated at time $s$.
Since $|\dot{v}_\beta(s)|^2\cong\eps^4$ but only $|\ddot{v}_\beta(s)|\cong\eps^{7/2}$,
the last term in (\ref{algor}) is better than the one next to the last,
and hence it is dropped. Using the remainder in (\ref{egyt}),
we find that
\begin{eqnarray*}
   \lefteqn{|T_{3+4, {\rm main}, 2}(t, t_1)|} \\ & \le & C\sum_{\beta=1}^N
   \int_{t_1}^t d\tau\int_{t_1}^{\tau} ds\,\bigg|\int d^3x\,\rho_\alpha(x)
   \Big[U(t-s)\Big((\dot{v}_\alpha(\tau)\cdot\nabla)([v_\alpha(s)-v_\beta(s)]\cdot\nabla)
   (\dot{v}_\beta(s)\cdot\nabla_v) \\ & & \hspace{15em}
   +\,(\dot{v}_\alpha(\tau)\cdot\nabla)
   (\ddot{v}_\beta(s)\cdot\nabla_v)\Big)\,\Phi_{v_\beta(s)}(\cdot-q_\beta(s))
   \Big](x+q_\alpha(t))\bigg|\,.
\end{eqnarray*}
Since $s\ge t_1\ge\tau_{\ast\ast}$, cf.~(\ref{T34-bd}),
we have $|\ddot{v}_\beta(s)|\le C\eps^{7/2}$ due to Lemma \ref{q3esti}.
Applying the standard method, we thus conclude from Lemma \ref{esti} that
\begin{eqnarray*}
   |T_{3+4, {\rm main}, 2}(t, t_1)| & \le & C\sum_{\stackrel{\beta=1}{\beta\neq\alpha}}^N
   \int_{t_1}^t d\tau\int_{t_1}^{\tau} ds\,\eps^{9/2}\,
   \bigg|\int\int d^3x\,d^3y\,\varphi(x)\varphi(y) \nonumber
   \\ & & \hspace{7em} \times\,\nabla^4\int d^3z\,\zeta_{v_\beta(s)}(z)
   \frac{1}{|\tilde{x}|}\,\delta(|\tilde{x}|-(t-s))
   \Big|_{\tilde{x}=x-y+q_\alpha(t)-q_\beta(s)-z}\bigg|
   \\ & & +\,C\sum_{\beta=1}^N
   \int_{t_1}^t d\tau\int_{t_1}^{\tau} ds\,\eps^{11/2}\,
   \bigg|\int\int d^3x\,d^3y\,\varphi(x)\varphi(y) \nonumber
   \\ & & \hspace{7em} \times\,\nabla^3\int d^3z\,\zeta_{v_\beta(s)}(z)
   \frac{1}{|\tilde{x}|}\,\delta(|\tilde{x}|-(t-s))
   \Big|_{\tilde{x}=x-y+q_\alpha(t)-q_\beta(s)-z}\bigg|\,.
\end{eqnarray*}
Using $\int_{t_1}^t d\tau\int_{t_1}^{\tau} ds
\le\int_0^t d\tau\int_0^{\tau} ds$ and observing
$\int_0^t d\tau\int_0^{\tau} ds=\int_0^t ds\int_s^t d\tau$,
we then introduce the change of variables $\tilde{s}=t-s$, $\tilde{\tau}
=t-\tau$, and omitting the tilde again we arrive at
\begin{eqnarray*}
   |T_{3+4, {\rm main}, 2}(t, t_1)| & \le &
   C\eps^{9/2}\,\sum_{\stackrel{\beta=1}{\beta\neq\alpha}}^N
   \int_0^t ds\int_0^s d\tau\,
   \bigg|\int\int d^3x\,d^3y\,\varphi(x)\varphi(y) \nonumber
   \\ & & \hspace{7em} \times\,\nabla^4\int d^3z\,\zeta_{v_\beta(t-s)}(z)
   \frac{1}{|\tilde{x}|}\,\delta(|\tilde{x}|-s)
   \Big|_{\tilde{x}=x-y+q_\alpha(t)-q_\beta(t-s)-z}\bigg|
   \\ & & +\,C\eps^{11/2}\,\sum_{\beta=1}^N
   \int_0^t ds\int_0^s d\tau\,
   \bigg|\int\int d^3x\,d^3y\,\varphi(x)\varphi(y) \nonumber
   \\ & & \hspace{7em} \times\,\nabla^3\int d^3z\,\zeta_{v_\beta(t-s)}(z)
   \frac{1}{|\tilde{x}|}\,\delta(|\tilde{x}|-s)
   \Big|_{\tilde{x}=x-y+q_\alpha(t)-q_\beta(t-s)-z}\bigg|
   \\ & \le & C\eps^{9/2}\,\sum_{\stackrel{\beta=1}{\beta\neq\alpha}}^N
   \int_0^t ds\,s\,
   \bigg|\int\int d^3x\,d^3y\,\varphi(x)\varphi(y) \nonumber
   \\ & & \hspace{7em} \times\,\nabla^4\int d^3z\,\zeta_{v_\beta(t-s)}(z)
   \frac{1}{|\tilde{x}|}\,\delta(|\tilde{x}|-s)
   \Big|_{\tilde{x}=x-y+q_\alpha(t)-q_\beta(t-s)-z}\bigg|
   \\ & & +\,C\eps^{11/2}\,\sum_{\beta=1}^N
   \int_0^t ds\,s\,
   \bigg|\int\int d^3x\,d^3y\,\varphi(x)\varphi(y) \nonumber
   \\ & & \hspace{7em} \times\,\nabla^3\int d^3z\,\zeta_{v_\beta(t-s)}(z)
   \frac{1}{|\tilde{x}|}\,\delta(|\tilde{x}|-s)
   \Big|_{\tilde{x}=x-y+q_\alpha(t)-q_\beta(t-s)-z}\bigg|\,.
\end{eqnarray*}
In the first term we have $\alpha\neq\beta$, and we estimate $s\le t\le C\eps^{-3/2}$
to be left with an ``$\eps^3\nabla^4$\,''. Exactly as in (\ref{alpsp}) we see that this part
is bounded by $C\eps^5$, whence
\begin{eqnarray}\label{celo}
   |T_{3+4, {\rm main}, 2}(t, t_1)| & \le &
   C\eps^5+\,C\eps^{11/2}\,\sum_{\beta=1}^N
   \int_0^t ds\,s\,
   \bigg|\int\int d^3x\,d^3y\,\varphi(x)\varphi(y) \nonumber
   \\ & & \hspace{7em} \times\,\nabla^3\int d^3z\,\zeta_{v_\beta(t-s)}(z)
   \frac{1}{|\tilde{x}|}\,\delta(|\tilde{x}|-s)
   \Big|_{\tilde{x}=x-y+q_\alpha(t)-q_\beta(t-s)-z}\bigg|\,.
   \nonumber \\ & &
\end{eqnarray}
For $\alpha\neq\beta$ we use $s\le t\le C\eps^{-3/2}$,
and by means of Lemma \ref{werdenf}(a), (b), and (\ref{surf-int}) we see that
\begin{eqnarray}\label{alpspwd}
   & & \eps^{11/2}\bigg(\int_0^{\tau_{\ast\ast}}+\int_{\tau_{\ast\ast}}^t\bigg)ds\,s
   \,\bigg|\int\int d^3x\,d^3y\,\varphi(x)\varphi(y) \nonumber
   \\ & & \hspace{11em} \times\,\nabla^3\int d^3z\,\zeta_{v_\beta(t-s)}(z)
   \frac{1}{|\tilde{x}|}\,\delta(|\tilde{x}|-s)
   \Big|_{\tilde{x}=x-y+q_\alpha(t)-q_\beta(t-s)-z}\bigg|
   \nonumber \\ & & \le C\eps^4\int_0^{\tau_{\ast\ast}} ds\,
   \int\int d^3x\,d^3y\,\varphi(x)\varphi(y)\,\eps^4\,\frac{1}{s}
   \int_{|z-[x-y+q_\alpha(t)-q_\beta(t-s)]|=s} d^2z
   \nonumber \\ & & \quad +\,C\eps^4\int_{\tau_{\ast\ast}}^t ds\,
   \int\int d^3x\,d^3y\,\varphi(x)\varphi(y)\,\eps^4\,
   \int_{|z-[x-y+q_\alpha(t)-q_\beta(t-s)]|=s} d^2z\,\frac{1}{|z|}
   \nonumber \\ & & \le C\eps^4\Big(\eps^4\tau_{\ast\ast}^2+\eps^4 t^2\Big)\le C\eps^5
\end{eqnarray}
for $t\le T\eps^{-3/2}$. Finally for $\alpha=\beta$ we can invoke Lemma \ref{werdenf}(c)
to find
\begin{eqnarray}\label{zugspwd}
   & & \eps^{11/2}\bigg(\int_0^{8R_\varphi}+\int_{8R_\varphi}^t\bigg)ds\,s
   \,\bigg|\int\int d^3x\,d^3y\,\varphi(x)\varphi(y)
   \nonumber \\ & & \hspace{11em}\times\,\nabla^3\int d^3z\,\zeta_{v_\alpha(t-s)}(z)
   \frac{1}{|\tilde{x}|}\,\delta(|\tilde{x}|-s)
   \Big|_{\tilde{x}=x-y+q_\alpha(t)-q_\alpha(t-s)-z}\bigg|
   \nonumber \\ & & \le C\eps^{11/2}\int_0^{8R_\varphi} ds\,s\,
   \int\int d^3x\,d^3y\,|\nabla^3\varphi(x)|\varphi(y)\,\frac{1}{s}
   \int_{|z-[x-y+q_\alpha(t)-q_\alpha(t-s)]|=s} d^2z\,\frac{1}{|z|}
   \nonumber \\ & & \quad +\,C\eps^{11/2}\int_{8R_\varphi}^t ds\,s\,
   \int\int d^3x\,d^3y\,\varphi(x)\varphi(y)\,\frac{1}{s^5}\,
   \int_{|z-[x-y+q_\alpha(t)-q_\alpha(t-s)]|=s} d^2z
   \nonumber \\ & & \le C\eps^{11/2}.
\end{eqnarray}
In view of (\ref{celo}), (\ref{alpspwd}), and (\ref{zugspwd}) we infer that
$|T_{3+4, {\rm main}, 2}(t, t_1)|\le C\eps^5$, and together with (\ref{T34data})
and (\ref{T34main1}) we conclude that indeed (\ref{T34-bd}) holds.

\subsection{Bounding $|\ddot{R}_\alpha(t)|$}
\label{ddotR-sect}

We are going to verify the estimate (\ref{ddotR-bd-2}).
A direct calculation starting from (\ref{R-def}) reveals that
\begin{eqnarray}\label{ibidevo}
   \ddot{R}_\alpha(t) & = & \bigg(\frac{d^2}{dt^2}\,{m_{0\alpha}(v_\alpha)}^{-1}
   \bigg)m_{0\alpha}(v_\alpha)R_\alpha(t)
   +\bigg(\frac{d}{dt}\,{m_{0\alpha}(v_\alpha)}^{-1}
   \bigg)\bigg(\frac{d}{dt}\,m_{0\alpha}(v_\alpha)\bigg)R_\alpha(t)
   \nonumber \\ & & +\,\bigg(\frac{d}{dt}\,{m_{0\alpha}(v_\alpha)}^{-1}
   \bigg)m_{0\alpha}(v_\alpha)\dot{R}_\alpha(t)
   \nonumber \\ & & +\,\bigg(\frac{d}{dt}\,{m_{0\alpha}(v_\alpha)}^{-1}
   \bigg)m_{0\alpha}(v_\alpha)\bigg[\dot{R}_\alpha(t)
   -\bigg(\frac{d}{dt}\,{m_{0\alpha}(v_\alpha)}^{-1}
   \bigg)m_{0\alpha}(v_\alpha)R_\alpha(t)\bigg]
   \nonumber \\ & & +\,{m_{0\alpha}(v_\alpha)}^{-1}\,
   \sum_{\stackrel{\beta=1}{\beta\neq\alpha}}^N S_{\alpha\beta}(t),
\end{eqnarray}
with the terms
\begin{eqnarray}\label{Salphbet-def}
   S_{\alpha\beta}(t) & = & \int d^3x\,\rho_\alpha(x-q_\alpha)\bigg\{
   (\ddot{v}_\beta\cdot\nabla_v) E_{v_\beta}
   +(\dot{v}_\beta\cdot\nabla_v)^2 E_{v_\beta}
   +2(\dot{v}_\beta\cdot\nabla_v)([v_\alpha-v_\beta]\cdot\nabla) E_{v_\beta}
   \nonumber \\ & & \hspace{8em}
   +\,([\dot{v}_\alpha-\dot{v}_\beta]\cdot\nabla) E_{v_\beta}
   +([v_\alpha-v_\beta]\cdot\nabla)^2 E_{v_\beta}
   +\ddot{v}_\alpha\wedge B_{v_\beta}
   \nonumber \\ & & \hspace{8em}
   +\,2\dot{v}_\alpha\wedge (\dot{v}_\beta\cdot\nabla_v) B_{v_\beta}
   +2\dot{v}_\alpha\wedge ([v_\alpha-v_\beta]\cdot\nabla) B_{v_\beta}
   +v_\alpha\wedge (\ddot{v}_\beta\cdot\nabla_v) B_{v_\beta}
   \nonumber \\ & & \hspace{8em}
   +\,v_\alpha\wedge (\dot{v}_\beta\cdot\nabla_v)^2 B_{v_\beta}
   +v_\alpha\wedge (\dot{v}_\beta\cdot\nabla_v)([v_\alpha-v_\beta]\cdot\nabla) B_{v_\beta}
   \nonumber \\ & & \hspace{8em}
   +\,v_\alpha\wedge ([\dot{v}_\alpha-\dot{v}_\beta]\cdot\nabla) B_{v_\beta}
   +v_\alpha\wedge (\dot{v}_\beta\cdot\nabla_v)([v_\alpha-v_\beta]\cdot\nabla) B_{v_\beta}
   \nonumber \\ & & \hspace{8em}
   +\,v_\alpha\wedge ([v_\alpha-v_\beta]\cdot\nabla)^2 B_{v_\beta}\bigg\}(x-q_\beta),
\end{eqnarray}
where all $q_\alpha$, $q_\beta$, etc., are evaluated at time $t$.
Invoking the bounds (\ref{cava1})--(\ref{haberl1}), (\ref{ibidevo}) yields
\begin{equation}\label{ddotR}
   |\ddot{R}_\alpha(t)|\le C\eps^6+C\max_{\beta\neq\alpha}|S_{\alpha\beta}(t)|,
   \quad t\in [\tau_{\ast\ast}, T\eps^{-3/2}].
\end{equation}
To estimate $S_{\alpha\beta}(t)$, we introduce for $\alpha\neq\beta$
the interaction terms
\begin{eqnarray}\label{dxdy}
   \nabla\Psi_{\alpha\beta}(t) & := & \int d^3x\,\rho_\alpha(x-q_\alpha(t))
   \nabla\phi_{v_\beta(t)}(x-q_\beta(t)) \nonumber \\ & = &
   (-i)\,e_\alpha e_\beta\int d^3k\,k\frac{{|\hat{\varphi}(k)|}^2}
   {k^2-{(k\cdot v_\beta(t))}^2}\,e^{ik\cdot [q_\beta(t)-q_\alpha(t)]}
   \nonumber \\ & = & \frac{e_\alpha e_\beta}{4\pi}
   \int\int d^3x d^3y\,\varphi(x-q_\alpha(t))\varphi(y-q_\beta(t))
   \nabla\zeta_{v_\beta(t)}(x-y),
\end{eqnarray}
with $\zeta_v(x)$ from Lemma \ref{zetav-esti}, and for $l, j\ge 0$
we will also need the derivatives
\begin{eqnarray}\label{dxdy-lj}
   \nabla_v^l\nabla^j\Psi_{\alpha\beta}(t) & := & \int d^3x\,\rho_\alpha(x-q_\alpha(t))
   \nabla_v^l\nabla^j\phi_{v_\beta(t)}(x-q_\beta(t))
   \nonumber \\ & = & \frac{e_\alpha e_\beta}{4\pi}
   \int\int d^3x d^3y\,\varphi(x-q_\alpha(t))\varphi(y-q_\beta(t))
   \nabla_v^l\nabla^j\zeta_{v_\beta(t)}(x-y).
\end{eqnarray}
To illustrate the method for bounding $S_{\alpha\beta}(t)$, let us first consider
\[ S_{\alpha\beta, 1}(t)=\int d^3x\,\rho_\alpha(x-q_\alpha(t))
   (\ddot{v}_\beta\cdot\nabla_v) E_{v_\beta(t)}(x-q_\beta(t)), \]
which comprises the first term in (\ref{Salphbet-def}). Recalling
$E_v(x)=-\nabla\phi_v(x)+(v\cdot\nabla\phi_v(x))v$ from (\ref{EBv-def})
and calculating $(\ddot{v}\cdot\nabla_v) E_v(x)$ explicitly, by means of (\ref{dxdy})
and (\ref{dxdy-lj}) we may rewrite $S_{\alpha\beta, 1}(t)$ as
\begin{eqnarray}\label{wiweht}
   S_{\alpha\beta, 1}(t) & = & -(\ddot{v}_\beta(t)\cdot\nabla_v)\nabla
   \Psi_{\alpha\beta}(t)+v_\beta(t)\,(\ddot{v}_\beta(t)\cdot\nabla)\Psi_{\alpha\beta}(t)
   \nonumber \\ & & +\,v_\beta(t)\,(\ddot{v}_\beta(t)\cdot\nabla_v)(v_\beta(t)\cdot\nabla)
   \Psi_{\alpha\beta}(t)+\ddot{v}_\beta(t)\,
   (v_\beta(t)\cdot\nabla)\Psi_{\alpha\beta}(t).
\end{eqnarray}
{}From Lemma \ref{zetav-esti} we obtain
\begin{eqnarray}\label{geoschl}
   \Big|\nabla_v^l\nabla^j\Psi_{\alpha\beta}(t)\Big|
   & \le & C\,\int\int d^3x d^3y\,\varphi(x-q_\alpha(t))\varphi(y-q_\beta(t))
   {|x-y|}^{-(j+1)} \nonumber \\ & = & C\,\int\int d^3x d^3y\,\varphi(x)\varphi(y)
   {|x-y+q_\alpha(t)-q_\beta(t)|}^{-(j+1)}.
\end{eqnarray}
Since $\alpha\neq\beta$ we have $|x-y+q_\alpha(t)-q_\beta(t)|\ge |q_\alpha(t)-q_\beta(t)|
-2R\varphi\ge C_\ast\eps^{-1}-2R_\varphi\ge (C_\ast/2)\eps^{-1}$
for $|x|, |y|\le R_\varphi$ and $t\in [0, T\eps^{-3/2}]$, due to Lemma \ref{esti}.
Therefore (\ref{geoschl}) yields
\begin{equation}\label{Psi-esti}
   \Big|\nabla_v^l\nabla^j\Psi_{\alpha\beta}(t)\Big|\le
   C\eps^{j+1},\quad l, j\ge 0,\quad\alpha\neq\beta,
   \quad t\in [0, T\eps^{-3/2}],
\end{equation}
and applied to (\ref{wiweht}) we find in view of Lemma \ref{esti}
and Lemma \ref{q3esti} that
\[ |S_{\alpha\beta, 1}(t)|\le C\eps^{7/2}\eps^2+C\sqrt{\eps}\eps^{7/2}\eps^2
   +C\sqrt{\eps}\eps^{7/2}\sqrt{\eps}\eps^2+C\eps^{7/2}\sqrt{\eps}\eps^2
   \le C\eps^{11/2},\quad t\in [\tau_{\ast\ast}, T\eps^{-3/2}]. \]
In principle, all other terms in (\ref{Salphbet-def}) may be
handled in the same manner, the rule of thumb being that one first
counts the powers of $\eps$ due to Lemma \ref{esti}, then one counts
the $\nabla$-derivatives, with $E_v, B_v\cong\nabla\phi_v$,
and a $\nabla^j$ gives an additional $\eps^{j+1}$, whereas the
$\nabla_v$-derivatives do not hurt the estimate. This way term by
term can be bounded, the worst ones being
\begin{eqnarray*}
   & & \bigg|\int d^3x\,\rho_\alpha(x-q_\alpha(t))
   \bigg\{([\dot{v}_\alpha(t)-\dot{v}_\beta(t)]\cdot\nabla)
   E_{v_\beta(t)}(x-q_\beta(t))
   \\ & & \hspace{10em} +\,([v_\alpha(t)-v_\beta(t)]\cdot\nabla)^2
   E_{v_\beta(t)}(x-q_\beta(t))\bigg|\le C\eps^5.
\end{eqnarray*}
Consequently, (\ref{ddotR}) shows that also
\begin{equation}\label{ddotR-bd}
   |\ddot{R}_\alpha(t)|\le C\eps^5,
   \quad\alpha=1, \ldots, N,\quad t\in [\tau_{\ast\ast}, T\eps^{-3/2}],
\end{equation}
holds.

\subsection{Conclusion of the proof of Lemma \ref{q4esti}}
\label{conclu-sect}

We recall from (\ref{tauastast-def}) that $\tau_{\ast\ast}=(C_\ast/8)\eps^{-1}$,
cf.~also Lemma \ref{werdenf}. To begin with, we fix some $\alpha\in \{1, \ldots, N\}$.
{}From (\ref{tengl}) and (\ref{gbma}) we know that
for $t\in [\tau_{\ast\ast}, T\eps^{-3/2}]$ we have
\begin{eqnarray*}
  |\stackrel{...}{v}_\alpha(t)| & \le & C\eps^5
   +C\,\bigg|\int d^3x\,\rho_\alpha(x)({\cal L}_\alpha(t)Z)
   (x+q_\alpha(t), t)\bigg| \\ & \le &
   C\eps^5+|T_{{\rm data}}(t, t_1)|+|T_1(t, t_1)|+|T_2(t, t_1)|+|T_3(t, t_1)+T_4(t, t_1)|,
\end{eqnarray*}
with $t_1$ still to be selected. If below we can moreover ensure that
\begin{equation}\label{sevts}
   t_1\in [\tau_{\ast\ast}, t],\quad t\in [t_1+\tau_{\ast\ast}, T\eps^{-3/2}],
   \quad t\ge\tau_{\ast\ast},
\end{equation}
then (\ref{data-bd}), (\ref{T2-bd}), and (\ref{T34-bd}) imply
\[ |\stackrel{...}{v}_\alpha(t)|\le C\eps^5+|T_1(t, t_1)|. \]
Concerning $T(t, t_1)$, we rely on the bounds obtained in Section \ref{T1-sect}.
Assuming (\ref{sevts}) it is found from (\ref{T1-esti-1}) that
for $t\in [\tau_{\ast\ast}, T\eps^{-3/2}]$
\begin{equation}\label{hbyasi}
   |\stackrel{...}{v}_\alpha(t)|\le C\eps^5
   +\bigg|\sum_{\beta=1}^N\,\int d^3x\,\rho_\alpha(x)\int_{t_1}^t ds\,
   \Big[U(t-s)\Big((\stackrel{...}{v}_\beta(s)\cdot\nabla_v)\Phi_{v_\beta(s)}
   (\cdot-q_\beta(s))\Big)\Big](x+q_\alpha(t))\bigg|\,.
\end{equation}
Now we fix $t\in [2\tau_{\ast\ast}, T\eps^{-3/2}]$
and set $t_1=\tau_{\ast\ast}$. Then (\ref{sevts}) holds, and (\ref{hbyasi}) implies
\begin{eqnarray*}
   |\stackrel{...}{v}_\alpha(t)| & \le & C\eps^5
   \\ & & +\,\bigg|\sum_{\beta=1}^N\,\int d^3x\,\rho_\alpha(x)
   \int_{\tau_{\ast\ast}}^{2\tau_{\ast\ast}} ds\,
   \Big[U(t-s)\Big((\stackrel{...}{v}_\beta(s)\cdot\nabla_v)\Phi_{v_\beta(s)}
   (\cdot-q_\beta(s))\Big)\Big](x+q_\alpha(t))\bigg|
   \\ & & +\,\bigg|\sum_{\beta=1}^N\,\int d^3x\,\rho_\alpha(x)
   \int_{2\tau_{\ast\ast}}^t ds\,
   \Big[U(t-s)\Big((\stackrel{...}{v}_\beta(s)\cdot\nabla_v)\Phi_{v_\beta(s)}
   (\cdot-q_\beta(s))\Big)\Big](x+q_\alpha(t))\bigg|\,.
\end{eqnarray*}
Utilizing (\ref{lampa}) with $t_2=\tau_{\ast\ast}$ and (\ref{wamoz-sum})
with $t_2=2\tau_{\ast\ast}$, it follows that
\[ |\stackrel{...}{v}_\alpha(t)|\le C\eps^4+C\Big(\max_{1\le\kappa\le N}|e_\kappa|^2\Big)
   \Big(\sup_{s\in [2\tau_{\ast\ast}, T\eps^{-3/2}]}
   \max_{1\le\kappa\le N}|\stackrel{...}{v}_\kappa(s)|\Big) \]
for $t\in [2\tau_{\ast\ast}, T\eps^{-3/2}]$, whence
\begin{equation}\label{eps4-bd}
   \sup_{t\in [2\tau_{\ast\ast}, T\eps^{-3/2}]}
   \max_{1\le\kappa\le N}|\stackrel{...}{v}_\kappa(t)|\le C\eps^4,
\end{equation}
if the $|e_\kappa|$ are chosen small enough.
Next we fix $t\in [3\tau_{\ast\ast}, T\eps^{-3/2}]$ and set $t_1=2\tau_{\ast\ast}$.
Then again (\ref{sevts}) is satisfied, therefore by (\ref{hbyasi})
\begin{eqnarray*}
   |\stackrel{...}{v}_\alpha(t)| & \le & C\eps^5
   \\ & & +\,\bigg|\sum_{\beta=1}^N\,\int d^3x\,\rho_\alpha(x)
   \int_{2\tau_{\ast\ast}}^{3\tau_{\ast\ast}} ds\,
   \Big[U(t-s)\Big((\stackrel{...}{v}_\beta(s)\cdot\nabla_v)\Phi_{v_\beta(s)}
   (\cdot-q_\beta(s))\Big)\Big](x+q_\alpha(t))\bigg|
   \\ & & +\,\bigg|\sum_{\beta=1}^N\,\int d^3x\,\rho_\alpha(x)
   \int_{3\tau_{\ast\ast}}^t ds\,
   \Big[U(t-s)\Big((\stackrel{...}{v}_\beta(s)\cdot\nabla_v)\Phi_{v_\beta(s)}
   (\cdot-q_\beta(s))\Big)\Big](x+q_\alpha(t))\bigg|\,.
\end{eqnarray*}
For the first part (\ref{lamperl}) applies with $t_3=2\tau_{\ast\ast}$,
whereas for the second part we can use (\ref{wamoz-sum})
with $t_2=3\tau_{\ast\ast}$. Accordingly we infer
\begin{eqnarray*}
   |\stackrel{...}{v}_\alpha(t)| & \le & C\eps^5
   +C\Big(\sup_{s\in [2\tau_{\ast\ast}, 3\tau_{\ast\ast}]}
   \max_{1\le\kappa\le N}|\stackrel{...}{v}_\kappa(s)|\Big)\eps^{1/4}
   \\ & & +\,C\Big(\max_{1\le\kappa\le N}|e_\kappa|^2\Big)
   \Big(\sup_{s\in [3\tau_{\ast\ast}, T\eps^{-3/2}]}
   \max_{1\le\kappa\le N}|\stackrel{...}{v}_\kappa(s)|\Big)
   \\ & \le & C\eps^{17/4}+C\Big(\max_{1\le\kappa\le N}|e_\kappa|^2\Big)
   \Big(\sup_{s\in [3\tau_{\ast\ast}, T\eps^{-3/2}]}
   \max_{1\le\kappa\le N}|\stackrel{...}{v}_\kappa(s)|\Big),
\end{eqnarray*}
where we have used (\ref{eps4-bd}). As this hold
for all $t\in [3\tau_{\ast\ast}, T\eps^{-3/2}]$,
by choosing the $|e_\kappa|$ small enough we can ensure
\begin{equation}\label{eps174-bd}
   \sup_{t\in [3\tau_{\ast\ast}, T\eps^{-3/2}]}
   \max_{1\le\kappa\le N}|\stackrel{...}{v}_\kappa(t)|\le C\eps^{17/4}.
\end{equation}
Now it is clear how this procedure is iterated to gain factors $\eps^{1/4}$.
{}From (\ref{eps174-bd}) we obtain the bound
\[ \sup_{t\in [4\tau_{\ast\ast}, T\eps^{-3/2}]}
   \max_{1\le\kappa\le N}|\stackrel{...}{v}_\kappa(t)|\le C\eps^{9/2}, \]
and so on, until the power $\eps^5$ is reached. Then no further improvement
is possible, since there are other error terms of order ${\cal O}(\eps^5)$
in (\ref{hbyasi}). This way we arrive at
\[ \sup_{t\in [6\tau_{\ast\ast}, T\eps^{-3/2}]}
   \max_{1\le\kappa\le N}|\stackrel{...}{v}_\kappa(t)|\le C\eps^5, \]
and this completes the proof of Lemma \ref{q4esti}. {\hfill$\Box$}\bigskip


\section{Appendix B: Proof of Lemma \ref{tayl}}
\label{tayl-sect}

The proof follows the lines of the proof of \cite[Lemma 3.2]{Nteil},
although some care has to be taken since the key estimate on $\stackrel{...}{v}_\alpha(t)$
from Lemma \ref{q4esti} does not hold for $t\in [0, T\eps^{-3/2}]$,
but only for $t\in [6\tau_{\ast\ast}, T\eps^{-3/2}]$. We will consider
only assertion (b) of Lemma \ref{tayl}, the other parts being
verified similarly. Recalling $\xi_{\alpha\beta}=q_\alpha(t)-q_\beta(t)$
and $\alpha\neq\beta$, we first introduce
\begin{eqnarray}\label{D1-form}
   D_1(t) & = & i\int_0^t d\tau\int d^3k\,
   {|\hat{\varphi}(k)|}^2 e^{-ik\cdot\xi_{\alpha\beta}}
   \bigg\{ e^{-ik\cdot [q_\beta(t)-q_\beta(t-\tau)]}
   - e^{-ik\cdot [\tau v_\beta-\frac{1}{2}\tau^2\dot{v}_\beta
   +\frac{1}{6}\tau^3\ddot{v}_\beta]}\bigg\}\,\frac{\sin|k|\tau}{|k|}k \nonumber \\
   & = & -\,\nabla_\xi\int_0^t d\tau\int d^3k\,
   {|\hat{\varphi}(k)|}^2 e^{-ik\cdot\xi_{\alpha\beta}}
   \bigg\{ e^{-ik\cdot [q_\beta(t)-q_\beta(t-\tau)]}
   - e^{-ik\cdot [\tau v_\beta-\frac{1}{2}\tau^2\dot{v}_\beta
   +\frac{1}{6}\tau^3\ddot{v}_\beta]}\bigg\}\,\frac{\sin|k|\tau}{|k|} \nonumber \\
   & = & -\,\nabla_\xi\int\int d^3x\,d^3y\,\varphi(x)
   \varphi(y)\int_0^t d\tau\,\Big\{\psi_{\tau}\Big(x-y+\xi_{\alpha\beta}+q_\beta(t)
   -q_\beta(t-\tau)\Big) \nonumber \\ & & \hspace{13em}
   -\psi_{\tau}\Big(x-y+\xi_{\alpha\beta}+\tau v_\beta-\frac{1}{2}\tau^2\dot{v}_\beta
   +\frac{1}{6}\tau^3\ddot{v}_\beta\Big)\Big\}, \quad
\end{eqnarray}
where $\psi_\tau(x)=(4\pi |x|)^{-1}\delta(|x|-\tau)$. To proceed further,
we need two technical lemmas.

\begin{lemma}\label{zero-lem-1} For $|x|, |y|\le R_\varphi$
and $t\in [t_0, T\eps^{-3/2}]$ fixed,
where $t_0=4(R_\varphi+C^\ast\eps^{-1})$, cf.~(\ref{t0-def}), we define the function
$\theta=\theta(\tau)=\theta(\tau; x-y, t, \alpha, \beta)$ through
\[ \theta(\tau)=\tau-|x-y+q_\alpha(t)-q_\beta(t-\tau)|,\quad\tau\in [0, t]. \]
Then $2\ge\theta'(\tau)\ge 3/4$, and there exists
a unique $\tau_0=\tau_0(x-y, t, \alpha, \beta)\in [0, t_0]$
such that $\theta(\tau_0)=0$. More precisely, the estimate
\begin{equation}\label{tau0}
   (C_\ast/2)\eps^{-1}\le\tau_0\le 2C^\ast\eps^{-1}
\end{equation}
holds.
\end{lemma}
{\bf Proof\,:} We have $\theta'(\tau)
=1-\frac{x-y+q_\alpha(t)-q_\beta(t-\tau)}{|x-y+q_\alpha(t)-q_\beta(t-\tau)|}
\cdot v_\beta(t-\tau)$, whence $2\ge\theta'(\tau)\ge 3/4$
in view of Lemma \ref{esti} for $\eps$ small enough.
For the other claims, we first note that $0\ge\theta(0)=-|x-y+q_\alpha(t)-q_\beta(t)|
\ge -(2R_\varphi+C^\ast\eps^{-1})\ge -2C^\ast\eps^{-1}$ and therefore also
$\theta(t_0)=\theta(0)+\int_0^{t_0}\theta'(\tau)\,d\tau
\ge -2C^\ast\eps^{-1}+3t_0/4=3R_\varphi+C^\ast\eps^{-1}>0$, whence
$\theta$ has a unique zero $\tau_0$ satisfying $\tau_0\in [0, t_0]$.
To verify (\ref{tau0}), we estimate
\begin{eqnarray*}
   |x-y+q_\alpha(t)-q_\beta(t-\tau)| & \ge &
   |q_\alpha(t)-q_\beta(t)|-|x|-|y|-|q_\beta(t)-q_\beta(t-\tau)|
   \\ & \ge & C_\ast\eps^{-1}-2R_\varphi-C\sqrt{\eps}\tau
\end{eqnarray*}
by (\ref{diff-bound}) and (\ref{v-bound}). Since $\tau_0\in [0, t_0]$
we have $\sqrt{\eps}\tau_0\le C\eps^{-1/2}$, thus
\[ \tau_0=|x-y+q_\alpha(t)-q_\beta(t-\tau_0)|\ge (C_\ast/2)\eps^{-1} \]
for $\eps$ sufficiently small. On the other hand,
in view of (\ref{diff-bound}) and (\ref{v-bound}) also
\begin{eqnarray*}
   \tau_0=|x-y+q_\alpha(t)-q_\beta(t-\tau_0)| & \le &
   |q_\alpha(t)-q_\beta(t)|+|x|+|y|+|q_\beta(t)-q_\beta(t-\tau_0)|
   \\ & \le & C^\ast\eps^{-1}+2R_\varphi+C\sqrt{\eps}\tau_0,
\end{eqnarray*}
and therefore $\tau_0\le 2C^\ast\eps^{-1}$ for $\eps$ small enough.
{\hfill$\Box$}\bigskip

\begin{lemma}\label{zero-lem-2} In the setting of Lemma \ref{zero-lem-1}
we now define $\bar{\theta}=\bar{\theta}(\tau)
=\bar{\theta}(\tau; x-y, t, \alpha, \beta)$ by
\[ \bar{\theta}(\tau)=\tau-\bigg|x-y+\xi_{\alpha\beta}(t)+\tau v_\beta(t)
   -\frac{1}{2}\tau^2\dot{v}_\beta(t)+\frac{1}{6}\tau^3\ddot{v}_\beta(t)\bigg|,
   \quad\tau\in [0, t]. \]
Then $2\ge\bar{\theta}'(\tau)\ge 3/4$, and there exists
a unique $\tau_1=\tau_1(x-y, t, \alpha, \beta)\in [0, t_0]$
such that $\bar{\theta}(\tau_1)=0$. Again we can arrange for
\begin{equation}\label{tau1}
   (C_\ast/2)\eps^{-1}\le \tau_1\le 2C^\ast\eps^{-1}
\end{equation}
to be satisfied.
\end{lemma}
{\bf Proof\,:} Here we have $\bar{\theta}'(\tau)=1-\frac{(\ldots)}{|\ldots|}
\cdot\Big(v_\beta(t)-\tau\dot{v}_\beta(t)+\frac{1}{2}\tau^2\ddot{v}_\beta(t)\Big)$.
Due to $t\ge t_0\ge\tau_{\ast\ast}=(C_\ast/8)\eps^{-1}$ we obtain $\tau^2|\ddot{v}_\beta(t)|
\le Ct^2\eps^{7/2}\le C\sqrt{\eps}$ by Lemma \ref{q3esti}, and also $|v_\beta(t)|
+\tau |\dot{v}_\beta(t)|\le C\sqrt{\eps}$ in view of Lemma \ref{esti}.
Since $\bar{\theta}(0)=\theta(0)$ we can proceed as before
in the proof of Lemma \ref{zero-lem-1}. {\hfill$\Box$}\bigskip

Returning to (\ref{D1-form}) and using Lemmas \ref{zero-lem-1} and
\ref{zero-lem-2}, we may thus simply write
\begin{equation}\label{coln}
   D_1(t)=-\frac{1}{4\pi}\,\int\int d^3x\,d^3y\,\varphi(x)
   \varphi(y)\,\nabla_\xi\Big(\tau_0^{-1}-\tau_1^{-1}\Big).
\end{equation}
Calculating $\nabla_\xi\tau_0^{-1}$ and $\nabla_\xi\tau_1^{-1}$
from the defining properties $\theta(\tau_0)=0$ and $\bar{\theta}(\tau_1)=0$,
we arrive at
\begin{eqnarray*}
   \nabla_\xi\tau_0^{-1} & = & -\tau_0^{-3}\bigg\{
   \Big(x-y+\xi_{\alpha\beta}+q_\beta(t)-q_\beta(t-\tau_0)\Big)
   \\ & & \hspace{3em}
   +\Big(x-y+\xi_{\alpha\beta}+q_\beta(t)-q_\beta(t-\tau_0)\Big)\cdot
   v_\beta(t-\tau_0)\nabla_\xi\tau_0\bigg\}, \\
   \nabla_\xi\tau_1^{-1} & = & -\tau_1^{-3}\bigg\{
   \Big(x-y+\xi_{\alpha\beta}+\tau_1 v_\beta-\frac{1}{2}\tau_1^2\dot{v}_\beta
   +\frac{1}{6}\tau_1^3\ddot{v}_\beta\Big)
   \\ & & \hspace{3em}
   +\Big(x-y+\xi_{\alpha\beta}+\tau_1 v_\beta-\frac{1}{2}\tau_1^2\dot{v}_\beta
   +\frac{1}{6}\tau_1^3\ddot{v}_\beta\Big)\cdot\Big(v_\beta-\tau_1\dot{v}_\beta
   +\frac{1}{2}\tau_1^2\ddot{v}_\beta\Big)\nabla_\xi\tau_1\bigg\}.
\end{eqnarray*}
Therefore
\begin{eqnarray*}
   \Big|\nabla_\xi\tau_0^{-1}-\nabla_\xi\tau_1^{-1}\Big| & \le &
   \Big|\tau_0^{-3}-\tau_1^{-3}\Big|\,\Big|x-y+\xi_{\alpha\beta}+\tau_1 v_\beta
   -\frac{1}{2}\tau_1^2\dot{v}_\beta+\frac{1}{6}\tau_1^3\ddot{v}_\beta\Big|
   \\ & & \hspace{1em}\times\,\Big[1+\Big(|v_\beta|+\tau_1|\dot{v}_\beta|
   +\frac{1}{2}\tau_1^2|\ddot{v}_\beta|\Big)|\nabla_\xi\tau_1|\Big]
   \\ & & +\,\tau_0^{-3}|q_\beta(t-\tau_0)-q_\beta(t-\tau_1)|
   \\ & & +\,\tau_0^{-3}\Big|q_\beta(t)-q_\beta(t-\tau_1)
   -\tau_1 v_\beta+\frac{1}{2}\tau_1^2\dot{v}_\beta-\frac{1}{6}\tau_1^3\ddot{v}_\beta\Big|
   \\ & & +\,\tau_0^{-3}\Big|x-y+\xi_{\alpha\beta}+q_\beta(t)-q_\beta(t-\tau_0)\Big|
   |v_\beta(t-\tau_0)||\nabla_\xi(\tau_0-\tau_1)|
   \\ & & +\,\tau_0^{-3}\Big|x-y+\xi_{\alpha\beta}+q_\beta(t)-q_\beta(t-\tau_0)\Big|
   |v_\beta(t-\tau_0)-v_\beta(t-\tau_1)||\nabla_\xi\tau_1|
   \\ & & +\,\tau_0^{-3}\Big|x-y+\xi_{\alpha\beta}+q_\beta(t)-q_\beta(t-\tau_0)\Big|
   \\ & & \hspace{1em} \times\,\Big|v_\beta(t-\tau_1)-v_\beta+\tau_1\dot{v}_\beta
   -\frac{1}{2}\tau_1^2\ddot{v}_\beta\Big||\nabla_\xi\tau_1|
   \\ & & +\,\tau_0^{-3}|q_\beta(t-\tau_0)-q_\beta(t-\tau_1)||v_\beta(t-\tau_1)|
   |\nabla_\xi\tau_1|
   \\ & & +\,\tau_0^{-3}\Big|q_\beta(t)-q_\beta(t-\tau_1)
   -\tau_1 v_\beta+\frac{1}{2}\tau_1^2\dot{v}_\beta
   -\frac{1}{6}\tau_1^3\ddot{v}_\beta\Big||v_\beta(t-\tau_1)||\nabla_\xi\tau_1|.
\end{eqnarray*}
Hence Lemma \ref{esti}, Lemma \ref{q3esti}, (\ref{tau0}), and (\ref{tau1}) yield
\begin{eqnarray}\label{jkhal}
   \Big|\nabla_\xi\tau_0^{-1}-\nabla_\xi\tau_1^{-1}\Big| & \le &
   C\eps^{-1}\Big|\tau_0^{-3}-\tau_1^{-3}\Big|\Big[1+\sqrt{\eps}|\nabla_\xi\tau_1|\Big]
   +C\eps^{7/2}|\tau_0-\tau_1| \nonumber \\ & &
   +\,C\eps^3\Big|q_\beta(t)-q_\beta(t-\tau_1)-\tau_1 v_\beta
   +\frac{1}{2}\tau_1^2\dot{v}_\beta-\frac{1}{6}\tau_1^3\ddot{v}_\beta\Big|
   +C\eps^{5/2}|\nabla_\xi(\tau_0-\tau_1)|
   \nonumber \\ & & +\,C\eps^4|\tau_0-\tau_1||\nabla_\xi\tau_1|
   +C\eps^2\Big|v_\beta(t-\tau_1)-v_\beta+\tau_1\dot{v}_\beta
   -\frac{1}{2}\tau_1^2\ddot{v}_\beta\Big||\nabla_\xi\tau_1|
   \nonumber \\ & & +\,C\eps^{7/2}\Big|q_\beta(t)-q_\beta(t-\tau_1)
   -\tau_1 v_\beta+\frac{1}{2}\tau_1^2\dot{v}_\beta
   -\frac{1}{6}\tau_1^3\ddot{v}_\beta\Big||\nabla_\xi\tau_1|.
\end{eqnarray}
To bound the right-hand side further we note that
$t-\tau_1\ge t_0-2C^\ast\eps^{-1}=4R_\varphi+2C^\ast\eps^{-1}
\ge (3C_\ast/4)\eps^{-1}=6\tau_{\ast\ast}$ according to (\ref{tau1}),
recall (\ref{tauastast-def}). Thus $|\stackrel{...}{v}_\beta(s)|\le C\eps^5$
for all $s\in [t-\tau_1, t]$ by Lemma \ref{q4esti}, and this implies that
\begin{eqnarray*}
   q_\beta(t) & = & q_\beta(t-\tau_1)
   +\tau_1 v_\beta-\frac{1}{2}\tau_1^2\dot{v}_\beta
   +\frac{1}{6}\tau_1^3\ddot{v}_\beta+{\cal O}(\eps^5\tau_1^4),
   \\ v_\beta(t-\tau_1) & = & v_\beta-\tau_1\dot{v}_\beta
   +\frac{1}{2}\tau_1^2\ddot{v}_\beta+{\cal O}(\eps^5\tau_1^3).
\end{eqnarray*}
Utilizing this observation and the definition of $\tau_0$ and $\tau_1$,
we moreover obtain
\begin{eqnarray*}
   |\tau_0-\tau_1| & \le & \Big|q_\beta(t)-q_\beta(t-\tau_0)
   -\tau_1 v_\beta+\frac{1}{2}\tau_1^2\dot{v}_\beta
   -\frac{1}{6}\tau_1^3\ddot{v}_\beta\Big|
   \le C\sqrt{\eps}|\tau_0-\tau_1|+C\eps^5\tau_1^4
   \\ & \le & C\sqrt{\eps}|\tau_0-\tau_1|+C\eps,
\end{eqnarray*}
whence $|\tau_0-\tau_1|\le C\eps$ and consequently also
$|\tau_0^{-3}-\tau_1^{-3}|\le C\eps^5$. Next it is verified that
$|\nabla_\xi\tau_1|\le C$, and with some more effort also
that $|\nabla_\xi(\tau_0-\tau_1)|\le C\eps^{3/2}$.
%
%
Invoking all these estimates on the right-hand side of (\ref{jkhal})
it follows that $|\nabla_\xi\tau_0^{-1}-\nabla_\xi\tau_1^{-1}|
\le C\eps^4$, and recalling (\ref{coln}), we finally obtain
\begin{equation}\label{D1-esti}
   \sup_{t\in [t_0,\,T\eps^{-3/2}]}|D_1(t)|\le C\eps^4.
\end{equation}
The next step is to consider
\begin{eqnarray*}
   D_2(t) & = & i\int_0^t d\tau\int d^3k\,
   {|\hat{\varphi}(k)|}^2 e^{-ik\cdot\xi_{\alpha\beta}}
   \bigg\{e^{-ik\cdot [\tau v_\beta-\frac{1}{2}\tau^2\dot{v}_\beta
   +\frac{1}{6}\tau^3\ddot{v}_\beta]}-\,\bigg(1-ik\cdot \Big[\tau v_\beta
   -\frac{1}{2}\tau^2\dot{v}_\beta+\frac{1}{6}\tau^3\ddot{v}_\beta\Big]
   \\ & & \hspace{10em}
   -\frac{1}{2}\Big[\tau^2 {(k\cdot v_\beta)}^2
   -\tau^3(k\cdot v_\beta)(k\cdot\dot{v}_\beta)\Big]
   +\frac{i}{6}\tau^3(k\cdot v_\beta)^3\bigg)\bigg\}\,\frac{\sin|k|\tau}{|k|}k.
\end{eqnarray*}
With $\psi_\tau(x)=(4\pi |x|)^{-1}\delta(|x|-\tau)$ this may be rewritten as
\begin{eqnarray*}
   D_2(t) & = & -\nabla_\xi\int\int d^3x\,d^3y\,\varphi(x)\varphi(y)
   \int_0^t d\tau\,\bigg\{\psi_\tau\Big(x-y+\xi_{\alpha\beta}+\tau v_\beta
   -\frac{1}{2}\tau^2\dot{v}_\beta+\frac{1}{6}\tau^3\ddot{v}_\beta\Big)
   \\ & & \hspace{6em} -\,\psi_\tau\Big(x-y+\xi_{\alpha\beta}\Big)
   -\,\nabla\psi_\tau\Big(x-y+\xi_{\alpha\beta}\Big)\cdot\Big[\tau v_\beta
   -\frac{1}{2}\tau^2\dot{v}_\beta+\frac{1}{6}\tau^3\ddot{v}_\beta\Big]
   \\ & & \hspace{6em} -\,\bigg[\frac{1}{2}\tau^2 {(v_\beta\cdot\nabla)}^2
   -\frac{1}{2}\tau^3(v_\beta\cdot\nabla)(\dot{v}_\beta\cdot\nabla)
   +\frac{1}{6}\tau^3(v_\beta\cdot\nabla)^3\bigg]\psi_\tau\Big(x-y+\xi_{\alpha\beta}\Big)
   \bigg\}.
\end{eqnarray*}
By expanding $\psi_\tau(\cdot)$ about $x-y+\xi_{\alpha\beta}$ and using a similar technique
as for $D_1(t)$ it can be verified that also
\begin{equation}\label{D2-esti}
   \sup_{t\in [t_0,\,T\eps^{-3/2}]}|D_2(t)|\le C\eps^4.
\end{equation}
Finally we introduce
\begin{eqnarray*}
   D_3(t) & = & i\int_t^\infty d\tau\int d^3k\,
   {|\hat{\varphi}(k)|}^2 e^{-ik\cdot\xi_{\alpha\beta}}
   \bigg(1-ik\cdot \Big[\tau v_\beta
   -\frac{1}{2}\tau^2\dot{v}_\beta+\frac{1}{6}\tau^3\ddot{v}_\beta\Big]
   \\ & & \hspace{10em}
   -\frac{1}{2}\Big[\tau^2 {(k\cdot v_\beta)}^2
   -\tau^3(k\cdot v_\beta)(k\cdot\dot{v}_\beta)\Big]
   +\frac{i}{6}\tau^3(k\cdot v_\beta)^3\bigg)\,\frac{\sin|k|\tau}{|k|}k,
\end{eqnarray*}
and it remains to notice that as in \cite[p.~466/467]{Nteil} we obtain
$D_3(t)=0$ for $t\in [t_0, T\eps^{-3/2}]$. Taking into account
(\ref{D1-esti}) and (\ref{D2-esti}), we see that the assertion of
Lemma \ref{tayl}(b) is satisfied. {\hfill$\Box$}\bigskip


\setcounter{equation}{0}

\section{Appendix C: Proof of Lemma \ref{C-diff-lem}}
\label{C-diff-sect}

First it will be convenient to transform $(\bar{r}, \bar{u})$ to the time scale
of $(q, v)$. To this purpose we introduce
\begin{equation}\label{trafo}
   r_\alpha(t)=\eps^{-1}\bar{r}_\alpha(\eps^{3/2}t),\quad
   u_\alpha(t)=\sqrt{\eps}\bar{u}_\alpha(\eps^{3/2}t),
\end{equation}
where the $(\bar{r}_\alpha(t), \bar{u}_\alpha(t))$ are the solution
to the system induced by (\ref{C-dyn}) with data $(\bar{r}_\alpha^0,
\bar{u}_\alpha^0)$ from (\ref{comp-data}). Then
\begin{equation}\label{trafo-data}
   r_\alpha(0)=q_\alpha^0,\quad u_\alpha(0)=v_\alpha^0,
\end{equation}
by (\ref{comp-data}), and the corresponding equations are
\begin{equation}\label{coul-eq}
   m_\alpha\dot{u}_\alpha=\sum_{\stackrel{\beta=1}{\beta\neq\alpha}}^N
   \frac{e_\alpha e_\beta}{4\pi}
   \frac{r_\alpha-r_\beta}{|r_\alpha-r_\beta|^3},\quad\alpha=1, \ldots, N,
\end{equation}
valid for $t\in [0, (\tau_{{\rm C}}-\delta_0)\eps^{-3/2}]$.

\begin{lemma} We have
\begin{equation}\label{diff-bound-vo}
   C_0\eps^{-1}\le\inf_{t\in [0, (\tau_{{\rm C}}-\delta_0)\eps^{-3/2}]}
   |r_\alpha(t)-r_\beta(t)|,\quad
   \sup_{t\in [0, (\tau_{{\rm C}}-\delta_0)\eps^{-3/2}]}
   |r_\alpha(t)-r_\beta(t)|\le C^0\eps^{-1}\quad (\alpha\ne\beta),
\end{equation}
and
\begin{equation}\label{v-bound-vo}
   \sup_{t\in [0, (\tau_{{\rm C}}-\delta_0)\eps^{-3/2}]}|u_\alpha(t)|
   \le C\sqrt{\eps},
\end{equation}
with constants $C_0$, $C^0$, and $C>0$,
depending only on $\tau_{{\rm C}}$, $\delta_0$, and the data.
\end{lemma}
{\bf Proof\,:} The bounds in (\ref{diff-bound-vo})
follow from (\ref{trafo}) and the definition of $\tau_{{\rm C}}$.
For (\ref{v-bound-vo}), note that the system (\ref{coul-eq})
is Hamiltonian with conserved energy
\[ {\cal H}_{{\rm C}}(r, u)
   =\sum_{\alpha=1}^N\frac{1}{2}m_\alpha u_\alpha^2
   +\frac{1}{2}\sum_{\stackrel{\alpha, \beta=1}{\alpha\neq\beta}}^N
   \frac{e_\alpha e_\beta}{4\pi |r_\alpha-r_\beta|}, \]
whence
\[ \frac{1}{2}m_\alpha u_\alpha^2(t)
   \le {\cal H}_{{\rm C}}(r(0), u(0))
   -\frac{1}{2}\sum_{\stackrel{\alpha, \beta=1}{\alpha\neq\beta}}^N
   \frac{e_\alpha e_\beta}{4\pi |r_\alpha(t)-r_\beta(t)|}
   \le C\eps, \]
the latter in view of (\ref{diff-bound-vo}), (\ref{trafo-data}),
and (\ref{v-ini2}). {\hfill$\Box$}\bigskip

To prove Lemma \ref{C-diff-lem} we set
\[ C_\ast=\min\bigg\{\frac{C_1}{2}, \frac{C_0}{2}\bigg\}, \]
with $C_1$ from (\ref{q-ini2}) and $C_0$ from (\ref{diff-bound-vo}),
and we introduce
\[ \hat{t}=\sup\Big\{t_1\in [0, \min\{\tau_{{\rm C}}-\delta_0, T_0\}
   \eps^{-3/2}]: C_\ast\eps^{-1}\le \inf_{t\in [0, t_1]}
   |q_\alpha(t)-q_\beta(t)|\,\,\,\mbox{for}\,\,\,\alpha\neq\beta\Big\}. \]
Hence we need to show that in fact
$\hat{t}=\min\{\tau_{{\rm C}}-\delta_0, T_0\}\eps^{-3/2}$.
Note that $\hat{t}>0$ according to (\ref{q-ini2}), and also
\begin{equation}\label{apforl}
   C_\ast\eps^{-1}\le \inf_{t\in [0, \hat{t}\,]}|q_\alpha(t)-q_\beta(t)|,
   \quad\alpha\neq\beta.
\end{equation}
Utilizing the method of \cite[Lemma 2.1]{Nteil}, cf.~also Lemma \ref{esti},
we know that a lower bound of type (\ref{apforl}) leads to the further bounds
\begin{equation}\label{low-bound3}
   \sup_{t\in [0, \hat{t}\,]}|q_\alpha(t)-q_\beta(t)|\le C_4\eps^{-1}
   \quad (\alpha\ne\beta),\quad\sup_{t\in [0, \hat{t}\,]}|v_\alpha(t)|
   \le C_4\sqrt{\eps},
\end{equation}
with some constant $C_4>0$ depending only on $C_1$, $C_2$, $C_3$,
and $C_\ast$, but not on $\hat{t}\le C\eps^{-3/2}$.
Observe that these estimates hold from $t=0$,
and not only from $t={\cal O}(\eps^{-1})=t_0$, since we only need
to wait for this time span in case that we have to deal with expressions
resulting from data terms. This happens only when in the course of proof
some time derivative of the difference function $Z(x, t)$, cf.~(\ref{Z-def}),
is to be estimated. However, for (\ref{low-bound3}) such terms
do not appear, cf.~\cite[Sect.~5.2]{Nteil}, they first come up when
we estimate $|\ddot{v}_\alpha(t)|$. In addition, (\ref{low-bound3})
is obtained without the assumption that the $|e_\alpha|$ be small,
cf.~\cite[Sect.~5.1]{Nteil}.

These bounds can be used to derive the appropriate
lower-order effective equation for the true solution.

\begin{lemma} For $t\in [0, \hat{t}\,]$ we have
\begin{equation}\label{coul-eq-eff}
   m_\alpha\dot{v}_\alpha=\sum_{\stackrel{\beta=1}{\beta\neq\alpha}}^N
   \frac{e_\alpha e_\beta}{4\pi}
   \frac{q_\alpha-q_\beta}{|q_\alpha-q_\beta|^3}+{\cal O}(\eps^{5/2}),
   \quad\alpha=1, \ldots, N.
\end{equation}
\end{lemma}
{\bf Proof\,:} In principle we can follow the argument of Section \ref{force-sect}
and expand the Lorentz force $F_\alpha(t)$ from (\ref{self-act}),
with the difference being that we need to get a bound right from $t=0$,
and not only from $t=t_0={\cal O}(\eps^{-1})$. This means that Lemma \ref{Falph0}
cannot be used as is to be expected since for small times of order
${\cal O}(\eps^{-1})$ (before interaction) the effective Coulomb force
in (\ref{coul-eq-eff}) will be due to the initial force $F^{(0)}_\alpha (t)$.
It is only at times after $t={\cal O}(\eps^{-1})$ that the retarded
part of the field $F^{(r)}_\alpha (t)$ makes its influence felt.
However, from the viewpoint of a proof there is no sharp time
$t_\ast={\cal O}(\eps^{-1})$ at which this transition does manifest itself,
whence a separation of the force as $F_\alpha (t)=F^{(0)}_\alpha (t)
+F^{(r)}_\alpha (t)$ will not lead to the optimal bound.
Instead of this we write $F_\alpha (t)$ as a single integral as follows.
We first continue the particle trajectories and velocities to $t=-\infty$
and define
\begin{equation}\label{qfort}
   \tilde{q}_\alpha(t)=\left\{\begin{array}{c@{\quad:\quad}l}
   q_\alpha(t) & t\in [0, \infty[ \\ q_\alpha^0+tv_\alpha^0 & t\in
   ]-\infty, 0]\end{array}\right.,\quad\tilde{v}_\alpha(t)
   =\left\{\begin{array}{c@{\quad:\quad}l}v_\alpha(t) & t\in [0, \infty[
   \\ v_\alpha^0 & t\in ]-\infty, 0]\end{array}\right.;
\end{equation}
for simplicity the tilde is omitted henceforth. Using the relation
\[ \int_{-\infty}^0 ds\,e^{ik\cdot q_\beta(s)}\,\frac{\sin |k|(t-s)}{|k|}
   =\frac{e^{ik\cdot q_\beta^0}}{k^2-(k\cdot v_\beta^0)^2}
   \bigg(\cos |k|t+i(k\cdot v_\beta^0)\,\frac{\sin |k|t}{|k|}\bigg) \]
and (\ref{ini-bed}), a straightforward calculation shows
that we have the representation
\begin{eqnarray*}
   F^{(0)}_\alpha (t) & = & \sum_{\beta=1}^N e_\alpha e_\beta\,
   \int_{-\infty}^0 ds\int d^3k\,{|\hat{\varphi}(k)|}^2
   e^{-ik\cdot[q_\alpha(t)-q_\beta(s)]}\,\bigg\{-\cos|k|(t-s)\,v_\beta(s)
   +i\,\frac{\sin |k|(t-s)}{|k|}\,k \\ & & \hspace{19.7em}
   -\,i\,\frac{\sin |k|(t-s)}{|k|}\,v_\alpha(t)\wedge (k\wedge v_\beta(s))\bigg\},
\end{eqnarray*}
hence also
\begin{eqnarray*}
   F_\alpha (t) & = & \sum_{\beta=1}^N e_\alpha e_\beta\,
   \int_{-\infty}^t ds\int d^3k\,{|\hat{\varphi}(k)|}^2
   e^{-ik\cdot[q_\alpha(t)-q_\beta(s)]}\,\bigg\{-\cos|k|(t-s)\,v_\beta(s)
   +i\,\frac{\sin |k|(t-s)}{|k|}\,k \\ & & \hspace{19.7em}
   -\,i\,\frac{\sin |k|(t-s)}{|k|}\,v_\alpha(t)\wedge (k\wedge v_\beta(s))\bigg\},
\end{eqnarray*}
in view of (\ref{linny}) and (\ref{fab}). This form of $F_\alpha(t)$
is appropriate to deduce (\ref{coul-eq-eff}). The argument
proceeds entirely along the lines of Section \ref{force-sect},
but its realization is much easier, since we only have to take into account
the contributions of order ${\cal O}(\eps^2)$. To illustrate how (\ref{qfort}) enters,
we pick out a single term, e.g.
\[ A(t)=i\,\int_{-\infty}^t ds\int d^3k\,{|\hat{\varphi}(k)|}^2
   e^{-ik\cdot[q_\alpha(t)-q_\beta(s)]}\,\frac{\sin |k|(t-s)}{|k|}\,k \]
for $\alpha\neq\beta$, in order to draw a parallel to Lemma \ref{tayl}(b),
cf.~Section \ref{tayl-sect}. We are going to show that
\begin{equation}\label{Agleich}
   A(t)=i\,\int_{-\infty}^t ds\int d^3k\,{|\hat{\varphi}(k)|}^2
   \,\frac{\sin |k|(t-s)}{|k|}\,k+{\cal O}(\eps^{5/2}),\quad
   t\in [0, \hat{t}\,].
\end{equation}
To verify this, we introduce the difference
\begin{eqnarray}\label{deweih}
   D(t) & = & i\,\int_{-\infty}^t ds\int d^3k\,{|\hat{\varphi}(k)|}^2
   \bigg(e^{-ik\cdot[q_\alpha(t)-q_\beta(s)]}-1\bigg)\,\frac{\sin
   |k|(t-s)}{|k|}\,k \nonumber \\
   & = & i\,\int_0^{\infty} d\tau\int d^3k\,{|\hat{\varphi}(k)|}^2
   \bigg(e^{-ik\cdot[q_\alpha(t)-q_\beta(t-\tau)]}-1\bigg)
   \,\frac{\sin |k|\tau}{|k|}\,k \nonumber \\
   & = & -\nabla_\xi\int\int d^3x\,d^3y\,\varphi(x)\varphi(y)
   \nonumber \\ & & \hspace{4em}\times\int_0^\infty d\tau\,
   \Big\{\psi_\tau\Big(x-y+\xi_{\alpha\beta}+q_\beta(t)-q_\beta(t-\tau)\Big)
   -\psi_\tau\Big(x-y+\xi_{\alpha\beta}\Big)\Big\},\qquad
\end{eqnarray}
with $\psi_\tau(x)=(4\pi |x|)^{-1}\delta(|x|-\tau)$ and $\xi_{\alpha\beta}=q_\alpha(t)
-q_\beta(t)$, cf.~(\ref{D1-form}). Next note that $q_\alpha\in C^1(\R)$ and moreover
\begin{equation}\label{deckste}
   |v_\alpha(t)|\le C\sqrt{\eps},\quad t\in ]-\infty, \hat{t}\,],
   \quad 1\le\alpha\le N,
\end{equation}
by (\ref{qfort}), (\ref{low-bound3}), and (\ref{v-ini2}). For fixed $|x|, |y|\le R_\varphi$
and $t\in [0, \hat{t}\,]$ we define the function
\[ \theta: [0, \infty[\to\R,\quad
   \theta(\tau)=\tau-|x-y+q_\alpha(t)-q_\beta(t-\tau)|. \]
Then $\theta(0)=-|x-y+q_\alpha(t)-q_\beta(t)|=-{\cal O}(\eps^{-1})$
by (\ref{apforl}) and (\ref{low-bound3}), and in addition $\theta'(\tau)
=1-\frac{x-y+q_\alpha(t)-q_\beta(t-\tau)}{|x-y+q_\alpha(t)-q_\beta(t-\tau)|}
\cdot v_\beta(t-\tau)={\cal O}(1)$ due to (\ref{deckste}),
as $t-\tau\in ]-\infty, t]\subset ]-\infty, \hat{t}\,]$. Hence $\theta(\cdot)$
has a unique zero $\tau_0=\tau_0(x-y, t, \alpha, \beta)={\cal O}(\eps^{-1})$
in $[0, \infty[$. Setting $\tau_1=|x-y+\xi_{\alpha\beta}(t)|={\cal O}(\eps^{-1})$,
we see that (\ref{deweih}) can be rewritten as
\begin{equation}\label{4piD}
   4\pi D(t)=-\int\int d^3x\,d^3y\,\varphi(x)\varphi(y)
   \,\nabla_\xi(\tau_0^{-1}-\tau_1^{-1}).
\end{equation}
{}From the definitions we find that
\begin{eqnarray*}
   \nabla_\xi\tau_0^{-1} & = & -\tau_0^{-3}\bigg\{
   \Big(x-y+\xi_{\alpha\beta}+q_\beta(t)-q_\beta(t-\tau_0)\Big)
   \\ & & \hspace{3em}
   +\Big(x-y+\xi_{\alpha\beta}+q_\beta(t)-q_\beta(t-\tau_0)\Big)\cdot
   v_\beta(t-\tau_0)\nabla_\xi\tau_0\bigg\}, \\
   \nabla_\xi\tau_1^{-1} & = & -\tau_1^{-3}(x-y+\xi_{\alpha\beta}).
\end{eqnarray*}
Since $|\nabla_\xi\tau_0|\le C$, it follows by means of (\ref{deckste}),
and recalling $\tau_0^{-1}={\cal O}(\eps)$, that
\[ \nabla_\xi\tau_0^{-1}=-\tau_0^{-3}
   \Big(x-y+\xi_{\alpha\beta}+q_\beta(t)-q_\beta(t-\tau_0)\Big)
   +{\cal O}(\eps^{5/2}). \]
Next observe that $|q_\beta(t)-q_\beta(t-\tau_0)|\le C\sqrt{\eps}\tau_0
\le C\eps^{-1/2}$ in case that $t-\tau_0\ge 0$, due to (\ref{low-bound3}).
However, if $t-\tau_0\le 0$, then $|q_\beta(t)-q_\beta(t-\tau_0)|
=|q_\beta(t)-q_\beta^0|\le C\sqrt{\eps}t\le C\sqrt{\eps}\tau_0\le C\eps^{-1/2}$.
Thus
\[ \nabla_\xi\tau_0^{-1}=-\tau_0^{-3}(x-y+\xi_{\alpha\beta})
   +{\cal O}(\eps^{5/2}) \]
in all cases, and also
\[ |\tau_0-\tau_1|\le |q_\beta(t)-q_\beta(t-\tau_0)|\le C\eps^{-1/2}. \]
Therefore
\[ |\nabla_\xi(\tau_0^{-1}-\tau_1^{-1})|
   \le C\eps^{-1}|\tau_0^{-3}-\tau_1^{-3}|+C\eps^{5/2}
   \le C\eps^5\max\{\tau_0^2, \tau_1^2\}|\tau_0-\tau_1|+C\eps^{5/2}
   \le C\eps^{5/2}, \]
and in view of (\ref{4piD}) this completes the proof of (\ref{Agleich}).
Since it can be verified that all remaining terms can be handled
in an analogously manner, we deduce that (\ref{coul-eq-eff}) holds.
{\hfill$\Box$}\bigskip

Using (\ref{coul-eq-eff}) it is possible to complete
the proof of Lemma \ref{C-diff-lem}. In view of (\ref{coul-eq-eff}),
(\ref{coul-eq}), (\ref{diff-bound-vo}), and (\ref{apforl}), we have
\begin{eqnarray*}
   m_\alpha|\dot{v}_\alpha-\dot{u}_\alpha|
   & \le & \bigg|\sum_{\stackrel{\beta=1}{\beta\neq\alpha}}^N
   \frac{e_\alpha e_\beta}{4\pi}
   \frac{q_\alpha-q_\beta}{|q_\alpha-q_\beta|^3}
   -\sum_{\stackrel{\beta=1}{\beta\neq\alpha}}^N
   \frac{e_\alpha e_\beta}{4\pi}
   \frac{r_\alpha-r_\beta}{|r_\alpha-r_\beta|^3}\bigg|+C\eps^{5/2}
   \\ & \le & C\eps^3\sum_{\beta=1}^N |q_\beta-r_\beta|+C\eps^{5/2}
\end{eqnarray*}
for $t\in [0, \hat{t}\,]$. By the argument given in \cite[p.~449/450]{Nteil}
this yields
\begin{equation}\label{lowar}
   |q_\alpha(t)-r_\alpha(t)|\le C\eps^{5/2-3}=C\eps^{-1/2},
   \quad |v_\alpha(t)-u_\alpha(t)|\le C\eps^{5/2-3/2}=C\eps,
   \quad t\in [0, \hat{t}\,],
\end{equation}
for $\alpha=1, \ldots, N$. Consequently, by (\ref{diff-bound-vo})
and (\ref{lowar}) we obtain for $t\in [0, \hat{t}\,]$
\begin{eqnarray*}
   |q_\alpha(t)-q_\beta(t)| & \ge & |r_\alpha(t)-r_\beta(t)|
   -|q_\alpha(t)-r_\alpha(t)|-|q_\beta(t)-r_\beta(t)|
   \\ & \ge & C_0\eps^{-1}-C\eps^{-1/2}\ge (3C_0/4)\eps^{-1},
\end{eqnarray*}
the latter if $\eps>0$ is chosen small enough. Since $3C_0/4>C_\ast$,
this leads to a contradiction to the definition of $\hat{t}$
in case that $\hat{t}<\min\{\tau_{{\rm C}}-\delta_0, T_0\}\eps^{-3/2}$,
whence we must have $\hat{t}=\min\{\tau_{{\rm C}}-\delta_0, T_0\}\eps^{-3/2}$
as was to be shown. {\hfill$\Box$}\bigskip


\bigskip
\noindent {\bf Acknowledgements:} MK acknowledges support through
a Heisenberg fellowship of DFG.

\end{document}